\documentclass[mathleft
]{an}
\usepackage{graphicx}
\usepackage{times}
\begin{document}

% The following seven commands are intended for editorial usage and should be ignored by
% the author(s).
\Pagespan{789}{}% Document's page range. 
% If second parameter is left empty, the last page is computed automatically.
\Yearpublication{2006}%
\Yearsubmission{2005}%
\Month{11}%   
\Volume{999}%  
\Issue{88}% 
% \DOI{This.is/not.aDOI}% 

\title{Search for transiting exoplanets and variable stars in the open cluster NGC 7243}

\author{Z. Garai\inst{1}\fnmsep\thanks{Corresponding author:
  \email{zgarai@ta3.sk}\newline}
%Example 
%for footnote, note the usage of the \texttt{fnmsep}
%command as separator between institute number and footnote mark} 
\and T. Pribulla\inst{1}
\and \v{L}. Hamb\'{a}lek\inst{1}
\and R. Errmann\inst{2}\fnmsep\inst{3}\fnmsep\inst{11}
\and Ch. Adam\inst{2}
\and S. Buder\inst{2}
\and T. Butterley\inst{5}
\and V.S. Dhillon\inst{7}
\and B. Dincel\inst{2}
\and H. Gilbert\inst{2}
\and Ch. Ginski\inst{2}\fnmsep\inst{9}
\and L.K. Hardy\inst{7}
\and A. Kellerer\inst{4}
\and M. Kitze\inst{2}\fnmsep\inst{12}
\and E. Kundra\inst{1}
\and S.P. Littlefair\inst{7}
\and M. Mugrauer\inst{2}
\and J. Nedoro\v{s}\v{c}\'{i}k\inst{6}
\and R. Neuh\"auser\inst{2}
\and A. Pannicke\inst{2}
\and S. Raetz\inst{2}\fnmsep\inst{10}
\and J.G. Schmidt\inst{2}
\and T.O.B. Schmidt\inst{2}\fnmsep\inst{8}
\and M. Seeliger\inst{2}
\and M. Va\v{n}ko\inst{1}
\and R.W. Wilson\inst{5}
}
\titlerunning{Search for transiting exoplanets and variable stars in the open cluster NGC 7243}
\authorrunning{Z. Garai et al.}
\institute{
Astronomical Institute, Slovak Academy of Sciences, 059 60 Tatransk\'a
Lomnica, Slovak Republic
\and 
Astrophysical Institute and University Observatory,
Friedrich Schiller University, Schillergaesschen 2-3, 07745 Jena, Germany 
\and 
Abbe Center of Photonics, Friedrich Schiller University, Max-Wien-Platz 1, 07743 Jena, Germany
\and
University of Cambridge, Cavendish Laboratory, JJ Thomson Avenue, Cambridge, CB30HE, UK
\and
Dep. of Physics, Centre for Advanced Instrumentation, University of Durham, South Road, Durham DH1 3LE, UK
\and
Thuringia State Observatory in Tautenburg, 07778 Tautenburg, Germany
\and
Department of Physics and Astronomy, University of Sheffield, Sheffield S3 7RH, UK
\and
Hamburg Observatory, Gojenbergsweg 112, 21029 Hamburg, Germany
\and
Leiden Observatory, Niels Bohrweg 2, NL-2333 CA Leiden,	The Netherlands
\and
European Space Agency, Keplerlaan 1, 2200 AG, Noordwijk, The Netherlands
\and
Isaac Newton Group of Telescopes, P.O. Box 321, E-38700 Santa Cruz de La Palma, Canary Islands, Spain
\and
Institute of Physics, University of Rostock, D-18051 Rostock, Germany
} 
\received{dd mmm yyyy}
\accepted{dd mmm yyyy}
\publonline{dd mmm yyyy}


\abstract{We report results of the first five observing campaigns for the open stellar cluster NGC 7243 in the frame of project Young Exoplanet Transit Initiative (YETI). The project focuses on the monitoring of young and nearby stellar clusters, with the aim to detect young transiting exoplanets, and to study other variability phenomena on time-scales from minutes to years. After five observing campaigns and additional observations during 2013 and 2014, a clear and repeating transit-like signal was detected in the light curve of J221550.6+495611. Furthermore, we detected and analysed 37 new eclipsing binary stars in the studied region. The best fit parameters and light curves of all systems are given. Finally, we detected and analysed 26 new, presumably pulsating variable stars in the studied region. The follow-up investigation of these objects, including spectroscopic measurements of the exoplanet candidate, is currently planned.}

\maketitle

%%%%%%%%%%%%%%%%%%%%%%%%%%%%%%%%%%%%%%%%%%%%%%%%%%%%%%%%%%%%%%%%%%%%%%%%%%%%%%%%%%%%%%%%%%%%%%%%%%%%%%%%%%%%%%%%%%%%%%%%%%%%%%%%%%%%%%%%%%%%%%%%%%%%%%

\section{Introduction}

In the present decade several space missions and ground-based surveys discovered thousands of new transiting extrasolar-planet candidates: e.g., the Kepler mission (see e.g., Borucki et al. 2011; Batalha et al. 2013), the HATNet project (e.g., Bakos et al. 2007), the TrES project (e.g., Alonso et al. 2004), or the SuperWASP project (e.g., Collier Cameron et al. 2007). The transit technique (Seager \& Mall\'en-Ornelas 2003) is a very useful tool to determine the planet-to-star radius ratio, the planetary orbit inclination and the mean stellar density. In combination with the radial-velocity technique (Mayor \& Queloz 1995) we can also obtain the true mass of the planet (see e.g., Charbonneau et al. 2000). 

Observations of young transiting exoplanets are very important to test evolutionary models of planets. Such models have been described by several authors. Early works focused on the evolution of the solar-system giants (e.g., Grossman et al. 1972; Graboske et al. 1975). Saumon et al. (1996) presented a broad suite of models of extrasolar giant planets, ranging in mass from 0.3 to 15 Jupiter masses. Burrows et al. (1997) described the evolution of objects with masses from 0.3 to 200 Jupiter masses for effective temperatures below 1300 K. Case studies of evolutionary models for irradiated planets were presented e.g., by Barraffe et al. (2003) and Baraffe, Chabrier \& Barman (2008). Since planets loose memory of their initial conditions over time, these conditions were selected rather arbitrarily in the models (Marley et al. 2007). One way to set initial conditions of planets in evolutionary models more precisely is to discover young transiting exoplanets with well determined parameters, like radius, density and age. The youngest transiting exoplanets discovered so far are WASP-10b with an age up to 350 Myr (Maciejewski et al. 2011), CoRoT-Exo-2b with an age up to 500 Myr (Bouchy et al. 2008) and CoRoT-20b with an age up to 900 Myr (Deleuil et al. 2012). Recently, CoRoT-32b, which is a member of the 25-Myr old open cluster NGC 2232, was announced to be the youngest transiting exoplanet\footnote{http://corot3-kasc7.sciencesconf.org/37184/document}. The youngest exoplanets ($\sim$ 1 - 2 Myr) have been found via direct imaging (LkCa 15 b, Kraus \& Ireland 2012; a sub-stellar companion of GQ Lup, which can still be a planet, Neuh\"auser et al. 2005) and spectroscopy (a young planetary-mass object in the $\rho$ Oph cloud core; Marsh, Kirkpatrick \& Plavchan 2010).  

Several surveys have been undertaken to search for close-in young transiting exoplanets. The Monitor project (Aigrain et al. 2007) is a photometric survey of nine young (1 - 200 Myr) clusters to search for eclipses at very low mass stars, brown dwarfs and for planetary transits. So far, within the project, only Miller et al. (2008) reported six candidates observed in the NGC 2362 cluster field ($\sim$ 5 Myr), with occultation depths compatible with a planetary companion. Subsequently, the authors discussed that it is unlikely that they would have detected a planet in the cluster. The Palomar Transient Factory Orion project (van Eyken et al. 2011) aimed at searching for photometric variability in the young Orion region, with the primary goal of finding young extrasolar planets. A transit candidate (PTFO 8-8695b) around the weak-lined T Tauri star CVSO30 in the 8 Myr old cluster 25 Ori was reported by van Eyken et al. (2012) and Barnes et al. (2013).      

Since we observed NGC 7243 within the framework of the Young Exoplanet Transit Initiative (hereafter YETI), we describe this survey in more detail. Project YETI (Neuh\"auser et al. 2011) focuses on the monitoring of young ($\leq$ 100 Myr) and nearby ($\leq$ 1 kpc) stellar clusters, with the aim to detect young transiting exoplanets, and to study other variability phenomena on time-scales from minutes to years. The network uses several metre-class telescopes around the world for continuous monitoring of the clusters. For a summary of the telescope network we refer the reader to the work of Neuh\"auser et al. (2011) and Errmann et al. (2014a). The first target cluster of YETI was Trumpler 37, where a periodic (1.364894(15) days) transit-like signal ($\Delta R = $ 0.054 mag) was detected at a late F-type star. Subsequently, the follow-up observations showed that the candidate is an astrophysical false positive rather than a true planet (Errmann et al. 2014b). The 25 Ori cluster was also monitored by the project. It was the second YETI target, whose international campaign started in 2010. First results are shown in Errmann et al. (2014a). The monitoring of the clusters Trumpler 37 and 25 Ori with YETI has already finished, however, IC348, Collinder 69, NGC 1980 and NGC 7243 are currently observed and investigated by the YETI. 

In this paper we report the results of YETI's first five observing campaigns on NGC 7243. The layout of the paper is as follows. In Section 2, we briefly describe the NGC 7243 open stellar cluster. In Section 3, we present the observations and data analysis. Section 4 describes our results, including our exoplanet transit candidate, eclipsing binaries and pulsating variable stars. Our findings are discussed in Section 5.                                 

%%%%%%%%%%%%%%%%%%%%%%%%%%%%%%%%%%%%%%%%%%%%%%%%%%%%%%%%%%%%%%%%%%%%%%%%%%%%%%%%%%%%%%%%%%%%%%%%%%%%%%%%%%%%%%%%%%%%%%%%%%%%%%%%%%%%%%%%%%%%%%%%%%%%%%

\begin{table*}[t!]
\centering
\caption{The parameters of the telescopes and CCD cameras. The table gives observatory name, its geographical coordinates, telescope diameter, CCD camera type, chip size in pixels, and field of view (FoV).}
\label{Table 1}
\begin{tabular}{lllllll}
\hline
\hline
Observatory & Long. & Lat.  & Mirror $\oslash$ & CCD camera & No. of pixels & FoV\\
            & [deg] & [deg] & [cm]             &            &               & [min $\times$ min]\\
\hline
\hline
Star\'a Lesn\'a  (G1)                 & 20.28916 E  & 49.15166 N & 60 & FLI ML 3041  & 2048 $\times$ 2048 & 14 $\times$ 14\\
Star\'a Lesn\'a  (G2)                 & 20.29027 E  & 49.15222 N & 60 & MI G4-9000   & 3048 $\times$ 3048 & 17 $\times$ 17\\
University Observatory Jena (GSH)     & 11.48416 E  & 50.92888 N & 60/90 & STK$^5$   & 2048 $\times$ 2048 & 53 $\times$ 53\\
La Palma Observatory        (LPO)     & 17.88138 W  & 28.76055 N & 50 & KAF-3200 ME  & 2184 $\times$ 1472 & 10 $\times$ 7 \\
\hline
\end{tabular}
\end{table*} 

\section{The open stellar cluster NGC 7243}

The NGC 7243 open stellar cluster is located in the Lacerta constellation. It has the following center coordinates: $\alpha = 22^h15^m08^s$ and $\delta = +49^{\circ}53'51''$ (J2000.0). Trumpler (1930) was the first investigator: he determined distances and classification of 100 open clusters, including NGC 7243. In his work it is classified as class III 2p\footnote{Class III 2p: detached clusters with no noticeable concentration, in which the stars are more or less thinly but nearly uniformly scattered; medium range in the brightness of the stars; poor clusters with less than 50 stars.} and the cluster's distance is 750 pc. Based on proper motions of 814 stars in the cluster area, Lengauer (1937) identified 39 cluster members. Later, Artyukhina \& Kholopov (1966) investigated the cluster members. Based on \textit{B} versus \textit{B-V} diagram, obtained from the data of Mianes \& Daguillon (1956), as well as Becker \& Stock (1954), the authors found 86 members. Polarization measurements of NGC 7243 were published by Krzemi\'nski (1961). Later, Hill \& Barnes (1971) presented a spectroscopic and photometric investigation, where spectral types and radial velocities, as well as \textit{UBV} and \textit{uvby}, H$\beta$ photometry for most Lengauer's proper motion members brighter than 12.5 mag were included. Two stars out of 38 members were definitely excluded based on their position in the color-magnitude (hereafter CM) diagrams. A distance modulus of 9.4 $\pm$ 0.1 mag was found and a cluster age of 76 Myr was determined from the brightest star. This distance modulus was confirmed later by Hill, Fisher \& Allison (1974). Maitzen \& Pavlovski (1987) investigated 20 stars from the list of Hill \& Barnes (1971), in their search for CP2-stars. These authors found three peculiar stars, other two stars were near the detection level. In the Lyng\aa\hspace{0.1mm} (1987) catalogue NGC 7243 is classified as type II 2m\footnote{Class II 2m: detached clusters with little central concentration; medium range in the brightness of the stars; moderately rich clusters with 50 - 100 stars.} (classification scheme of Trumpler 1930) with 40 cluster members and an angular diameter of 21'.

Recently, Jilinski et al. (2003) presented the results of astrometric and photometric investigations of NGC 7243. Based on the star positions on the cluster CM diagram authors identified 209 members down to $V \sim 15.5$ mag. Subsequently, two bright stars were added to the list of members according to the results of previous determinations of their radial velocities and photoelectric \textit{UBV} magnitudes. The cluster age ($\sim$ 250 Myr) was estimated based on the $(B-V)_0$ value of the turn-off point. The total cluster mass was estimated to 348 - 522 $M_\odot$. The magnitude of the brightest star is $V \sim 8.43$ mag. The distance of the cluster was estimated to about 698 pc ($V_0-M_v=9.22$ mag). 

The key requirement for the successful realization of the global aim of YETI, described in Section 1, is the target selection for observations. It follows several criteria described in Neuh\"auser et al. (2011). Based on these criteria the open cluster NGC 7243 was chosen in 2013. The cluster with its 698 pc distance, 21' angular diameter and 211 cluster members is an ideal YETI target, observable from most YETI observatories. On the other hand, compared to Trumpler 37, also observed within YETI (Errmann et al. 2014b), we can expect less discoveries on NGC 7243 since it has fewer members (614 versus 211). The estimate of the number of expected transiting planets $N_p$ can be parameterized as (Neuh\"auser et al. 2011):
\begin{equation}
\label{equation1}
N_p=N_{\ast}~f_p~\rho_t~\rho_{eff},
\end{equation}
where $N_{\ast}$ is the number of cluster members included in the transit search, $f_p$ is the fraction of stars with close-in planets within 0.1 AU ($\sim$ 0.012; Butler et al. 2006), $\rho_t$ is the probability to view the orbit nearly edge-on ($\sim$ 0.1 for close-in planets; Neuh\"auser et al. 2011) and $\rho_{eff}$ is a measure of the efficiency of the observation ($\sim$ 0.7 - 1.0; Neuh\"auser et al. 2011). For $N_{\ast}=211$ and $\rho_{eff}=1.0$, the number of expected transiting planets is $N_p=0.25$. This means that in the best case we can expect discovery of one cluster member planet.              

%%%%%%%%%%%%%%%%%%%%%%%%%%%%%%%%%%%%%%%%%%%%%%%%%%%%%%%%%%%%%%%%%%%%%%%%%%%%%%%%%%%%%%%%%%%%%%%%%%%%%%%%%%%%%%%%%%%%%%%%%%%%%%%%%%%%%%%%%%%%%%%%%%%%%%

\section{Observations and data analysis}
\label{analysis}

The monitoring of NGC 7243 started in summer 2013 with two observing campaigns. Additional data were obtained also before and after the campaigns. The observations were performed at the Star\'a Lesn\'a Observatory (G1 and G2 pavilions of the Astronomical Institute of the Slovak Academy of Sciences) by two 60cm, f/12.5 Cassegrain telescopes\footnote{http://www.astro.sk/l3.php?p3=slo}, at the University Observatory Jena in Gro\ss schwabhausen (hereafter GSH) with a 60/90cm, f/3 Schmidt telescope\footnote{http://www.astro.uni-jena.de/index.php/gsh-home.html; for STK see Mugrauer \& Berthold (2010)} and on the La Palma Observatory (hereafter LPO) with a robotic 50cm, f/10 Ritchey-Chr\'etien telescope\footnote{https://sites.google.com/site/point5metre/home} operated by the Universities of Durham and Sheffield. The parameters of the telescopes and CCD cameras are listed in Table 1. 

\begin{table}[b!]
\centering
\caption{Observation log of NGC 7243 campaigns. Number of observing nights, sorted by observatories and campaigns, is given in the 4th column.}
\label{Table 2}
\begin{tabular}{llll}
\hline
\hline
Observatory & Camp.    & Camp.	 & Obs.\\
name        & number   & date    & nights\\
\hline
\hline
G1          & 1st & 2013.08.03 - 2013.08.14 & 8\\
G1          & 2nd & 2013.08.31 - 2013.09.12 & 6\\
G1          & 3rd & 2014.09.02 - 2014.09.14 & 4\\
G1          & 5th & 2014.10.10 - 2014.10.19 & 2\\
G1          & --  & other observations      & 9\\
\hline
G1          & --  & total observations      & 29\\
\hline
G2          & 4th & 2014.09.19 - 2014.09.26 & 3\\
G2          & 5th & 2014.10.10 - 2014.10.19 & 1\\
G2          & --  & other observations      & 18\\
\hline
G2          & --  & total observations      & 22\\
\hline
GSH	    & 1st & 2013.08.03 - 2013.08.14 & 4\\
GSH         & 2nd & 2013.08.31 - 2013.09.12 & 5\\
GSH         & --  & other observations      & 17\\
\hline
GSH         & --  & total observations      & 26\\
\hline
LPO	    & 1st & 2013.08.03 - 2013.08.14 & 12\\
LPO         & 2nd & 2013.08.31 - 2013.09.12 & 7\\
\hline
LPO         & --  & total observations      & 19\\
\hline
\end{tabular}
\end{table} 

In 2013, we obtained data over 47 nights. The first observing night was July 22 and the last was December 21. Because in a few cases we detected only one eclipse/transit, we observed NGC 7243 also in year 2014 during three additional campaigns, but also before and after the campaigns. In 2014, we obtained data over 26 nights. The first observing night was July 18 and the last was December 30. To extend the magnitude range we alternated short and long exposures. The CCD observations always used the Bessell \textit{R} filter (Bessell 1990). The telescopes were pointed to the star at J221508.0+495354. The observations log is given in Table 2.

The data were reduced using IRAF\footnote{http://iraf.noao.edu} as described in Errmann et al. (2014b). The first step was the dark and flat correction (and an overscan correction in the case of the data obtained from Jena). The pixel coordinates of the stars in the frames were found using Source Extractor -- SExtractor\footnote{http://www.astromatic.net/software/sextractor} (Bertin \& Arnouts 1996). The final list contained 1020, 1819, 317 and 19461 stars for the G1, G2, LPO and GSH observatory, respectively. The tracking offsets were removed by extracting the star positions in each image with SExtractor, and by comparing that list to the reference list of stars. The optimal size of the aperture, used for the photometric measurements, was determined separately for each night to account for changes in seeing and telescope focus. Aperture photometry was then performed with the optimized aperture, followed by differential photometry as described by Broeg, Fern\'andez \& Neuh\"auser (2005). All stars in the field were used to create an artificial standard star, but the more photometrically stable stars were assigned higher weights. This artificial standard star was used to calculate the differential magnitudes of all objects.

Due to different number of stars included in night-to-night FoV (telescope pointing) and used to create the artificial standard star, extracted light curves for individual objects showed night-to-night magnitude offsets within certain telescope data, as well. These offsets were determined using a bright and photometrically stable star near the center of telescope FoV (star J221508.1+495603 for G1, G2, LPO and short cadence GSH data, and star J221514.3+500421 for long cadence GSH data). After the proper shifting of the light curves, we used a 5$\sigma$ clipping to clean the outliers from the data. The 5$\sigma$ clipping was selected according to effective cleaning as well as in order to avoid cleaning eclipse events. In the next step we visually inspected the data and selected the objects showing variable photometry. We mainly focused on the search for exoplanet transits-like signals, eclipsing binaries and pulsating variables. We further analyzed these objects as follows. 

We first fitted exoplanet transit-like signals using the IDL-based Transit Analysis Package -- TAP\footnote{http://ifa.hawaii.edu/users/zgazak/IfA/TAP.html} (Gazak et al. 2012) to obtain initial parameters. The software uses Markov Chain Monte Carlo (MCMC) techniques to explore the fitting parameter space and it is based on the analytic transit light-curve models of Mandel \& Agol (2002). The package incorporates white- and red-noise parametrisation (Carter \& Winn 2009) allowing for robust estimates of parameter uncertainty distributions. First, the optimal ephemeris ($P_{\rm orb}$ and $T_c$) of the exoplanet candidate was found by fitting a trigonometric polynomial to all the data. In this way we excluded a lot of period aliases. Subsequently, using TAP we determined preliminary parameters by fitting the transits simultaneously. In this step only light curves when whole transit was covered were included. Remaining outliers were cleaned "manually". These parameters were the orbital period ($P_{\rm orb}$), the inclination angle ($i$), the relative semi-major axis ($a/R_s$), the ratio of the radii ($R_p/R_s$) and the mid-transit time ($T_c$). The quadratic limb darkening coefficients were first adjusted during the fitting procedure. The resulting coefficients were found far from the theoretical predictions for an F-type star (see Subsection 4.1). Hence we adopted the coefficients 0.2689 and 0.3225 calculated for the \textit{R} filter, $T_{eff}=6500$ K, solar metallicity and $\log g$, using the on-line applet EXOFAST - limbdark\footnote{http://astroutils.astronomy.ohio-state.edu/exofast/}. This software interpolates the Claret \& Bloemen (2011) quadradic limb darkening tables. We then used the software JKTEBOP\footnote{http://www.astro.keele.ac.uk/jkt/codes/jktebop.html} (Southworth, Maxted \& Smalley 2004) for the final fitting. The code is used to fit a model to the light curves of detached eclipsing binary stars in order to derive the radii of the stars as well as various other quantities. It is very stable and has a lot of additional goodies, including extensive Monte Carlo or bootstrapping error analysis algorithms. It is also excellent for transiting extrasolar planetary systems. Using JKTEBOP we simultaneously fitted the transits from all nights and telescopes including incomplete transits. Light-curve magnitude offsets among different telescopes were corrected adjusting the light-scale factor parameter only. Subsequently, six free parameters were adjusted during the fitting procedure, including the orbital period ($P_{\rm orb}$), the mid-transit time ($T_c$), the inclination angle ($i$), the ratio of the radii ($R_p/R_s$) and the sum of fractional radii ($(R_p+R_s)/a$). The best TAP-fit parameter values were used here as initial parameters. The initial value for the sum of fractional radii ($(R_p+R_s)/a$) was calculated from the parameters $a/R_s$ and $R_p/R_s$. The quadratic limb darkening coefficients were fixed during the fitting procedure as in TAP. To estimate the uncetranities in parameters we used JKTEBOP-task No. 8, which executes Monte-Carlo simulations. To obtain final results we applied $10~000$ iteration steps.  

In the case of the eclipsing binaries we used the code Roche (Pribulla 2004; 2012), which is devoted to the modeling of multi-dataset observations of close eclipsing binaries such as radial velocities, multi-wavelength light curves and broadening functions. The code is based on Roche geometry and includes circular surface spots, eccentric orbits, asynchronous rotation and third light. In the case of asynchronous rotation and eccentric orbits the assumptions of Wilson (1979) are used (Pribulla 2012). The routines of Roche are, however, completely independent of the Wilson \& Devinney code (see Wilson 1994 and references therein). The Roche code has been successfully used to analyze observations of several eclipsing systems (e.g., Pribulla et al. 2008, 2011; Zasche, Svoboda \& Uhl\'a\v{r} 2012). First, as in the case of the exoplanet candidates, the optimal ephemeris ($P_{\rm orb}$ and $T_0$) was found by fitting a trigonometric polynomial to the data. The light curve of each star per telescope was then phase-folded with the best period and inspected. Light-curve magnitude offsets among different telescopes were corrected as follows. The phased light curves of each star were spread into 200 bins. The average brightness was determined for each bin and the data points were then shifted according to the average differences. After the proper shifting the light curves, we repeated the period search on the data from all telescopes. Finally, the data were phased with this new orbital period and remaining outliers were cleaned "manually". Several eclipsing binary stars were analyzed only using the GSH data, because of the wider FoV of the Schmidt telescope.

The light-curve analysis was limited by lack of spectroscopic observations and single filter used for the photometry. Thus it was impossible to determine temperatures of the components, their mass ratio and the size of the orbit. Because the light-curve shape primarily depends on the temperature ratio and not on the absolute values of the component's temperatures, for all objects the primary temperature was set to 6000 K. Hence, convective envelopes were assumed. The mass ratio affects the shape of light curves of detached binaries only slightly and can reliably be determined for totally-eclipsing contact binaries without third light only (see Mochnacki \& Doughty 1972; Hamb\'alek \& Pribulla 2005). Therefore, the mass ratio was fixed for detached systems to 0.9 (see Paragraph 4.2.1).  For contact binaries with partial eclipses the amplitude of the light-curve changes provides low mass-ratio estimate (Hamb\'alek \& Pribulla 2005). To assess the effect of the mass ratio, it was fixed to several values for $q < 1$ with the 0.05 step. When insignificant variation of $\chi^2$ with the mass ratio was found (typically partially-eclipsing systems), the parameters were listed for $q$ = 0.50 (see Paragraph 4.2.2).

Usually six free parameters were adjusted during the Roche fitting procedure: the orbital period ($P_{\rm orb}$), the mid-eclipse time ($T_0$), the inclination angle ($i$), the surface equipotencials ($\Omega_1,_2$), the mass ratio ($q$) and the effective temperature of the secondary component ($T_2$). The mass ratio and the surface equipotentials were then used to compute the fractional volume radii ($r_1, r_2$) for detached systems and the fill-out factor $f$ for contact binaries. Several systems show clear light-curve asymmetry -- the O'Connell effect (Milone 1968). We can explain this feature assuming a spot on the primary component at about 90 deg of longitude (1), and/or assuming a spot on the secondary component at about 270 deg of longitude (2), if we consider dark stellar spots. In the case of bright stellar spots, a spot on the primary component is at about 270 deg of longitude (3), and/or a spot on the secondary component is at about 90 deg of longitude (4). Having photometry in single passband, the determination of the spot parameters (latitude, longitude, spot temperature factor) is rather cursory and serves only to fit the data. We modelled these $R$ passband light curves assuming the first alternative -- a single spot on stellar equator with temperature factor $k$ = 0.7 (the ratio of the spot and photoshere temperature, adopted from a typical solar value). Using larger temperature factor provides fit of practically the same goodness, but results in larger spot radius.

To extract individual frequencies of the pulsating stars we have used the code Period04\footnote{https://www.univie.ac.at/tops/Period04/} by Lenz \& Breger (2005). The code handles the statistical analysis of large astronomical time series containing gaps.  The period analysis is done by a standard discrete Fourier transform algorithm producing a power spectrum. Then frequencies with highest amplitudes were selected and investigated. The highest frequency peak is then multiplied/convolved with a spectral window and subtracted. A new period analysis is performed on these residuals to find another high frequency. This procedure was repeated up until all significant frequencies were found. Empirical results from observational analyzes by Breger et al. (1993) and numerical simulations from Kuschnig et al. (1997) have shown that the ratio between signal and noise (SNR) in amplitude should not be lower than 4.0 for high significance. However, our data have many large gaps which generates red noise and lowers the amplitudes of significant frequencies while reallocates their power into large spectral window. To limit the number of significant frequencies, we have applied a moderate criterion of SNR$~>2.0$. The noise value was calculated from a small interval around the frequency we were investigating. The remaining frequencies are used in fitting the dataset with a Levenberg-Marquardt non-linear least-squares fitting procedure (Bevington 1969). The method produces also an error matrix for frequencies, amplitudes and phases of found sinusoidal signals. Afterwards a Monte-Carlo method was used to refine the statistical errors for all fit parameters.
 
\begin{figure}[t!]
\centering
\includegraphics[width=80mm]{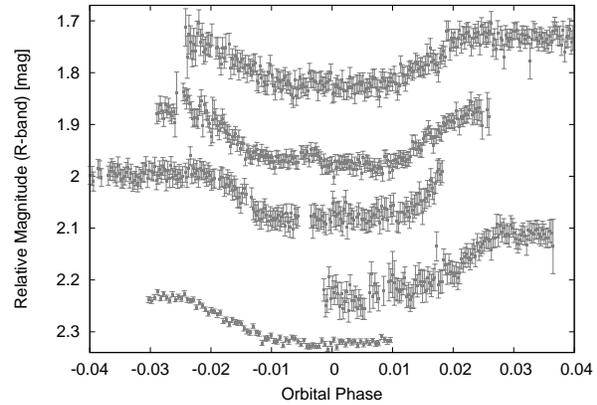}
\caption{The transit-like signals of J221550.6+495611 detected in the $R$ passband. From top to bottom there are the following detections: August 6, 2013 (G1); July 18, 2014 (G1); October 18, 2014 (G1); October 5, 2014 (G2) and October 2, 2013 (GSH). Each event is phased with the orbital period $P_{\rm orb}=4.38023$ days, mid-transit time $T_c=2456568.3506$ HJD and arbitrarily shifted vertically for clarity.}
\label{Fig. 1}
\end{figure}
  
\begin{table}[b!]
\centering
\caption{An overview of the exoplanet transit candidate best fit parameters. Fixed parameters are listed without an error. A comparison between TAP and JKTEBOP analysis.}
\label{Table 3}
\begin{tabular}{lll}
\hline
\hline
Parameter & TAP   & JKTEBOP\\
          & value & value\\
\hline
\hline
$P_{\rm orb}$ [days]        & 4.3801(2)      	& 4.380383(12)\\
$i$ [deg]                   & 88.2(2)        	& 87.7(4)\\
$a/R_s$                     & 8.3(4)         	& --\\
$(R_p+R_s)/a$		    & --	       	& 0.144(3)\\	
$R_p/R_s$                   & 0.263(5)       	& 0.262(2)\\
$T_c$ [2456000+ HJD]        & 511.4092(13) 	& 511.4145(7)\\
LD linear coeff.	    & 0.2689		& 0.2689\\		
LD non-linear coeff.	    & 0.3225		& 0.3225\\
Light-scale ft. [mag]       & --	       	& -0.0036(9)\\
\hline
\end{tabular}
\end{table} 

%%%%%%%%%%%%%%%%%%%%%%%%%%%%%%%%%%%%%%%%%%%%%%%%%%%%%%%%%%%%%%%%%%%%%%%%%%%%%%%%%%%%%%%%%%%%%%%%%%%%%%%%%%%%%%%%%%%%%%%%%%%%%%%%%%%%%%%%%%%%%%%%%%%%%%

\begin{figure*}[t!]
\centering
\centerline{
\includegraphics[width=80mm]{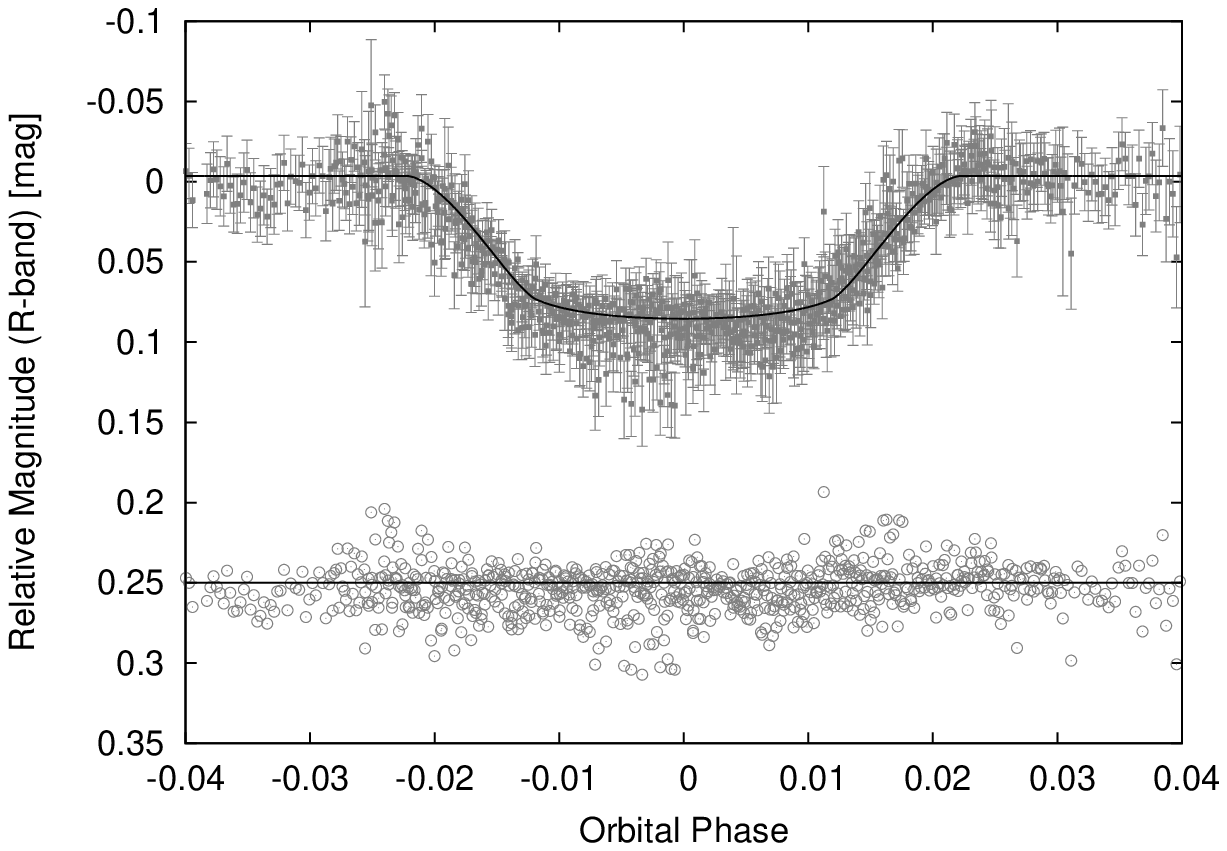}
\includegraphics[width=80mm]{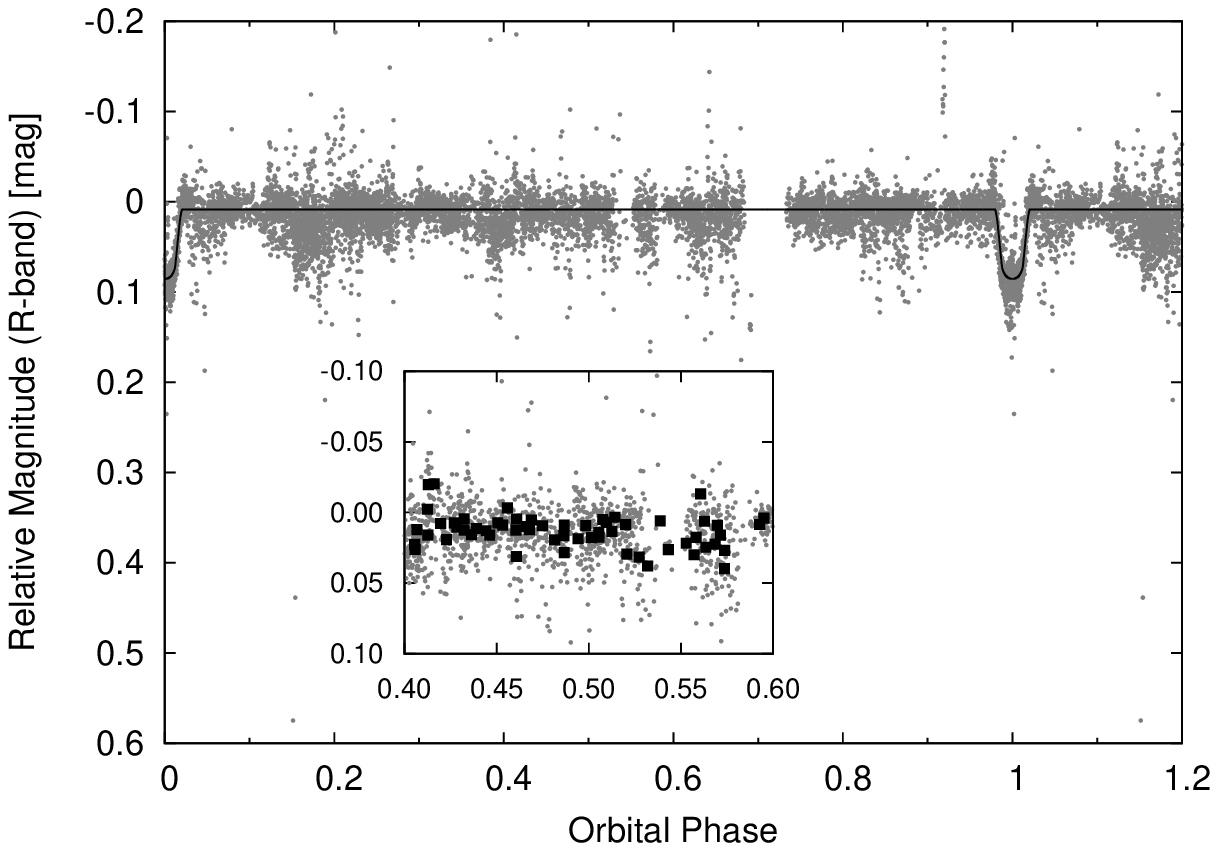}
}
\caption{Left: The phase-folded transit light curve of the exoplanet candidate in the $R$ passband for all nights and telescopes combined (top). The solid black line is our best JKTEBOP fit model described in Subsection 4.1. Corresponding residuals are also shown (bottom). Right: The full phase-folded light curve of the exoplanet transit candidate in the $R$ passband for all nights and telescopes combined. The solid black line is our best JKTEBOP fit model described in Subsection 4.1. The inset shows the phase-folded and binned light curve from 0.4 to 0.6 in phase. The transit is seen at phase $0=1$ and there are no additional eclipses seen.}
\label{Fig. 2}
\end{figure*}

\begin{figure}
\centering
\includegraphics[width=80mm]{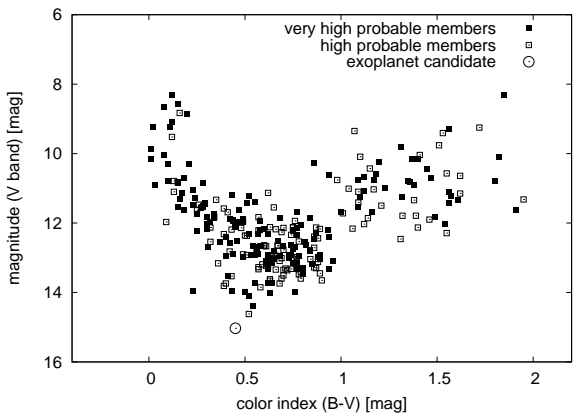}
\caption{CM diagram for probable cluster members with very high (99 - 96 \% -- filled squares) and high (95 - 91 \% -- open squares) membership probabilities. The open circle marks the position of the parent star candidate.}
\label{Fig. 3}
\end{figure}

\section{Results}

\subsection{The exoplanet candidate J221550.6+495611}
\label{candidate}
A clear transit-like signal was detected first on August 6, 2013 at G1. Four more detections were registered later: July 18, 2014 and October 18, 2014 at G1, October 5, 2014 at G2 and October 2, 2013 at GSH. Only the first three observations covered a complete transit event (Fig. 1). We found that the transit-like signal repeats about every 4.380 days and has a depth of about 0.08 mag in the $R$ passband. Subsequently, we fitted the transit light curves as described in Section~\ref{analysis}. The best fit parameters are summarized in Table 3 and the fitted light curve is shown in Fig 2. The transit light curve in Fig 2 (left) has a U-shape, which is typical for exoplanetary transits, however the transit depth, 0.08 mag in the $R$ passband, and the corresponding value of $R_p/R_s \sim 0.2$, is unusually large for an extrasolar planet. One of the deepest transit has WASP-10b, a 3 Jupiter-mass gas-giant planet transiting a late-type K star: about 0.03 mag (Christian et al. 2009). 

Subsequently, we investigated the parent star of the exoplanet candidate from the viewpoint of color information and the effective star temperature. For this purpose we observed J221550.6+495611 in the $B$ and $V$ passbands at G2 on January 28, 2015. We obtained 15 frames in each filter. The data were reduced using IRAF. The reduction included dark and flat correction, astrometric calibration and differential photometry. As a comparison star (J221547.3+495737; $B=14.67$ mag; $V=13.96$ mag; $B-V=0.71$ mag) and a check star (J221547.1+495512; $B=14.25$ mag; $V=13.45$ mag; $B-V=0.80$ mag) were selected the closest photometrically stable stars of similar brightness. After the transformation to the international photometric system we got for the parent star candidate the following values: $B=15.48(2)$ mag; $V=15.03(1)$ mag; $B-V=0.45(2)$ mag. We further calculated the corresponding effective star temperature ($T_{eff}\sim6500$ K), based on an on-line application\footnote{http://www.uni.edu/morgans/stars/b\_v.html}, which uses relations from Flower (1996) and Torres (2010). The effective temperature $\sim6500$ K indicates an F-spectral-type star.

Finally, we investigated the membership probability of the parent-star candidate. The star position on the cluster CM diagram was used as a photometric criterion of the cluster membership. For this purpose we used the membership catalogue compiled by Jilinski et al. (2003). The complete catalogue contains 2623 stars in the region of NGC 7243, listing their positions, proper motions, $B$ magnitudes, $B-V$ color indices and membership probabilities. From this catalogue we selected stars with very high (99 - 96 \%) and high (95 - 91 \%) membership probabilities. The $B-V$ versus $V$ CM diagram is shown in Fig. 3. Subsequently, using the $B-V$ and $V$ values, described in the previous paragraph, we re-plotted the parent star candidate onto the same CM diagram. We can see the marginal position of our target object in the CM diagram. This indicates that this star is possibly a field star only, but according to the other outliers, which are visible in the left side of the diagram, it can be also a cluster member.

The confirmation process of the planet candidate, which is currently planned, would require direct imaging, multi-passband photometry and high-dispersion spectroscopy (see Subsection 5.1).    

\subsection{Eclipsing binary stars}

According to the General Catalogue of Variable Stars\footnote{http://www.sai.msu.su/gcvs/gcvs/} (Samus et al. 2012; hereafter GCVS) there are two known Algol-type variables in our studied region: TZ Lac (J221412.3+495618) and XX Lac (J221710.2+500111). Apart from that, in the studied region there is one more Algol-type eclipsing binary presented in the International Variable Star Index\footnote{http://www.aavso.org/vsx/index.php} catalogue (Watson, Henden \& Price 2014; hereafter VSX): NSVS6038647 (J221230.9+500233), which is also described in the Northern Sky Variability Survey\footnote{http://skydot.lanl.gov/nsvs/nsvs.php}. Our observations led to discovery of 37 new eclipsing binary stars in the studied region. The $R$ passband magnitude range of the stars is from $17.8$ mag to $11.7$ mag. The light-curve analysis was performed using the Roche code as described in Section~\ref{analysis}. In the following two Paragraphs 4.2.1 and 4.2.2 we describe the light-curve analysis of selected binaries only. Light-curve solutions for all systems are shown in Appendix A and B.

\subsubsection{Detached eclipsing binaries}

As discussed in Section~\ref{analysis}, for detached eclipsing binary stars the mass ratio does not affect the light-curve shape significantly and only fractional radii of the components can be determined (in favourable cases). When both minima are visible, the luminosities of the componets and their masses must be similar. Hence, $q$ = 0.90 was adopted for most systems. Only in the case of J221200.9+494442, J221428.8+495509 and J221514.4+500059, $\chi^2$ was found to significantly depend on the mass ratio. Resulting photometric elements for all detached systems are given in Table~\ref{detached}.

J221515.2+495017 is a close but detached eclipsing binary with 0.33 mag deep minima in the \textit{R} passband. For the best orbital period, about $P_{\rm orb} = 2.6597$ days, the secondary minimum is not visible (Fig. A2 - bottom left). We can explain this fact in several ways. (1) The orbit is eccentric and the secondary eclipses do not occur. (2) The orbit is circular and the secondary component has a very low temperature, and therefore secondary eclipses are too shallow. (3) The true orbital period is two times larger and the minima have the same depth. Unfortunately, the phase interval $<0.4 - 0.6>$ for the double period was not covered with observations. The $R$ passband light curve was modelled assuming the first alternative. It was found that the secondary eclipse stops to occur for eccentricities higher than about 0.38. The best fit, however, follows the data rather poorly due to the proximity effects predicted by the model but not visible in the light curve. The resulting parameters for the fixed eccentricity of $e=0.4$ are listed in Table~\ref{detached}. The second alternative is not very probable: the secondary component would have to be very cold for the secondary minima to be absent, and at the same time large ($\Omega_2 < \Omega_1$). The third alternative is the most probable but requires more observations to cover the missing phases.

J221227.5+501309 ($P$ = 0.564323 days; Fig. A1 - middle top right) and J221536+495726 ($P$ = 0.508895 days; Fig. A3 - top left) are two short-period systems and at the same time detached binary stars. This means that their components must be low-mass stars of late spectral type. In spite of that, no significant light-curve asymmetry caused by surface activity (usual in such systems) is visible in our data. Their further analysis would, however, require multi-passband photometry and high-dispersion spectroscopy. 

\begin{table*}[t]
\centering
\caption{Photometric elements of newly-detected detached eclipsing binary stars, sorted by RA. A dark spot on the primary component was assumed in the case of objects marked with a $\dag$. Total eclipse is probable in systems marked with a $\star$. Fixed parameters are listed without an error. $R_{\rm max}$ is the maximal brightness of the star in the $R$ passband. 
The table lists the following best fit parameters: the orbital period ($P_{\rm orb}$), the mid-eclipse time ($T_0$), the inclination angle ($i$), the volume fractional radii $r_1, r_2$, and the effective temperature of the secondary component ($T_2$). }
\label{detached}
\begin{tabular}{llllllll}
\hline
\hline	
Identifier 				& $R_{\rm max}$	& $P_{\rm orb}$    	& $T_0$  	& $i$   	& $r_1$   	& $r_2$ 	& $T_2$		\\
           				& [mag]		& [days] 	& [2456000+ HJD]  	& [deg] 	&              	&            	& [K]		\\
\hline
\hline
J221200.9+494442${\star}$         	& 14.8		& 1.1528557(14) & 545.02361(2)		& 79.94(3)  	& 0.47917(10)	& 0.27208(11)  	& 4528(4)	\\ %S2580  %20150727
%T_1 = 7000 K too cold secondary, q tested between 0.25 - 0.40, fixed at 0.30
J221201.8+501925 			& 16.4		& 0.5523058(13)	 & 545.20391(6)		& 67.34(15)	& 0.3766(15)    & 0.3670(38)	& 4854(13)	\\ %S8038  %20150727
%T_1 = 7000 K, q cannot be determined, secondary on the critical surface
J221202.9+500046 			& 15.0		& 0.4709892(4)   & 544.80441(2)		& 62.74(2)  	& 0.3758(2)	& 0.3712	& 4247(5)	\\ %S2895  %20150727
%T_1 = 6000 K, q cannot be determined, secondary on the critical surface
J221227.5+501309 			& 17.1		& 0.5643230(14)	 & 544.93811(4)		& 79.76(10)	& 0.241(6)	& 0.278(6)	& 5414(7)	\\ %12003  %20150727
%T_1 = 6000 K, q cannot be determined, well detached
J221230.4+500215 			& 11.7		& 1.2479213(22)	& 545.22260(3)		& 68.16(5)	& 0.2610(10)	& 0.2954(15)	& 4851(7)	\\ %ST137  %20150727
%T_1 = 8000 K, chi^2 slightluy improves with decreasing q, q=0.90 used anyway
J221249.9+500443${\dag}$ 		& 14.8		& 0.5799668(7)	& 545.07797(4)		& 56.88(4)	& 0.3749(7)	& 0.35826(11)	& 5241(7)	\\ %S2547  %20150815
%T_1 = 6000 K, q cannot be determined, chi^2 deteriorates with decreasing q
J221325.1+495445${\star}$ 		& 16.4		& 1.653228(6)	& 545.69308(9) 		& 88.9(2)	& 0.3515(3)	& 0.2038(2)	& 5805(4)	\\ %S7989  %20150727
%T_1 = 6000 K, q=0.75 gives minimum of chi^2, not very significant, total eclipses
J221339.2+494101 			& 14.2		& 1.554524(12)  & 544.02713(20 		& 75.58(5)	& 0.164(3)	& 0.1458(20)	& 5007(10)	\\ %S1617  %20150727 divergent
%T_1 = 6000 K, q = 0.90 fixed, minima not covered sufficiently, fit not reliable
J221426.1+493813 			& 15.3		& 1.2438502(22)	& 544.66881(4)		& 77.88(13)	& 0.3388(4)	& 0.1705(11)	& 3742(4)	\\ %S3694  %20150727
%T_1 = 6000 K, q=0.15 is the best, chi^2 improves from 2.976 (q=0.9) to 2.652 (q=0.15)
J221428.8+495509 			& 16.4		& 1.5376868(13)	& 688.66565(16)		& 88.4(3)	& 0.1294(13)	& 0.0639(9)	& 4450(40)	\\ %G1582  %20150727
%T_1 = 6000 K, q=0.90, q cannot be determined, very noisy LC
J221447.3+493352${\dag}$ 		& 12.9		& 0.479141(7)   & 541.72580(8)		& 50.2(9)	& 0.3856(3)	& 0.3653(15)	& 3891(9)	\\ %ST576  %20150727
%T_1 = 6000 K, spotted, q cannot be determined, q=0.90
J221455.4+495602 			& 17.0		& 0.8588607(8)  & 642.74768(16) 	& 73.2(5)	& 0.270(5)	& 0.343(10)	& 4030(33)	\\ %G1813  %20150727
%T_1 = 6000 K, very noisy LC, q cannot be determined
J221512.2+494334 			& 16.1		& 1.416057(4)   & 550.06289(7)		& 76.68(8)	& 0.1908(8)     & 0.3002(12)	& 3977(5)	\\ %S6376  %20150727
%T_1 = 6000 K, noisy LC not fully covered, mutually shifted phase segments, very poor model
J221514.4+500059 			& 14.4		& 1.2417858(7)	& 651.41040(11)		& 69.86(8)	& 0.2283(19)	& 0.1866(21)	& 4001(13)	\\ %G1155  %20150727
%T_1 = 6000 K, noisy LC, chi^2 improves with decreasing q, solution for q=0.35
J221515.2+495017     			& 12.8		& 2.659727(6) 	& 695.5718(4) 		& 74.19(8) 	& 0.1735(3)	& 0.1468(3)   	& 6423(20)	\\ %G1061  %20150727
%eccentric system, parameters not reliable at all !!!, just limiting case when eclipse occurs
J221526.0+494930 			& 15.4		& 0.53535518(18)& 693.38450(6)		& 74.16(16)	& 0.3805(15)	& 0.3673(29)	& 3874(16)	\\ %G1308  %20150727
%Secondary on the critical surface, q = 0.90 used
J221536.0+495726 			& 14.8		& 0.50889522(5) & 675.95920(3)		& 89.2(7)	& 0.2798(5)	& 0.2990(4)	& 5779(2)	\\ %G1195  %20150727
%T_1 = 6000 K, q=0.90 is best, chi^2 slowly increasing going go smaller q
J221539.2+494749 			& 16.2		& 1.4663885(9)	& 689.08219(13)		& 86.71(18)	& 0.1779(8)	& 0.1117(8)	& 4444(16)	\\ %G1514  %20150727
% very noisy data, q = 0.90 is the best, chi^2 slowly increases with decreasing q
J221629.9+501313 			& 13.9		& 1.6722987(16)	& 545.92461(4)		& 77.44(3)	& 0.2285(6)	& 0.1821(9)	& 5952(2)	\\ %S1292  %20150727
% chi62 decreases just very slightly witht the mass ratio, q = 0.90 used for the above elements
J221633.9+495228 			& 16.4		& 0.6109303(8)  & 544.44120(3)		& 80.99(4)	& 0.3156(8)	& 0.3711(13)	& 5713(3)	\\ %S7932  %20150727
% looks like SD system, for q=0.90, secondary is filling its critical surface, q cannot reliably be determined 
J221656.8+501735${\dag}$ 		& 16.4		& 0.28799121(22) & 544.37081(2)		& 68.50(5)	& 0.3880(5)	& 0.3444(13)	& 5771(4)	\\ %S7697  %20150727
% insignificant improvement of chi^2 with decreasing mass ratio, primary almost filling its roche lobe
J221711.8+495948${\dag}$ 		& 15.0		& 0.30566844(17) & 544.58412(1)		& 61.52(2)	& 0.3643(6)	& 0.3694(3)     & 5441(2)	\\ %S2908  %20150727
% chi^2 deteriorates with decreasing q, q = 0.90 is rather probable, secondary close to critical surface
\hline
\end{tabular}                                                                
\end{table*} 

\begin{table*}[t]
\centering
\caption{Photometric elements of newly-detected contact eclipsing binary stars, sorted by RA. A dark spot on the primary component was assumed in the case of objects marked with a $\dag$. Total eclipse is probable in systems marked with a $\star$. Fixed parameters are listed without an error. $R_{\rm max}$ is the maximal brightness of the star in the $R$ passband.  
The table lists the following best fit parameters: the orbital period ($P_{\rm orb}$), the mid-eclipse time ($T_0$), the inclination angle ($i$), the fill-out factor ($f$, $0 < f < 1$), the mass ratio ($q$, $q \leq 1$), and the effective temperature of the secondary component ($T_2$). }
\label{contact}
\begin{tabular}{llllllll}
\hline
\hline	
Identifier 				& $R_{\rm max}$	& $P$    	& $T_0$  		& $i$   	& $f$   	& $q$       	& $T_2$	     \\
           				& [mag]		& [days] 	& [2456000+ HJD]  	& [deg] 	&              	&            	& [K]	     \\
\hline
\hline
J221205.2+501000 			& 15.9		& 0.37524686(26) & 544.93080(2)		& 73.02(3)	& 0.0995(22)	& 0.50   	& 5978(3)    \\ %S5685  20150728 Omega = 2.84614(64)
%q cannot be determined
J221217.3+495829${\dag}$${\star}$ 	& 15.5		& 0.3981976(3)	& 544.86808(2)		& 78.23(6)	& 0.250(12)     & 0.1929(4)     & 6128(3)    \\ %S4308  20150728 Omega = 2.18415(110)
%q well defined - total eclipses
J221232.5+501441${\dag}$ 		& 14.7		& 0.34139817(22)& 544.64931(2)		& 57.43(2)	& 0.00      	& 0.50       	& 5728(6)    \\ %S2414  20150728 Omega = 2.8758
%spotted but partial eclipses, ellipsoidal, just in contact
J221248.5+500527                        & 16.6	       & 0.3426652(7)  & 549.01270(5)          & 71.18(6)      & 0.00          & 0.50          & 5495(7)    \\ %08695  20150730 Omega = 2.8758
%Noisy LC, q not searched but fixed to 0.5, fill-out=0 fixed
J221249.5+493616 			& 14.8		& 0.27666140(20)& 544.76489(2)		& 40.73(10)	& 0.283(5)	& 2.0           & 5808(5)    \\ %S2596  20150729 Omega = 4.98628(492)
%chi^2 is decreasing with q, W-type very probable, miminum is close to q=2
J221319.1+494538                        & 17.7		& 0.3774138(15) & 546.23431(11)         & 65.89(9)      & 0.00          & 0.50          & 5722(16)   \\ %16513  20150730 Omega = 2.8758
%Noisy LC, q not seaerched but fixed to 0.5, fill-out=0.0, fixed
J221322.1+501628${\dag}$${\star}$ 	& 16.4		& 0.4422081(4)  & 545.38557(2)          & 85.70(15)  	& 0.219(10)	& 0.34200(67)   & 5994(3)    \\ %S8085  20150728 Omega = 2.51090(170)
%total eclipses, strong dependence of chi^2 on q, 
J221332.2+493046${\star}$		& 17.0		& 0.3584150(4)  & 546.32166(3)		& 84.6(3)	& 0.334(21)	& 0.20746(68)   & 6013(4)    \\ %10921  20150728 Omega = 2.20701(236)
%total eclipse, noisy LC, q well defined, 
J221348.8+493605${\star}$ 		& 17.1		& 0.3566974(5)	& 544.97208(4) 		& 80.51(8)	& 0.215(6)	& 0.50    	& 5828(6)    \\ %11667  20150728 Omega = 2.81158(172)
%q cannot be determined, noisy LC, partial eclipses
J221409.2+495507 			& 15.4		& 0.3359822(5)  & 544.87089(4)		& 34.99(16)	& 0.774(20)	& 0.50    	& 5269(12)   \\ %S4020  20150728 Omega = 2.64480(588)
%ellipsoidal variable, q cannot be determined
J221430.7+493102                        & 17.8		& 0.3364902(7)  & 548.32368(6)          & 76.10(8)      & 0.00          & 0.50          & 5550(8)    \\ %16629  20150730 Omega = 2.8758
%Noisy LC, q not searched but fixed to 0.5, fill-out=0.0, fixed
J221553.6+493011 			& 17.4		& 0.28941901(29)& 545.24009(3)		& 90        	& 0.168(17)	& 0.46803(177)  & 5908(5)    \\ %13856  20150728 Omega = 2.76669(398)
%possibly total eclipses, q optimized, i fixed to 90 degrees
J221609.1+500925${\dag}$		& 15.4		& 0.2823547(4)	& 544.58264(24)		& 38.61(22)	& 0.312(12)	& 0.50   	& 5540(100)  \\ %S4058  20150727 Omega = 2.78277(344)
% Ellipsoidal variation with some spot, impossible to determine q, parameters for q=0.50
J221622.9+501325 			& 16.8		& 0.3766157(9)  & 546.59791(6)		& 58.11(4)	& 0.00          & 0.50  	& 5726(11)   \\ %10206  20150729 Omega = 2.8758
%Noisy LC, q cannot be determined, fillout fixed to zero
J221628.3+495212${\dag}$ 		& 16.7		& 0.3183525(5)  & 544.57348(5)		& 57.40(10)	& 0.098(5)	& 0.50    	& 5712(17)   \\ %S9379  20150730 Omega = 2.84666(151)
%Moved from detached, spotted, q cannot be determined, 0.50 used
\hline
\end{tabular}                                                                
\end{table*}       

\subsubsection{Contact binary stars}

For contact binary stars mass ratios close to unity are not observed (see e.g., Rucinski 1993). For part of the systems $\chi^2$ was found to significantly depend on the mass ratio (mostly totally eclipsing). For systems, where $\chi^2$ was not varying significantly with the mass ratio, the solution listed in Table~\ref{contact} corresponds to $q = 0.50$.

J221609.1+500925 (Fig. B2 - middle bottom left) shows an O'Connell effect. The maxima difference is about 0.025 mag in the \textit{R} passband. The mass ratio was fixed to the value of 0.4 giving the lowest $\chi^2$. The resulting parameters are listed in Table~\ref{contact}. Fixing the spot to the equator and assuming spot temperature factor of $k$ = 0.7, its radius is $r_{\rm spot}$ = 20$\degr$. In the case of 8 more objects (marked in Table~\ref{detached} and ~\ref{contact}) we also assumed one spot. The light curve of these objects is less asymmetric than that of J221609.1+500925. 

J221332.2+493046 is a contact eclipsing binary with 0.43 and 0.40 mag deep minima in the $R$ passband. It shows total eclipses (Fig. B1 - bottom right). In this case the mass ratio was fixed to several values between 0.1 and 0.4. The minimum of $\chi^2$ was found  close to $q=0.2$. In the next step we modelled the light curve with mass ratio as a free parameter. The resulting photometric elements are listed in Table~\ref{contact}. Total eclipses are observed in 5 more cases (marked in Table~\ref{detached} and ~\ref{contact}).

For J221249.5+493616 the $\chi^2$ was found to decrease with increasing fixed mass ratio. The modelling showed that the most probable is the W-type configuration for the system with mass ratio close to 2. Thus the deeper minimum is the eclipse of the hotter but less massive secondary component. Similar possibility is probable for more systems (especially for short-period contact binaries) but the quality of data is insufficient. Conclusive analysis would require spectroscopic follow-up (see e.g., the DDO radial-velocity program, Pribulla et al. 2008).

\subsection{Pulsating variable stars}                                        
According to the GCVS and VSX catalogues there are three known pulsating variable stars in our studied region. Two of them, KN Lac (J221407.0+495543) and QQ Lac (J221655.4+500523), are slow irregular variables of late spectral types, and PZ Lac (J221557.4+493434) is a semi-regular late-type giant with poorly defined periodicity. Two other variable stars in our region, NSV 14080 (J221443.0+495139) and NSV 14071 (J221213.2+493109), are unique variable stars outside the range of the standard GCVS classifications. 

\begin{figure}[tbf]
  \centering
	\includegraphics[width=\columnwidth]{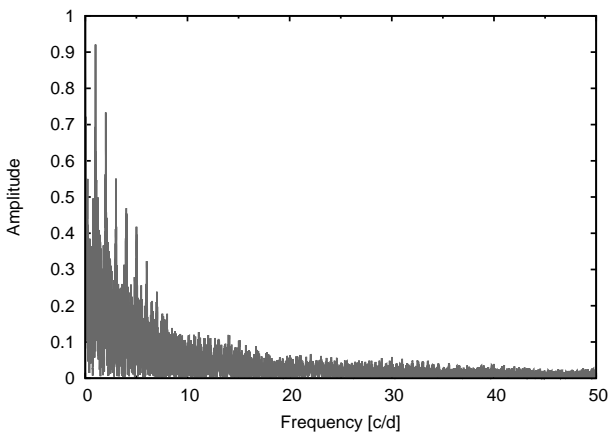}
	\caption{The spectral window of our pulsating star observation. Since for this purpose we used only the data obtained at GSH which have the lowest scatter, the data were spread with several wide gaps which produced a specific spectral window.}
	\label{fig:SW}
\end{figure}

Apart from that, we detected and analysed 26 new, presumably pulsating variable stars in the studied region. The $R$ passband magnitude range of the stars was from $17.5$ mag to $11.9$ mag. We used only the data obtained at GSH which have the lowest scatter. The observation window covers nearly 117 days. Data are spread with several wide gaps which is reflected in the corresponding spectral window (Fig. \ref{fig:SW}). The initial investigation was always carried for frequencies up to the Nyquist frequency $f_{N}\cong344$~c/d. Then we reduced the frequency interval to $\langle 0,50\rangle$~c/d because no significant frequencies larger than 50 cycles/day were found. This value roughly corresponds to 14 data points. Considering the noise in our data, every signal found in this mode would be dubious at best. The frequency step was set to $\Delta f=0.000428$~c/d. 

Non-phased light curves of each star were then inspected. Subsequently, we determined the $R$ passband magnitude amplitude. According to these amplitudes ($\Delta$mag is from interval $<0.02 - 0.2>$), the dominant frequencies and the light-curve features (Sterken \& Jaschek 1996), we assume $\delta$ Scuti and/or $\beta$ Cephei classification for these objects (see Table 6). Since the observations were taken only in the $R$ passband, we could not rule out the W UMa eclipsing binary stars possibility in some cases (see e.g., Duerbeck 1997). These objects need follow-up investigation to determine the true type of variability. In the following paragraphs we describe the analysis of selected objects only. The periodograms and non-phased one-night-run light curves are shown in Appendix C.

\begin{table*}[t!]
\centering
\caption{An overview of the frequencies of all detected and analyzed, presumably pulsating variable stars, sorted by RA. If more significant frequencies were found, they are listed in subsequent rows sorted by their amplitude. $R_{\rm max}$ is the maximal brightness of the star in the $R$ passband. $N$ is the number of points used in the period analysis. The notes column also shows an alternate explanation for the nature of the object variability: $\delta$ for $\delta$ Scuti-type star, $\beta$ for $\beta$ Cephei-type variable and $W$ for EW-type close binary star (i.e. W~UMa type).}
\label{tab:fpulse}
\begin{tabular}{llllllll}
\hline\hline
Identifier       & $R_{\rm max}$& $N$    & Frequency           & SNR    & Amplitude      & Period               & Notes 		\\
                 & [mag]	&        & [c/d]               &        & [mag]          & [days]               &  			\\
\hline\hline
J221142.1+501416 & $13.4$	&  $651$ & $16.9033(3)$        & $2.58$ & $0.008(5)$     & $0.0591600(12)$      & $\delta$      	\\%866
                 &        	&	 &  $4.005(14)$        & $2.44$ & $0.0060(8)$    & $0.2497(9)$          & few data - spurious 	\\
J221216.8+495442 & $14.0$	& $1458$ & $16.8076(4)$        & $3.20$ & $0.0067(4)$    & $0.0594967(13)$      & $\delta$ 		\\%1441
                 &		&        &  $7.986(4)$         & $3.18$ & $0.0046(4)$    & $0.12522(6)$         &                      	\\ 
J221232.9+493551 & $17.5$	& $1485$ &  $6.6085(3)$        & $3.71$ & $0.0446(2)$    & $0.151320(7)$        & $\delta$, $\beta$, $W$\\%14943
J221238.3+495437 & $13.1$	& $1208$ & $32.37926(2)$       & $3.64$ & $0.0071(2)$    & $0.0308840(1)$       & $\delta$ 		\\%666
J221242.8+500651 & $17.4$	& $1432$ &  $5.50320(13)$      & $3.99$ & $0.131(3)$     & $0.181712(4)$        & $W$ - low inclination	\\%14116
J221258.7+494342 & $13.2$	& $1465$ & $12.5571(3)$        & $3.07$ & $0.0063(3)$    & $0.0796359(17)$      & $\delta$ 		\\%747
                 & 		&        &  $0.8504(4)$        & $2.04$ & $0.0058(3)$    & $1.1759(5)$          & few data - spurious 	\\
                 &		&        & $23.11(14)$         & $2.36$ & $0.0027(7)$    & $0.0433(3)$          &  			\\
J221259.4+493457 & $14.0$	& $1515$ &  $0.7460(2)$        & $2.84$ & $0.0150(6)$    & $1.3404(4)$          & $\delta$ 		\\%1438
                 &		&        & $28.0705(5)$        & $3.30$ & $0.0096(5)$    & $0.0356246(6)$       &  			\\
                 &		&        &  $2.3799(4)$        & $2.18$ & $0.0087(5)$    & $0.42018(7)$         & window 		\\
                 &		&        & $25.0118(3)$        & $2.30$ & $0.0065(9)$    & $0.0399811(5)$       &  			\\
                 &		&        &  $3.6391(8)$        & $2.40$ & $0.0061(6)$    & $0.27479(6)$         & window 		\\
J221306.9+495538 & $12.3$	& $1500$ & $10.2463(5)$        & $2.91$ & $0.0095(8)$    & $0.097596(5)$        & $\delta$ 		\\%310
                 &		&        & $19.0348(14)$       & $2.10$ & $0.0033(6)$    & $0.052535(4)$        &  			\\
J221319.1+500259 & $12.8$	&  $710$ &  $1.87665(14)$      & $2.28$ & $0.00427(13)$  & $0.532865(4)$        & $\delta$, few data - spurious \\%534
                 &		&        &  $7.0692(2)$        & $2.49$ & $0.00267(13)$  & $0.141459(5)$        & few data - spurious 	\\
                 &		&        & $14.7547(2)$        & $2.33$ & $0.00252(13)$  & $0.0677751(11)$      & few data - spurious 	\\
J221324.7+501847 & $12.9$	& $1497$ &  $7.36097(9)$       & $3.81$ & $0.0184(4)$    & $0.1358515(17)$      & $\delta$, $\beta$ 	\\%548
J221337.1+500613 & $15.0$	& $1513$ &  $1.7463(3)$        & $2.89$ & $0.0253(11)$   & $0.572642(8)$        & $\beta$, window      	\\%2853
                 &		&        &  $5.7266(3)$        & $2.23$ & $0.0181(9)$    & $0.174623(10)$       &  			\\
J221346.6+501740 & $13.3$	& $1415$ & $11.8325(6)$	       & $2.50$ & $0.0038(4)$    & $0.084513(4)$        & $\delta$		\\%806
                 &		&        &  $2.97(12)$         & $3.56$ & $0.0038(13)$   & $0.337(15)$          & dubious		\\
                 &		&        & $11.4(2)$           & $2.51$ & $0.0032(11)$   & $0.0875(19)$         &  			\\
J221435.8+493557 & $14.0$	& $1507$ &  $8.9396(6)$        & $3.37$ & $0.0117(12)$   & $0.111862(7)$        & $\delta$, $\beta$    	\\%1408
                 &		&        &  $9.0(10)$          & $2.76$ & $0.007(3)$     & $0.111(14)$          & spot modulation\textrm{?} \\
                 &		&        & $16.398(4)$         & $2.43$ & $0.0064(10)$   & $0.060984(15)$       & unsure 		\\
J221457.8+500434 & $13.9$	& $1258$ &  $3.25633(15)$      & $3.08$ & $0.0088(2)$    & $0.307094(14)$       & $\beta$, $W$ - low inclination \\%1223
J221522.1+494441 & $15.6$	& $1517$ &  $2.57990(13)$      & $3.23$ & $0.0748(15)$   & $0.38761(2)$         & $\beta$, $W$ - low inclination \\%4471
J221556.7+501846 & $14.4$	& $1507$ &  $8.4597(4)$        & $3.24$ & $0.0082(4)$    & $0.118208(5)$        & $\delta$, $\beta$    	\\%1886
                 &		&        &  $3.7789(6)$        & $2.52$ & $0.0056(4)$    & $0.26462(4)$         &  			\\
                 &		&        & $11.6002(7)$        & $2.57$ & $0.0050(5)$    & $0.086206(5)$        &  			\\
J221557.8+493245 & $15.0$	& $1510$ &  $5.24803(14)$      & $4.13$ & $0.0270(6)$    & $0.190548(5)$        & $\delta$, $\beta$ 	\\%2998
J221559.8+500457 & $13.9$	& $1483$ & $22.9967(3)$        & $3.18$ & $0.0056(3)$    & $0.0434846(5)$       & $\delta$ 		\\%1259
                 &		&        & $14.0(6)$           & $2.35$ & $0.0020(5)$    & $0.072(3)$           &  			\\
J221606.2+494342 & $12.5$	& $1512$ &  $2.99145(13)$      & $3.81$ & $0.0144(3)$    & $0.334286(15)$       & $\beta$ 		\\%348
J221627.9+501640 & $14.0$	& $1508$ &  $9.2226(2)$        & $3.55$ & $0.0129(5)$    & $0.108429(3)$        & $\delta$, $\beta$   	\\%1435
                 &		&        & $14.46(5)$          & $3.23$ & $0.0040(12)$   & $0.0691(2)$          &  			\\
J221634.4+500546 & $11.9$	& $1506$ & $18.5588(5)$        & $2.91$ & $0.0073(5)$    & $0.0538827(13)$      & $\delta$ 		\\%168
J221636.9+500918 & $13.8$	& $1517$ & $10.9053(3)$        & $3.70$ & $0.0106(5)$    & $0.091698(3)$        & $\delta$ 		\\%1126
J221638.3+495824 & $12.9$	& $1509$ &  $1.56130(14)$      & $2.84$ & $0.0287(7)$    & $0.64049(6)$         & $\beta$, $W$ - low inclination \\%549
J221645.7+500534 & $15.4$	& $1500$ &  $1.9236(3)$        & $3.06$ & $0.0379(6)$    & $0.51986(7)$         & $\beta$, $W$ - low inclination \\%3859
J221657.3+495531 & $14.0$	& $1261$ & $12.1870(4)$        & $3.04$ & $0.0073(3)$    & $0.082055(3)$        & $\delta$ 		\\%1387
                 &		&        & $21.8924(12)$       & $2.86$ & $0.0029(3)$    & $0.045678(2)$        &  			\\
J221705.3+500838 & $17.3$	& $1131$ &  $5.7092(3)$        & $3.32$ & $0.047(3)$     & $0.175157(8)$        & $\delta$, $\beta$, $W$ low inclination \\%13091
\hline		 
\end{tabular}
\end{table*}

For the star at J221142.1+501416 (Fig. C1 - top) we have found two frequencies. Frequency $f_1=16.9033$~c/d ($P_1=0.05916$~days) corresponds to the sine wave function observed in the light curve. The precisely periodical nature could also point to a binary star variability type with $P_{\rm orb}\approx2\times P_1=0.11832$~day. However this is too short for a binary star, e.g., see Rucinski \& Pribulla (2008). The other frequency $f_2=4.005$~c/d ($P_2=0.2497$~days) is close to a $1/4$ harmonic alias of the diurnal period and considering the window of observation, the error of estimation is high. We cannot be certain if the frequency is intrinsic to the object. Other frequencies had smaller SNR, small amplitudes and were in the red noise frequency domain.

We interpret sharp peaks in the light curve of J221337.1+500613 (Fig. C3 - middle bottom) as a pulsation with frequency $f_2=5.7266$~c/d ($P_2=0.17462$~days). However, the most dominant frequency in the periodogram is $f_1=1.7463$~c/d ($P_1=0.57264$~day) which could be related to the observation window.

The object J221346.6+501740 (Fig. C3 - bottom) shows variations in peak heights in the light curve, thus a multi-periodic pulsations were assumed. The frequency $f_1=2.97$~c/d ($P_1=0.337$~days) has the highest SNR of all frequencies found above the SNR threshold. It has also the largest error of determination because of the period close to the observation window. Most likely this period should be ignored until more data with better coverage are provided. The frequency $f_2=11.8325$~c/d ($P_2=0.08451$~days) with the largest amplitude is also the most accurate one. Although close to it there another peak was found at $f_3=11.4$~c/d ($P_3=0.0875$~days) with period difference of $\sim4$ minutes. However, this period has a large uncertainty and we list it because of its SNR value. Frequencies $f_2$, $f_3$ should be investigated in more details, but to distinguish between them we would need longer observation window. If we assume that both frequencies start at the same phase, it will take $P_3/(P_3-P_2) \approx 29$ days until they shift in phase back to the original same phase. Also the time resolution should be better than $(P_3-P_2)/2 \approx 2$ minutes. The last found frequency $f_4=17.8112$~c/d ($P_4=0.05614$~days) has a much smaller uncertainty and can be considered intrinsic which causes the beat visible in the light curve.

Four frequencies were found at J221435.8+493557 (Fig. C4 - top) with SNR above the 2.0-limit. The most dominant is $f_1=8.9396$~c/d ($P_1=0.11186$~days). This is the visible pulsation period in the light curve. However, there are also modification of peaks that could be explained by a spot modulation. This may be the reason of that a similar frequency arises at $f_3=9.0$~c/d ($P_3=0.111$~days) with uncertainty larger than 1 c/d. The last noteworthy frequency is $f_4=16.398$~c/d ($P_4=0.06098$~days) which we cannot yet attribute to any process by a visual inspection of the light curve. The cadence of this beat may be a result of noise in the data, especially if it produces many sharp peaks of similar amplitude in this region of periodogram. There is also $f_2=0.0159$~c/d ($P_2=63.08114$~days) with uncertainty of $\sim3.4$~days and the lowest SNR$=2.02$ that is a result of a not-perfect alignment of time series from various nights. Thus we did not include it in the table of results.

At J221457.8+500434 (Fig. C4 - middle top) we have found only one significant frequency $f_1=3.25633$~c/d ($P_1=0.30709$~days). However, based on the smooth sine wave we could suspect that this object may be a binary star with $P_{\rm orb}=2\times P_1=0.61418$~day. 

The light curve of the object J221556.7+501846 (Fig. C4 - bottom) shows alternating peaks. The width of peaks with sharp maximum corresponds to the frequency $f_3=11.6002$~c/d ($P_3=0.0862$~days). On the other hand, the dominant frequency $f_1=8.4597$~c/d ($P_1=0.1182$~days) corresponds to the variability between the peaks with sharp and flat maxima, respectively. The last significant frequency $f_2=3.7789$~c/d ($P_2=0.26462$~days) arises from the period of peaks with flat maximum. Overall, we can interpret it as a multiperiodic variable star, the flat peaks may be the results of spot modulation.

Sharp peaks in the light curve of J221627.9+501640 (Fig. C5 - bottom) suggest a classical pulsation with strongest frequency $f_1=9.2226$~c/d ($P_1=0.10842$~days). The U-shaped minimum around JD 2456541.36 could be result of a presence of a photospheric spot. After pre-whitening low-amplitude, low-SNR frequencies, another $f_4=14.46$~c/d ($P_4=0.0691$~days) was found in the residuals. It has also a strong SNR albeit amplitude three-times smaller than $f_1$.

At J221705.3+500838 (Fig. C7 - bottom) the most significant frequency $f_1=5.7092$~c/d ($P_1=0.17515$~days) corresponds to the peak-to-peak changes. However, this object could be also interpreted as close binary star with low inclination and orbital period $P_{\rm orb}\sim 2\times P_1=0.35031$~days. After the pre-whitening another period with SNR$>2.0$ was found at $f_2=1.9122$~c/d ($P_2=0.52295$~days), but is likely an artefact of the window function.

\section{Discussion}

\subsection{The exoplanet candidate}

Its light curve has a U-shape, which is typical for exoplanetary transits, however the transit depth is unusually large for an extrasolar planet. Therefore, we can conclude that this object is (1) a hot Jupiter or a brown dwarf orbiting a K or M dwarf star, or (2) a grazing eclipsing binary star, or (3) an eclipsing binary star with a dominant third star/binary in the aperture. The obtained effective temperature for the parent star $\sim 6500$ K indicates an F-spectral-type star, which rules out the alternative (1) of a planet/brown dwarf orbiting a K or M dwarf star. On the other hand, we can take into account theoretical studies (Baraffe et al. 1998, 2002, 2003) that young planets and stars are much larger than old planets and stars, but planets contract slower than their parent stars, hence the transit depth, 0.08 mag in the $R$ passband, and the corresponding value of $R_p/R_s \sim 0.2$ are still possible for young systems (see Fig 2. in Neuh\"auser et al. 2011). This means that this object still can be a planet or brown dwarf, orbiting a young F-spectral-type star.              

Transit surveys already found a lot of variable objects that exhibit exoplanet transit-like signals. However, many of these candidates have turned out to be eclipsing binaries and not planets due to the astrophysical mimics (see e.g., Santerne et al. 2012 and references therein). A grazing eclipsing binary star causes a small dip in brightness which is approximately planet sized, however, the transit is V-shaped. Since our candidate has a U-shape, the alternative (2) is unlikely. Another astrophysical mimic is generated, when eclipse depth of a binary star is diluted with a dominant third star/binary in the aperture. It can mimic a planetary transit shape as well as depth. This alternative (3) can be ruled out with follow-up imaging. Another possibility is follow-up photometry with different filters, i.e. getting the change in depth and shape of the transit with wavelength (see e.g., Knutson et al. 2007).         

To check whether the candidate is a low-mass or stellar companion, we also searched for a secondary eclipse in the data. In the case of the stellar companion we should probably also see a small dip roughly at phase 0.5, if the orbit is circular (see Fig 3. in Errmann et al. 2014b). Fig. 2 (right) shows the full phase-folded light curve of the exoplanet candidate. The secondary eclipse is not visible on this plot. The inset shows the phase-folded light curve from 0.4 to 0.6 in phase, binned to better visualize the lack of detection of the secondary eclipse. The absence of the secondary eclipse is in favor of the low-mass companion alternative (1).                              
   
\subsection{Other variable stars}
      
We discovered and analysed 37 new eclipsing binary stars in the studied region. The light-curve analysis was performed using the Roche code (Pribulla 2004; 2012). We described in detail the light-curve analysis of 6 selected binaries. The best fit parameters and light curves of other objects were summarized as well. The resulting photometric elements are rather preliminary because only one photometric passband was used. Multi-passband photometry would improve modelling of spotted systems where temperature contrast of spots ($k$) can not be now determined. Colour information would help to discern pulsating variable stars from low-amplitude contact binaries. Analysis of most systems (except the four fortuitous cases of contact binaries showing total eclipses) is affected by lack of spectroscopic observations and no information about the mass ratio. With the maximum brightness range from 17.8 to 11.7 the eclipsing systems would require 4-meter class telescope with medium-resolution spectrograph ($R \sim 15000$) to arrive at spectroscopic orbits.

Moreover, we detected and analysed 26 new, presumably pulsating variable stars in the studied region (according to light-curve amplitudes, the period range and the light-curve features). Having single-passband photometry we can not exclude W UMa eclipsing binary stars in some cases. The observations are affected by lack of longer runs which resulted in poor spectral window (see Fig.~\ref{fig:SW}). This means that usually only one or two strongest periods can conclusively be found in the data and analysis is limited by severe aliasing. Further study of the objects would require coordinated un-interrupted observations from several observatories (see e.g., the WET project, Solheim 2003) 

The follow-up investigation of these objects, including spectroscopic measurements of the exoplanet candidate, is currently planned.

\vspace{0.5cm}
\acknowledgements
The authors thank J. Lopatovsk\'{y}, V. Koll\'{a}r, F. Eysoldt, A. Ide, S. Schoenfeld, L.L. Kiss, L. Moln\'{a}r, K. S\'{a}rneczky, Chen Wen-Ping, Po-Chieh Huang and all members of the YETI team for the technical assistance, comments and discussions. This study is partially based on observations obtained with telescopes of the University Observatory Jena, which is operated by the Astrophysical Institute of the Friedrich-Schiller-University. RoEr would like to thank DFG in the Priority Programme SPP 1385 in project NE 515 / 34-1 and the Abbe-School of Photonics. RaNe and MaKi would like to thank DFG for support in the Priority Programme SPP 1385 on the "First Ten Million Years of the Solar System" in projects NE 515 / 34-1 and 34-2. BaDi, AnPa, JaSc, and RaNe thank DFG in SFB TR 7 in TPs C2, C7 and B9. MaMu and ChGi thank DFG in project MU 2695 / 13-1. ChAd thanks DFG in projects NE 515 / 35-1 and 35-2 in SPP 1385. BaDi also thanks DFG in project RN 515 / 36-1. StRa thanks DFG in project NE 515 / 33-1 and 33-2 in SPP 1385.
MaKi would also like to thank DFG in project RE 882 / 12-1 and 12-2. We would like to acknowledge financial support from the Thuringian government (B 515-07010) for the STK CCD camera used in this project at Jena observatory. This work was supported by the VEGA grant of the Slovak Academy of Sciences No. 2/0143/14, by the Slovak Research and Development Agency under the contract No. APVV-0158-11 and by the realization of the Project ITMS No. 26220120009, based on the Supporting Operational Research and Development Program financed from the European Regional Development Fund.   

\clearpage
\newpage%%%%%%%%%%%%%%%%%%%%%%%%%%%%%%%%%%%%%%%%%%%%%%%%%%%%%%

\onecolumn
\clearpage
\newpage%%%%%%%%%%%%%%%%%%%%%%%%%%%%%%%%%%%%%%%%%%%%%%%%%%%%%%
\appendix
\section{An overview of the phase-folded detached eclipsing binary light curves with the best fit model overplotted (top) and corresponding residuals (bottom)}

\begin{figure*}[!h]
\centering
\centerline{
\includegraphics[width=70mm]{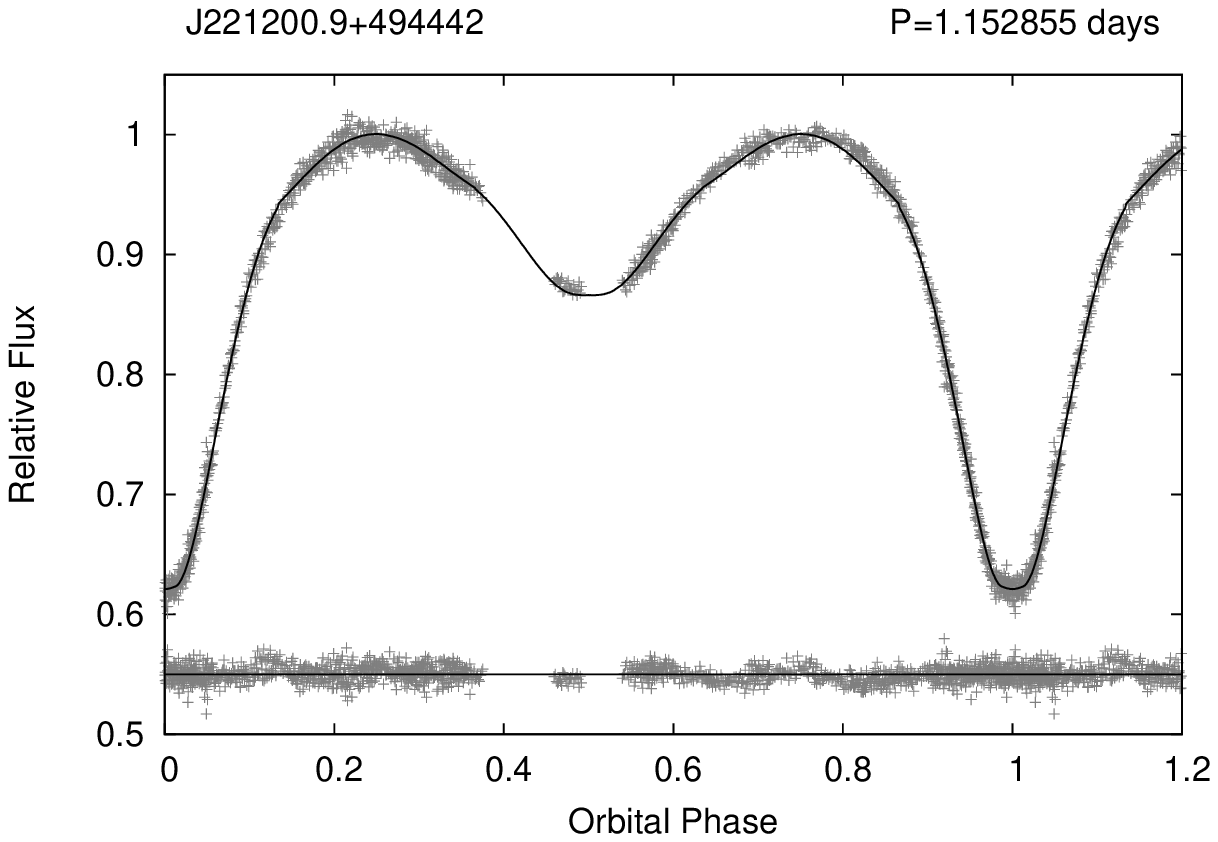} 
\includegraphics[width=70mm]{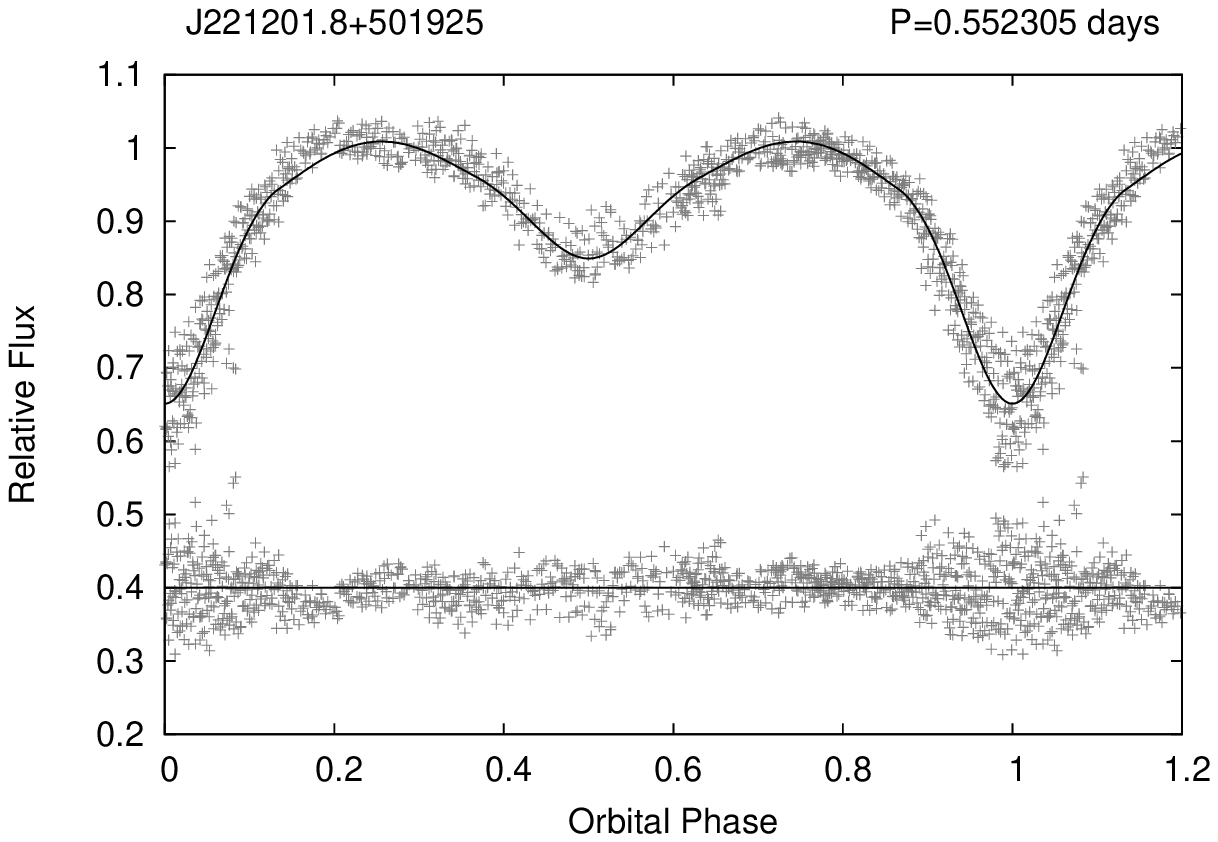} 
}
\centerline{
\includegraphics[width=70mm]{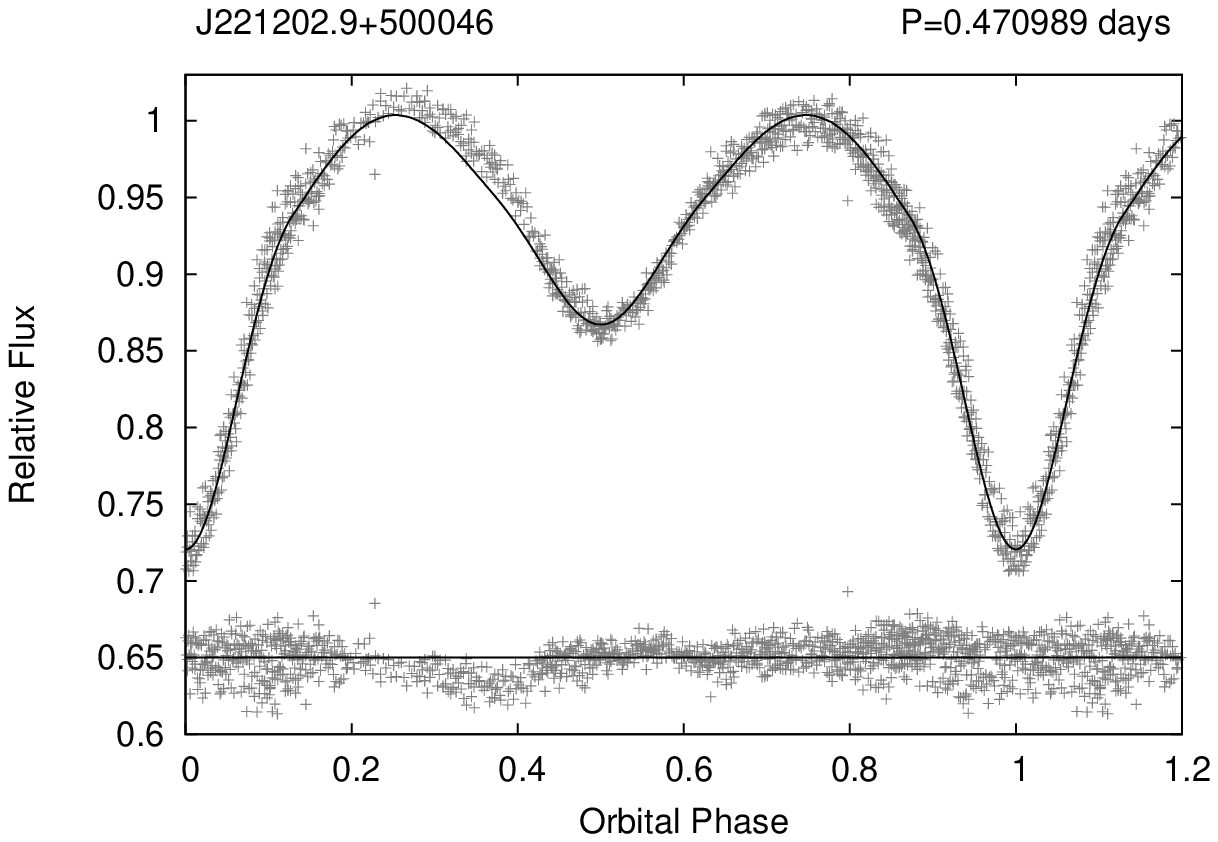}
\includegraphics[width=70mm]{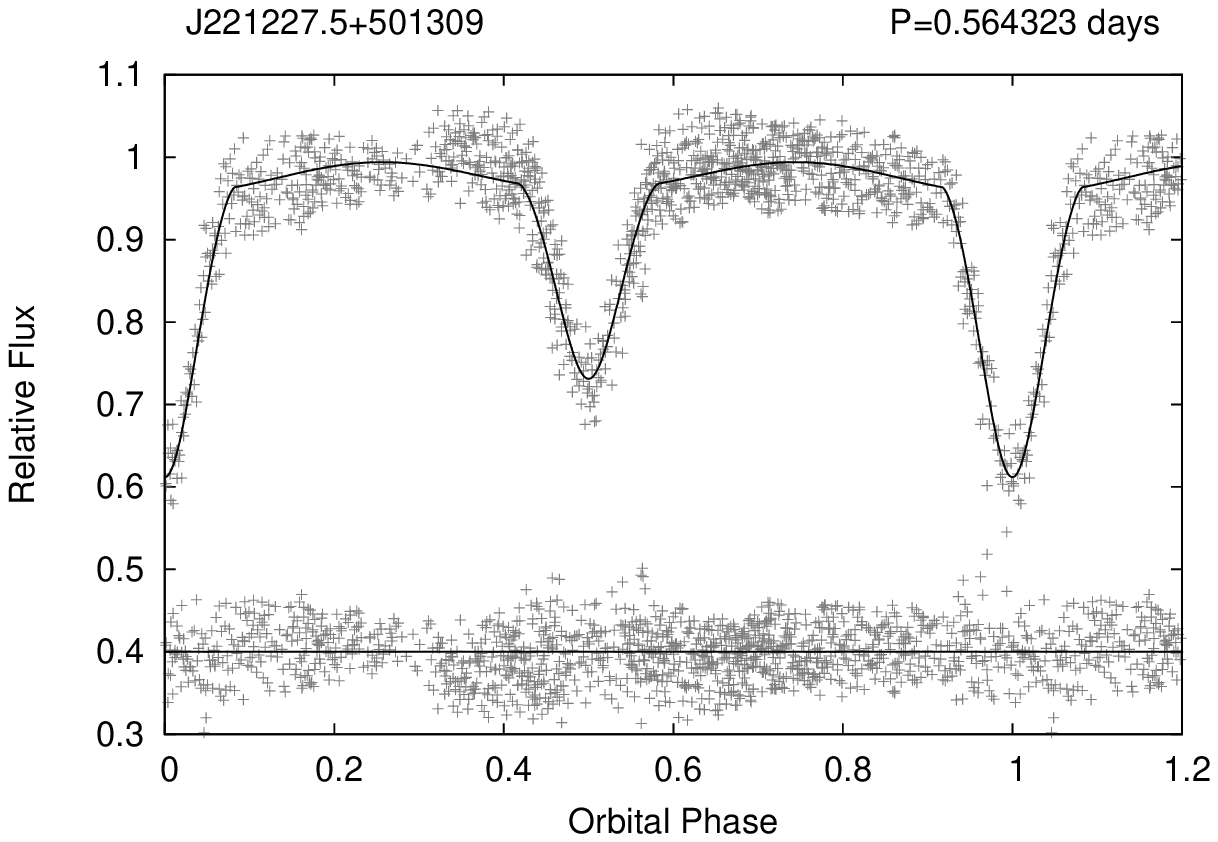}
}
\centerline{
\includegraphics[width=70mm]{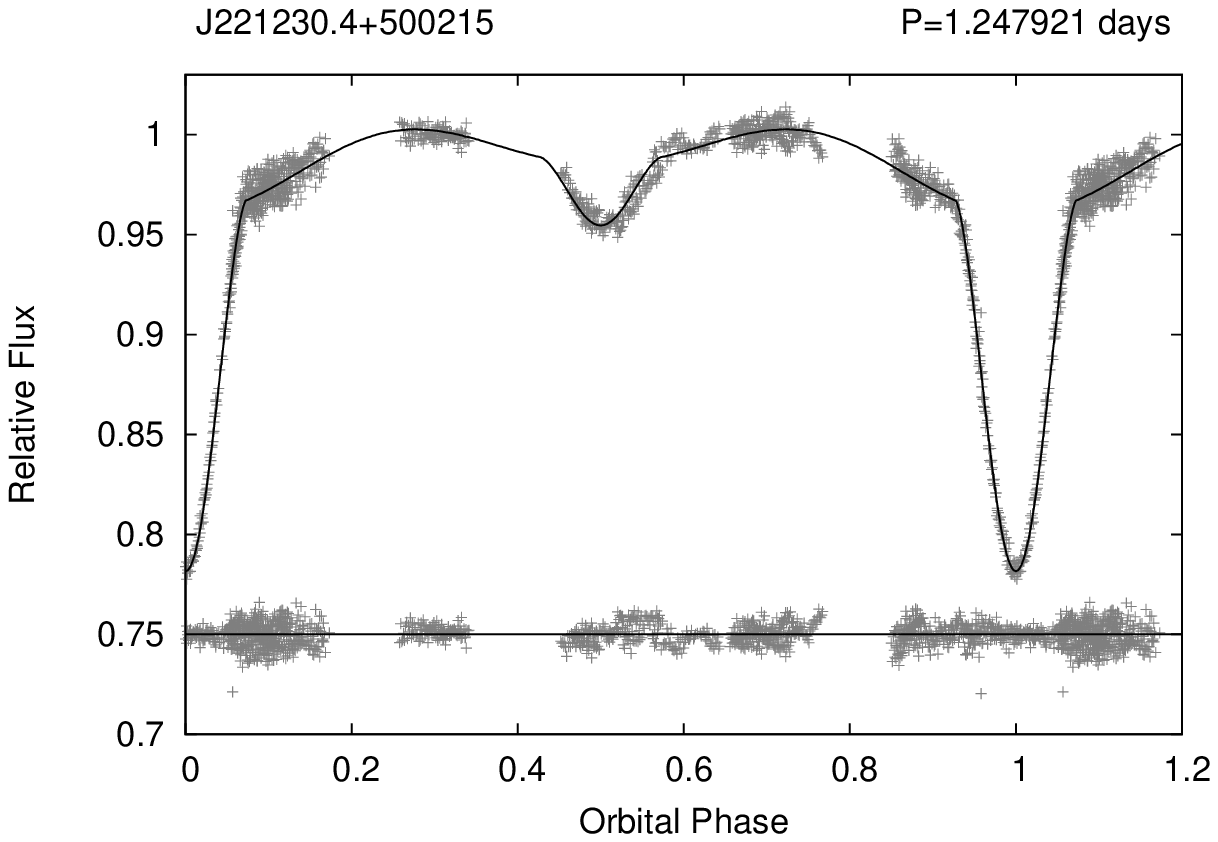}
\includegraphics[width=70mm]{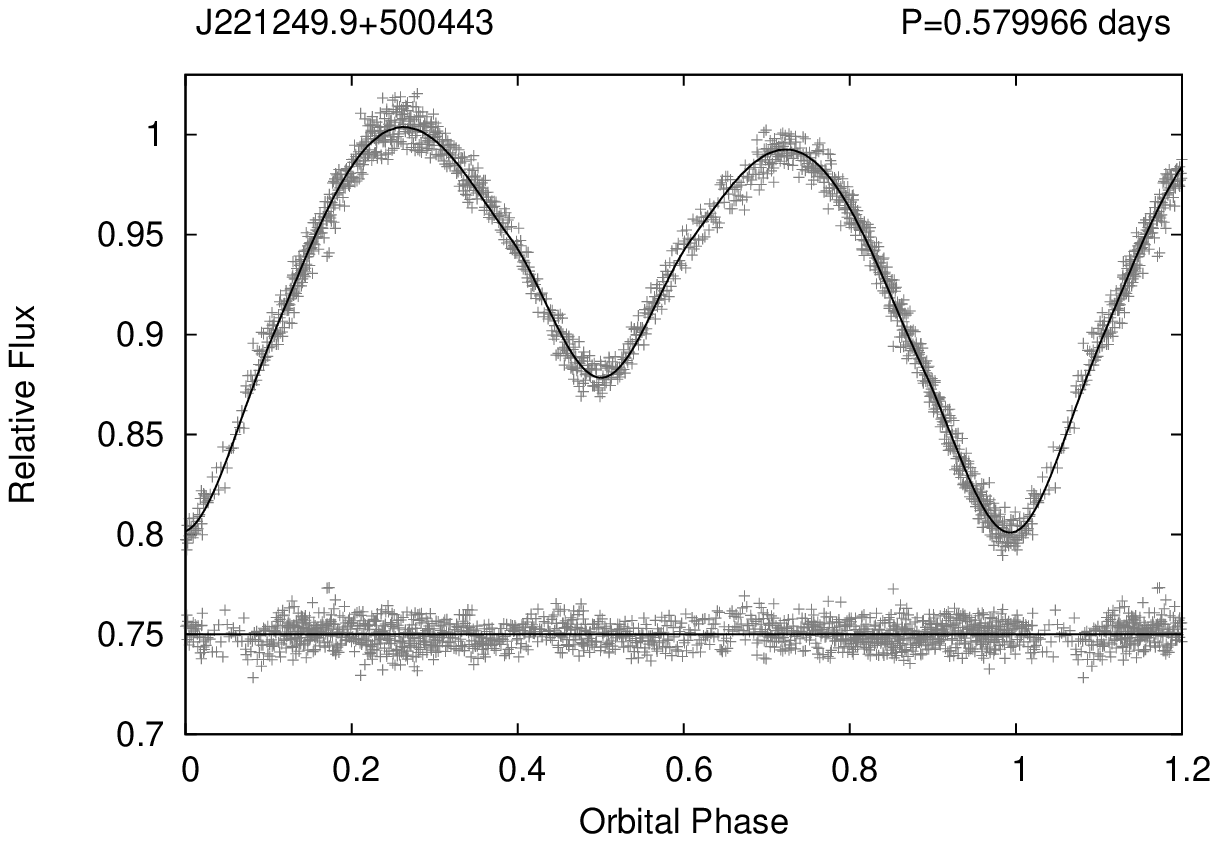} 
}
\centerline{
\includegraphics[width=70mm]{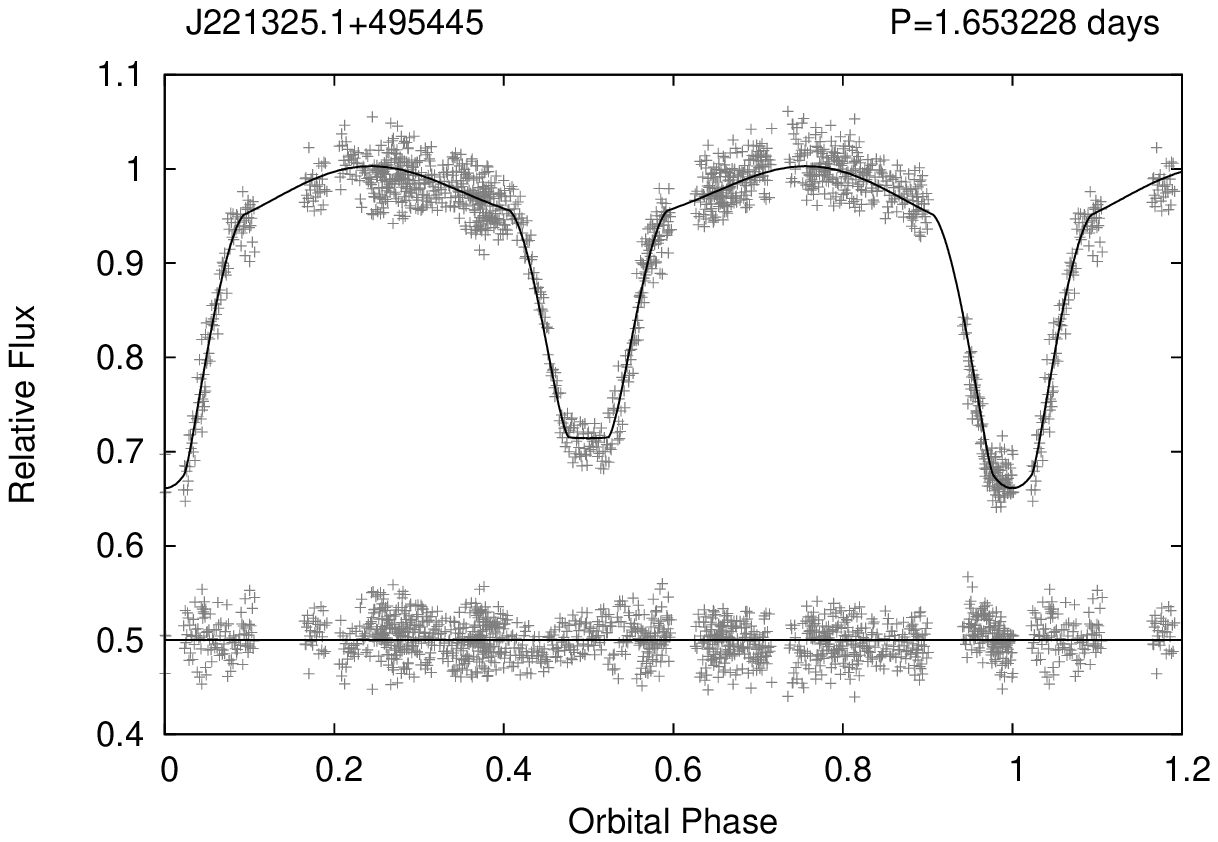} 
\includegraphics[width=70mm]{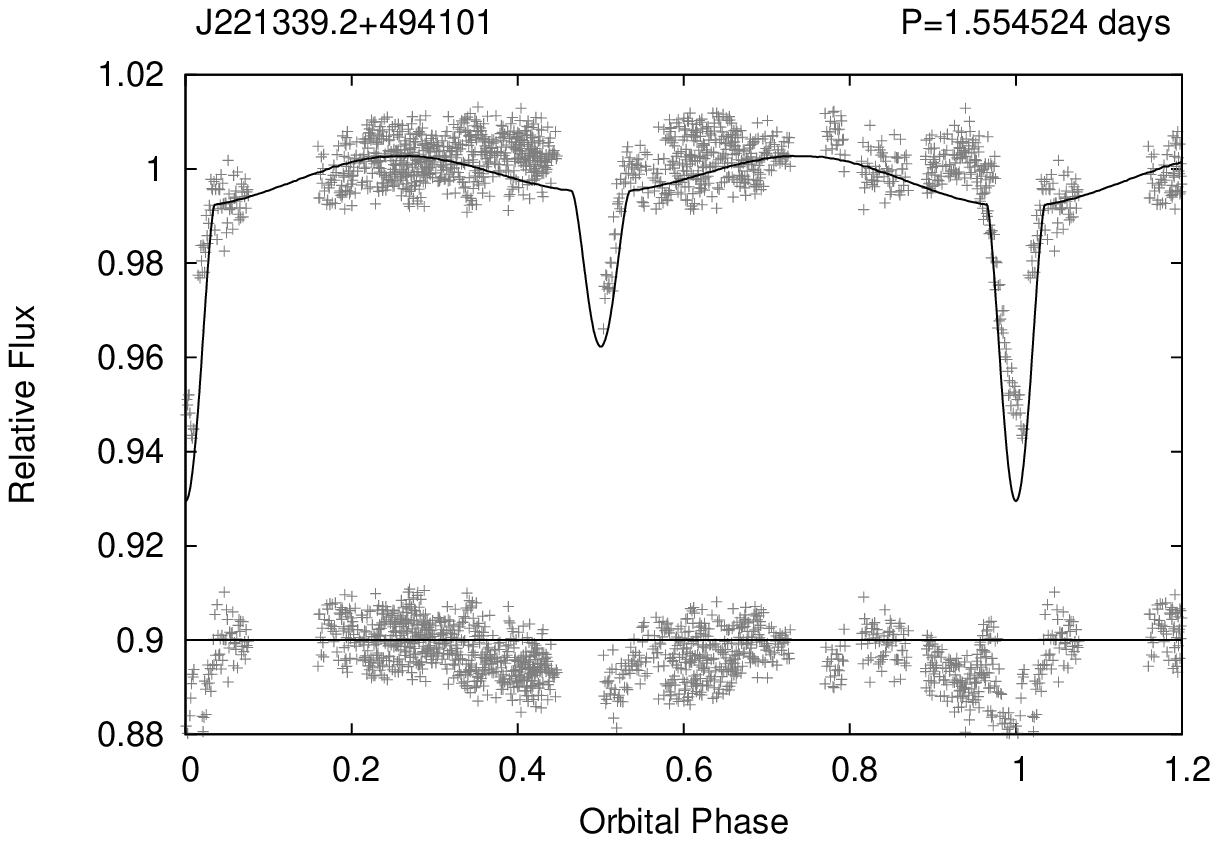} 
}
\caption{}
\label{Figure 1}
\end{figure*}

\begin{figure*}[!h]
\centering
\centerline{
\includegraphics[width=70mm]{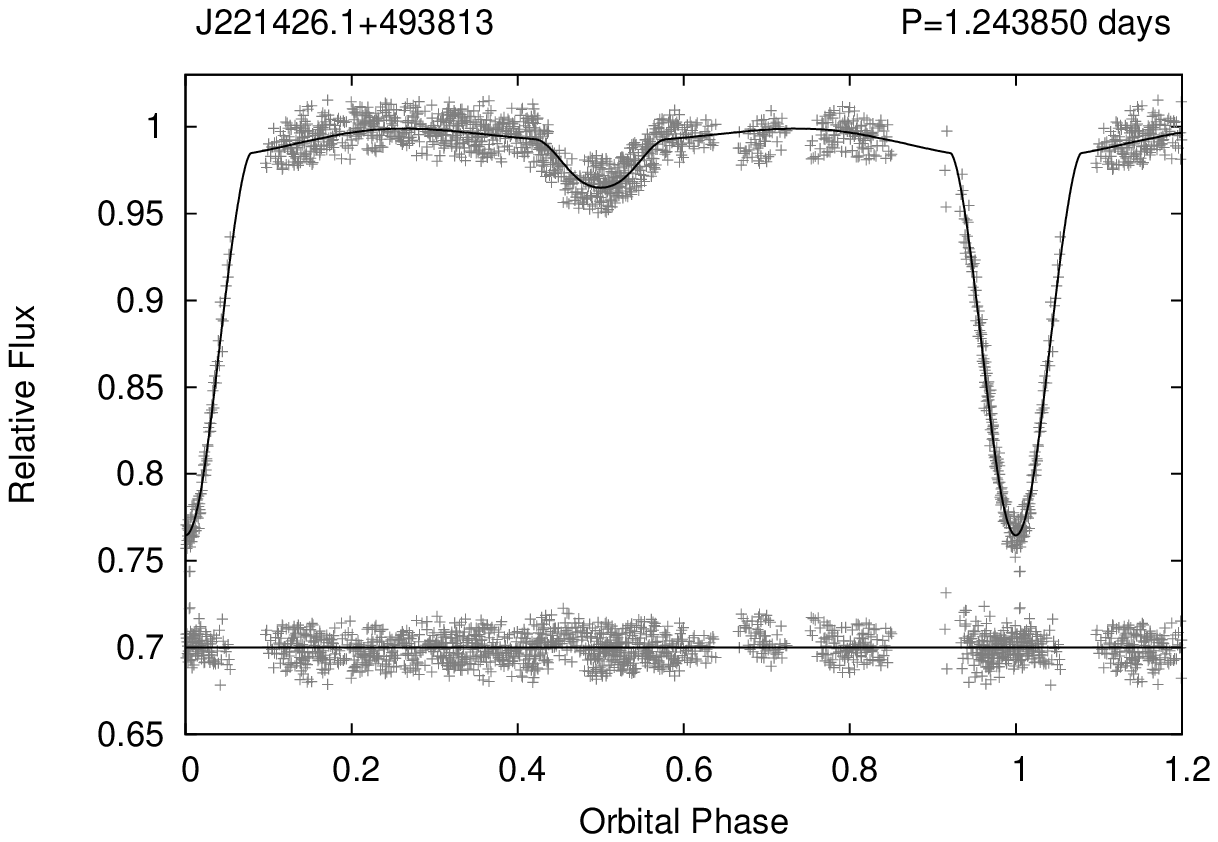} 
\includegraphics[width=70mm]{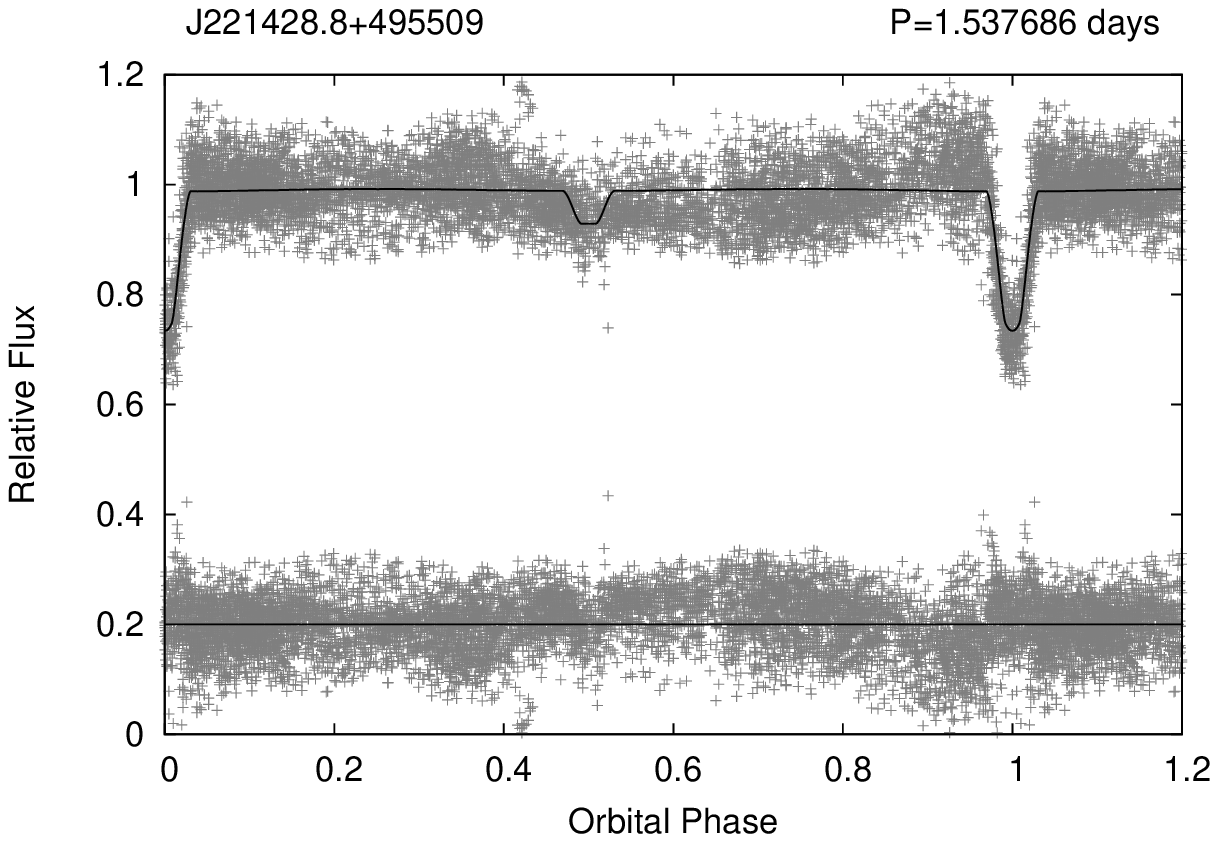} 
}
\centerline{
\includegraphics[width=70mm]{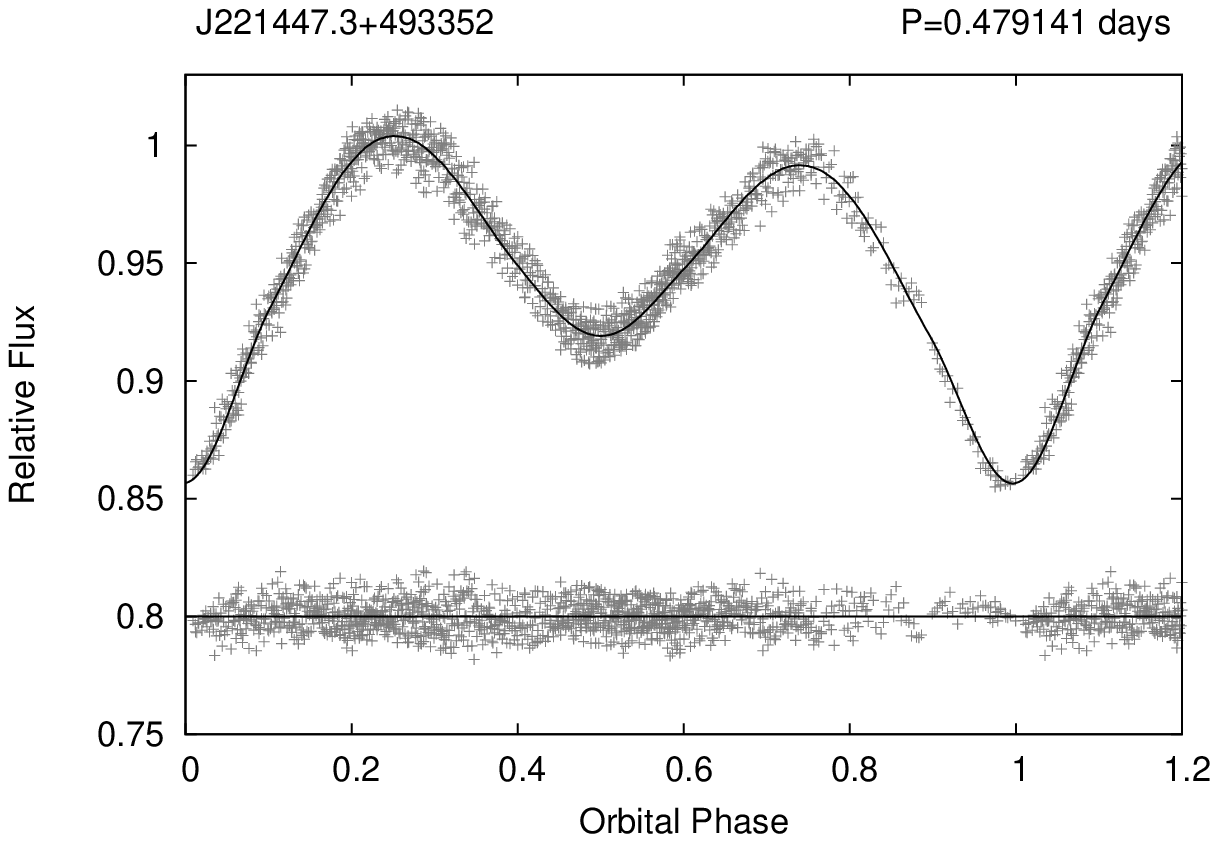} 
\includegraphics[width=70mm]{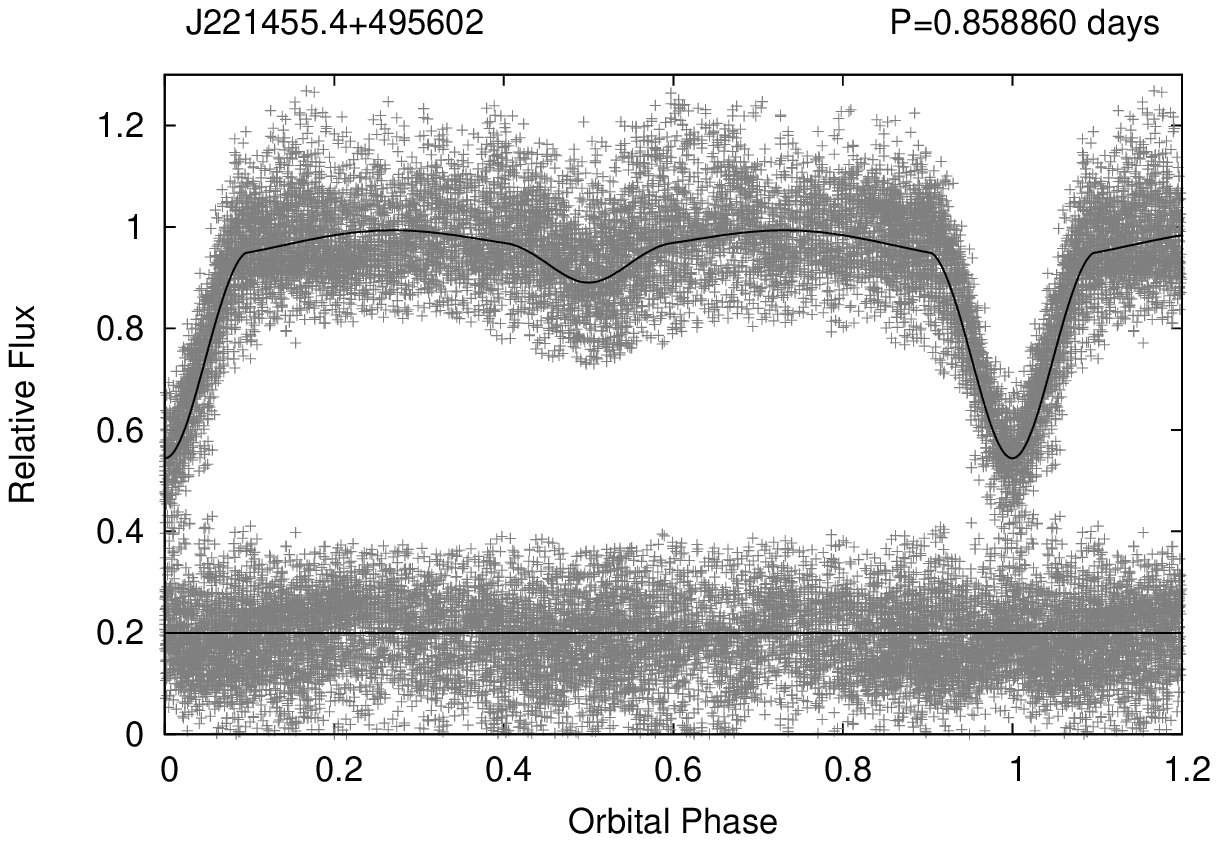}
}
\centerline{
\includegraphics[width=70mm]{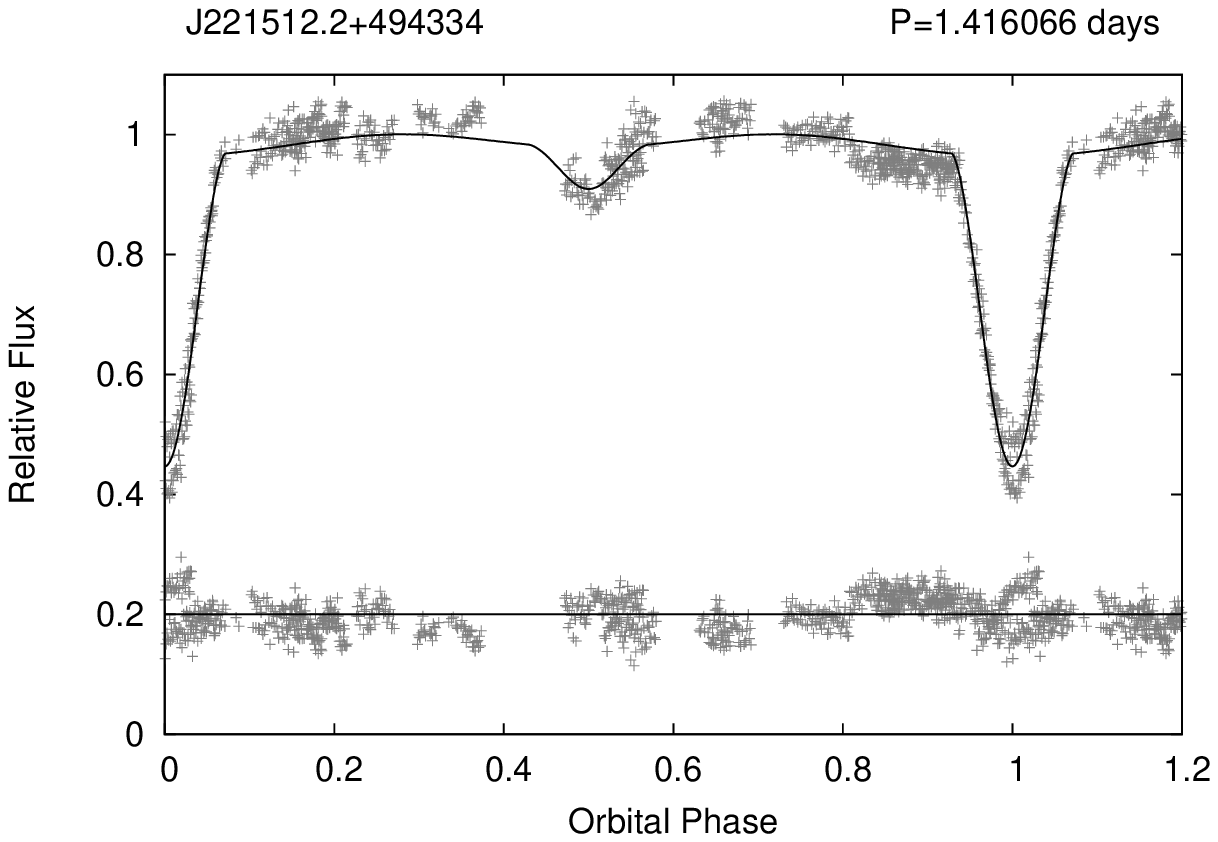}
\includegraphics[width=70mm]{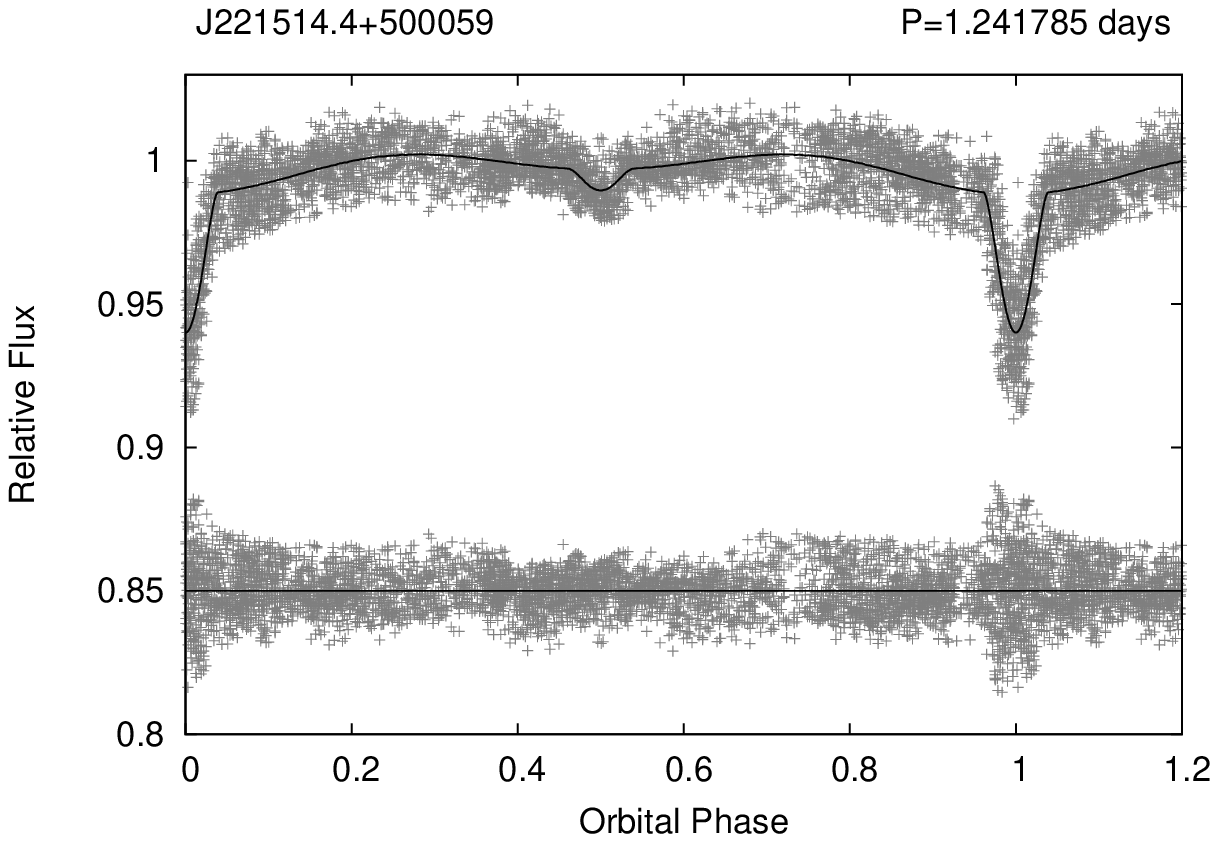} 
}
\centerline{
\includegraphics[width=70mm]{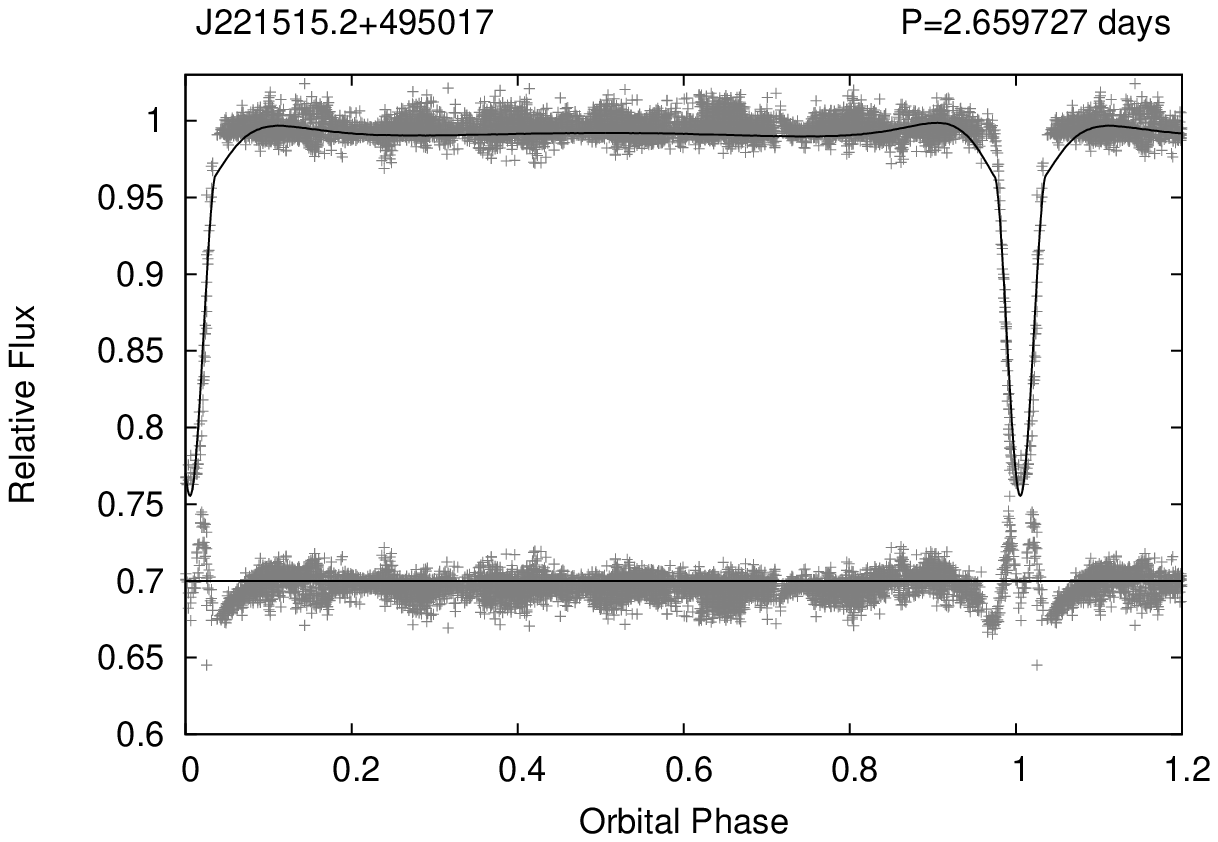} 
\includegraphics[width=70mm]{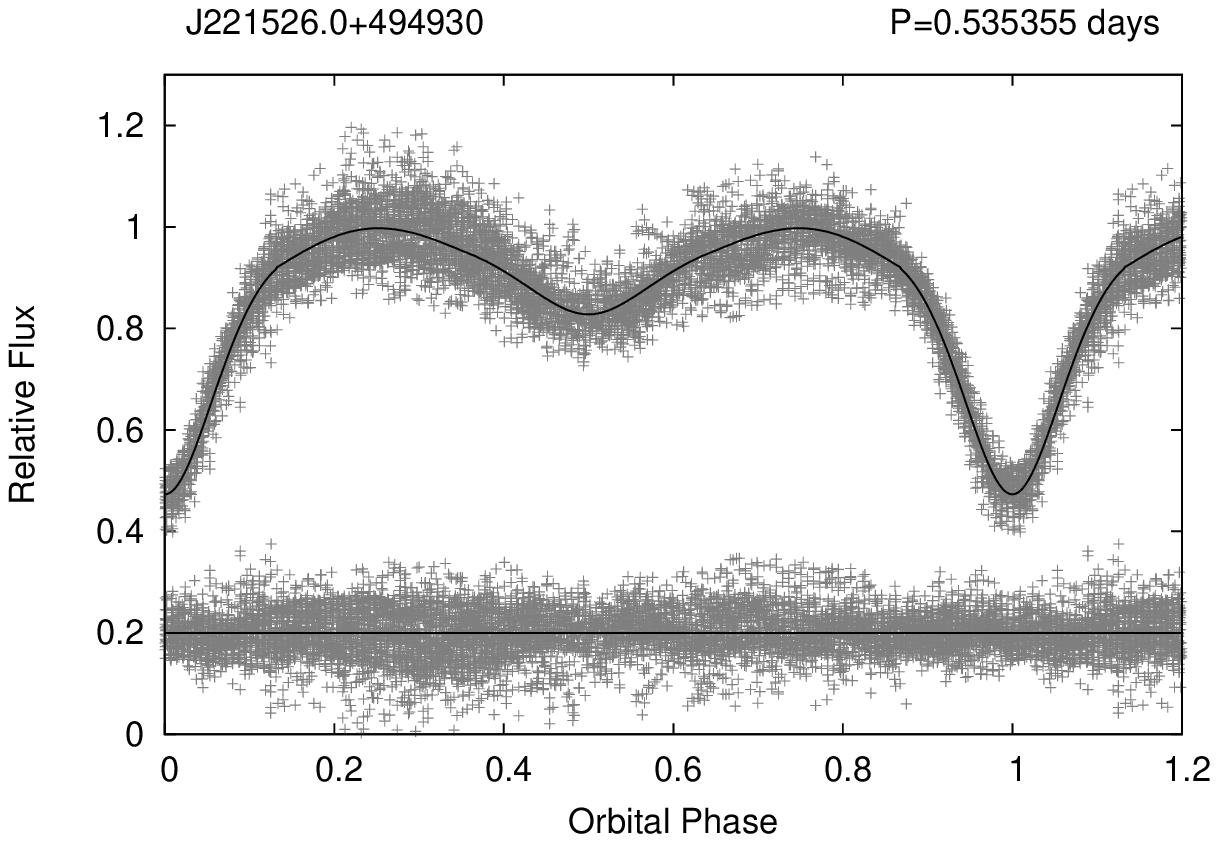} 
}
\caption{}
\label{Figure 1}
\end{figure*}

\begin{figure*}[!h]
\centering
\centerline{
\includegraphics[width=70mm]{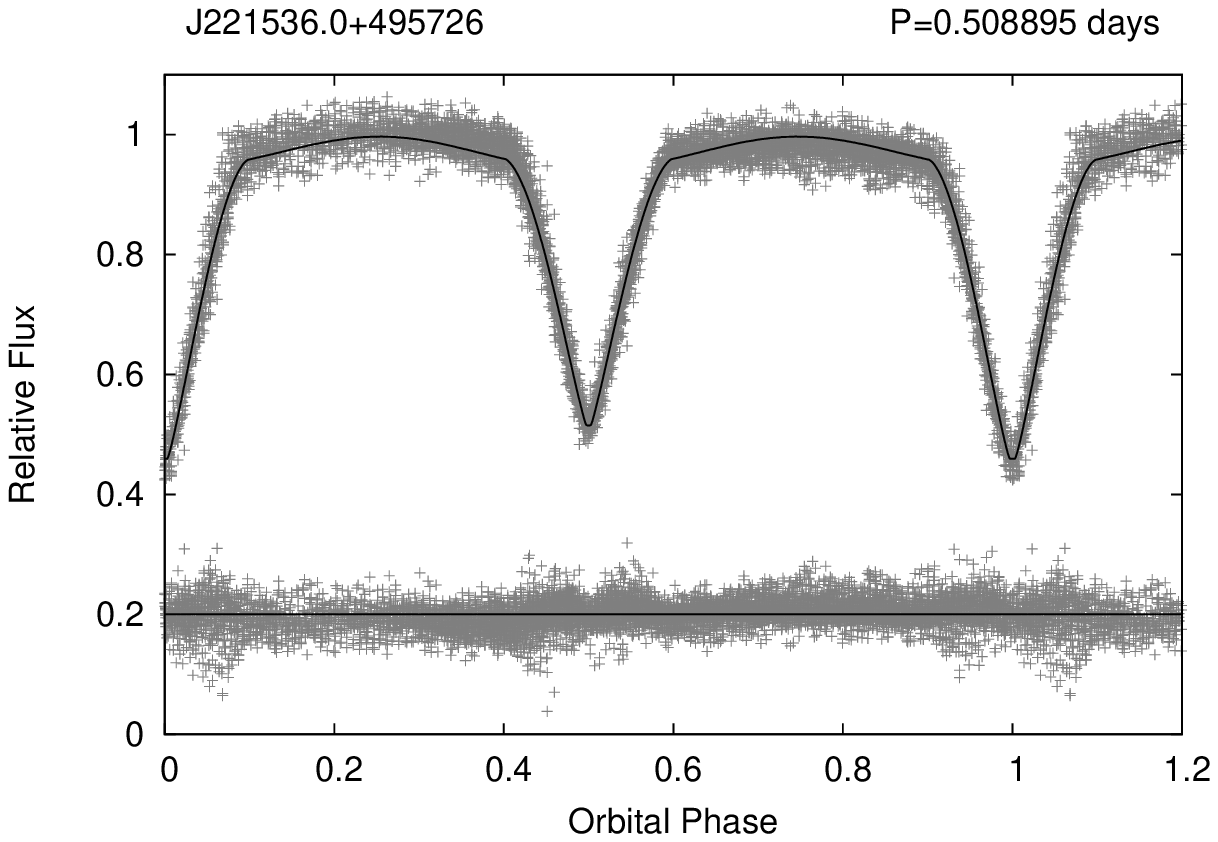} 
\includegraphics[width=70mm]{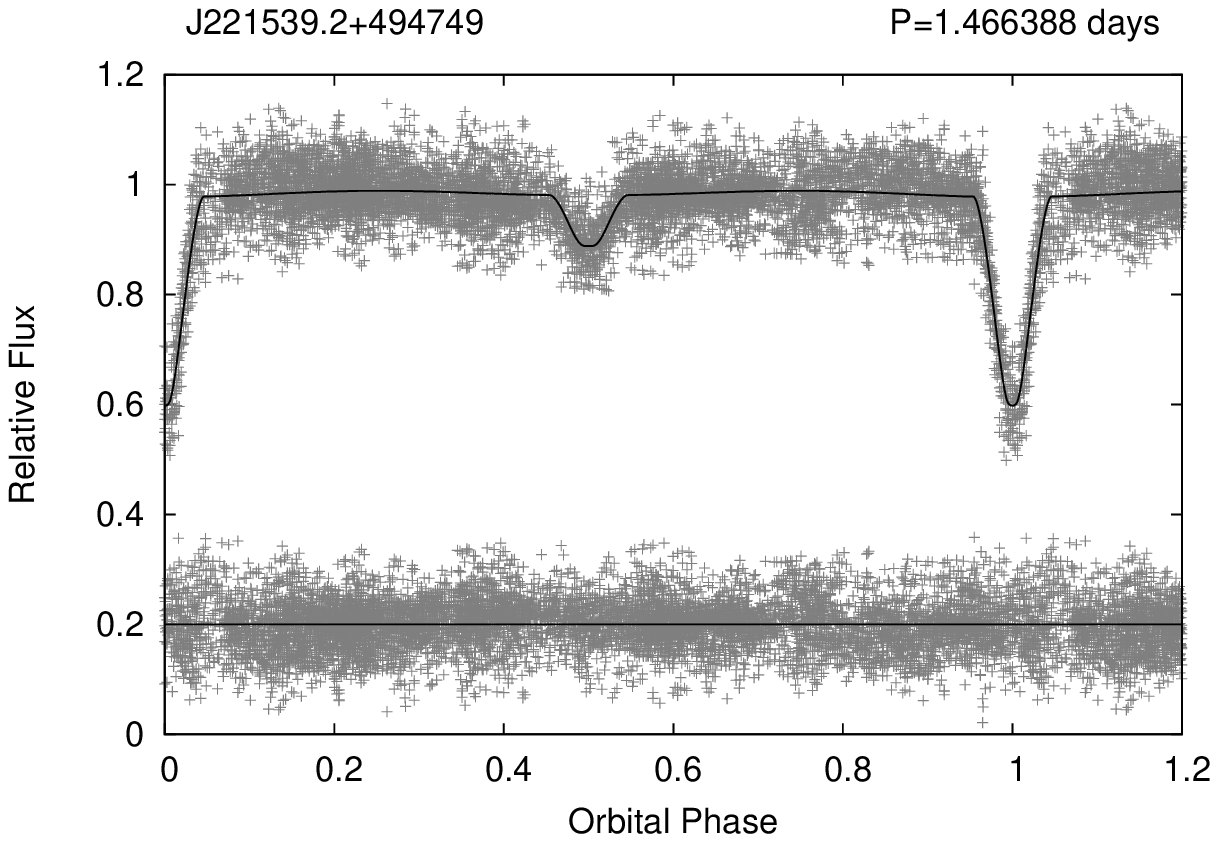} 
}
\centerline{
\includegraphics[width=70mm]{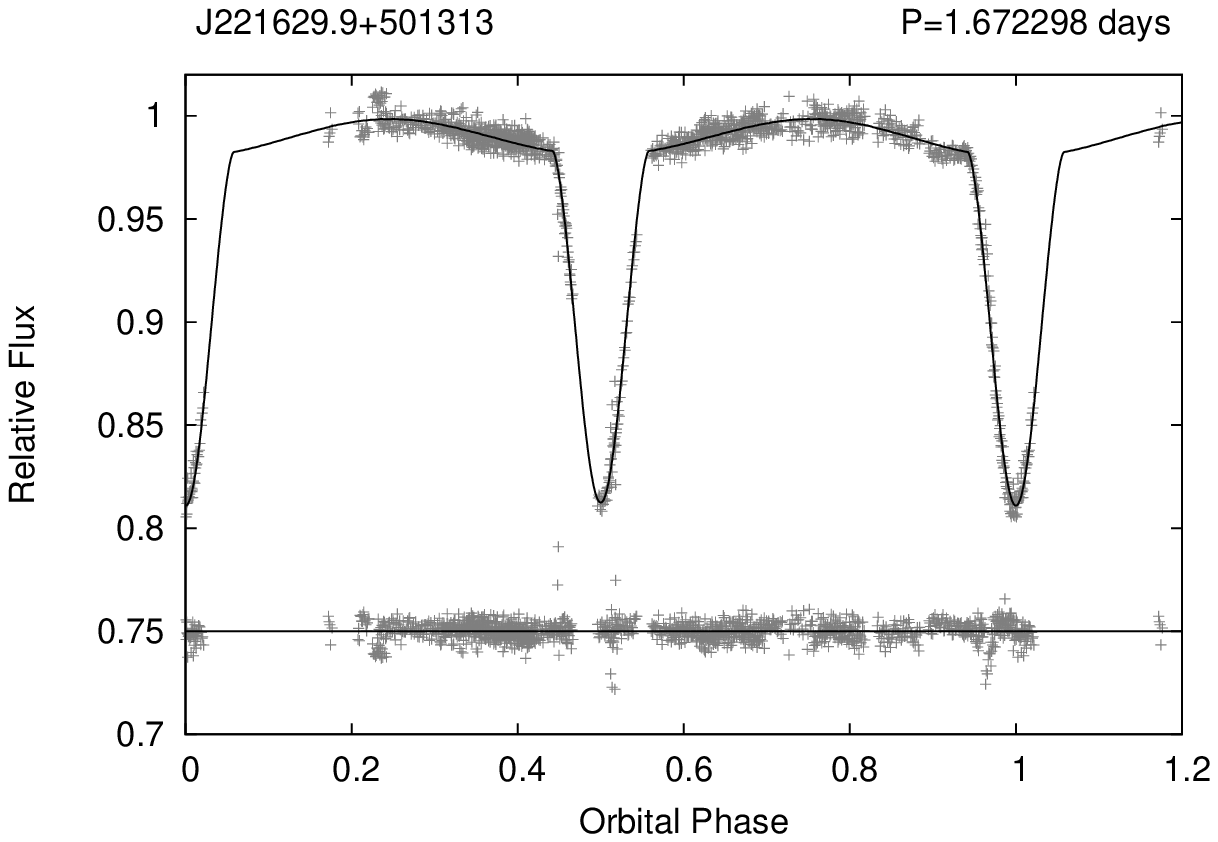} 
\includegraphics[width=70mm]{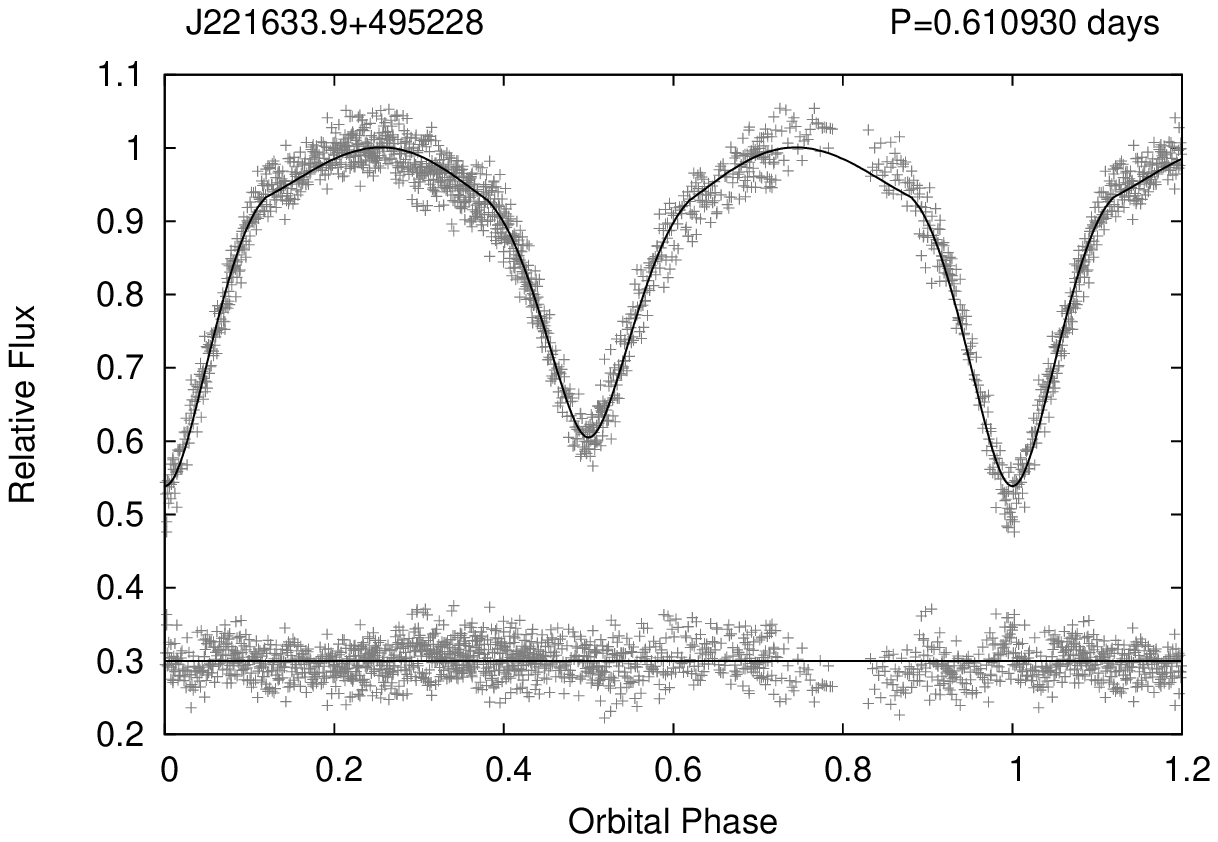} 
}
\centerline{
\includegraphics[width=70mm]{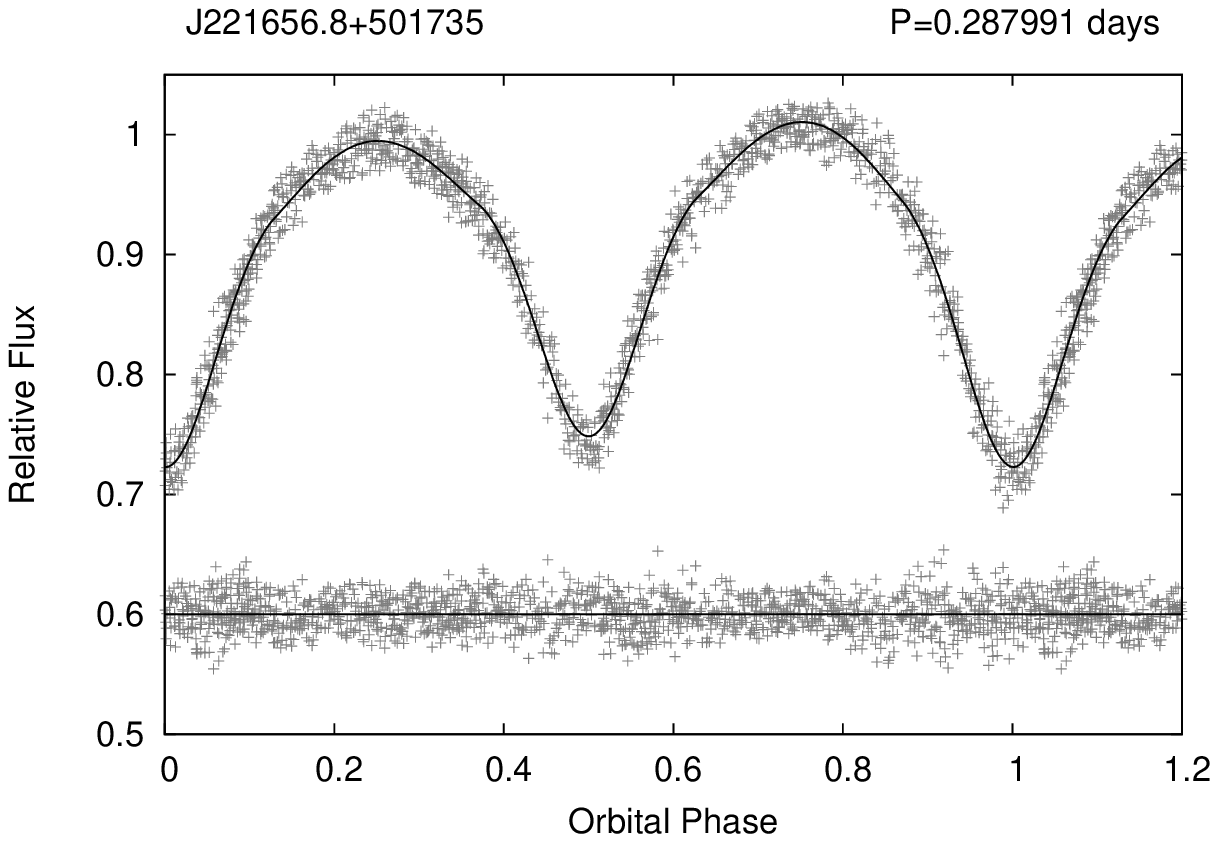} 
\includegraphics[width=70mm]{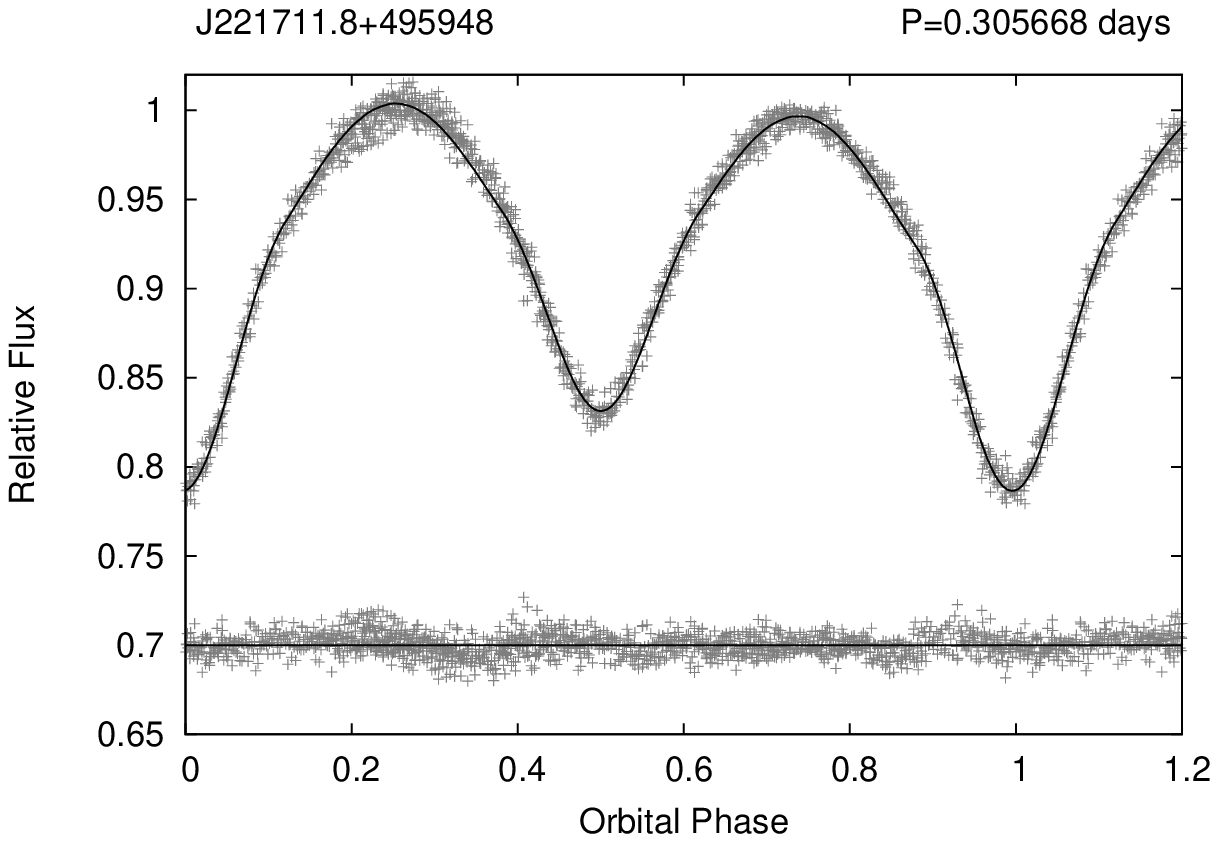} 
}
\caption{}
\label{Figure 1}

\end{figure*}

\clearpage
\newpage
\section{An overview of the phase-folded contact eclipsing binary light curves with the best fit model overplotted (top) and corresponding residuals (bottom)}

\begin{figure*}[!h]
\centering
\centerline{
\includegraphics[width=70mm]{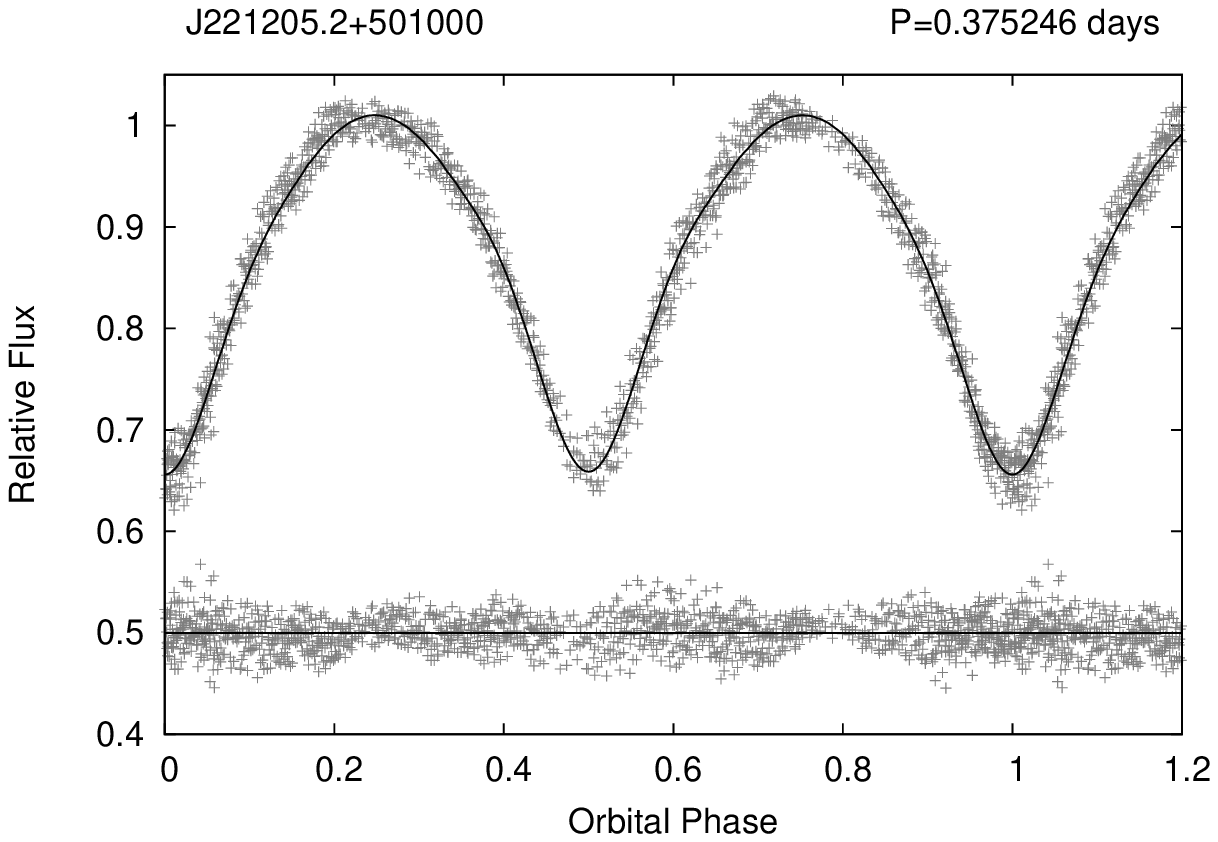} 
\includegraphics[width=70mm]{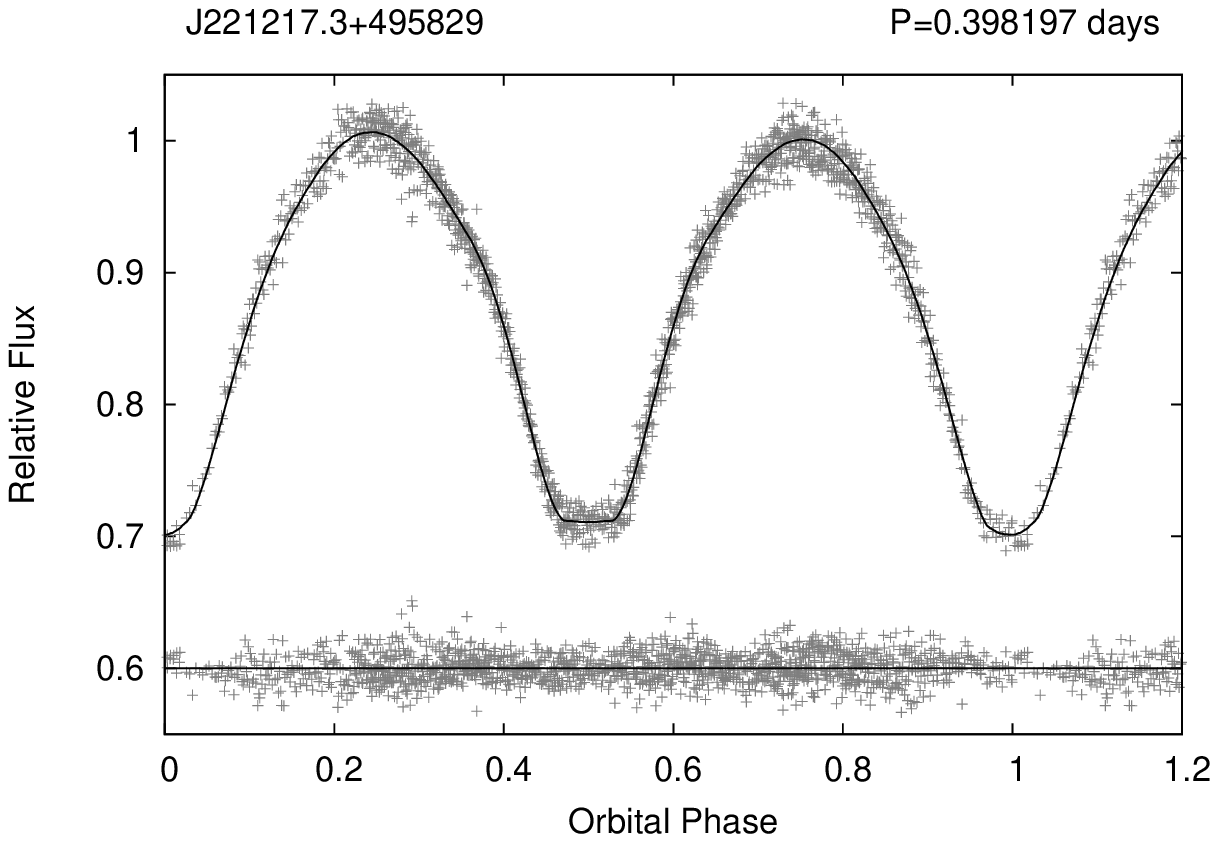}
}
\centerline{
\includegraphics[width=70mm]{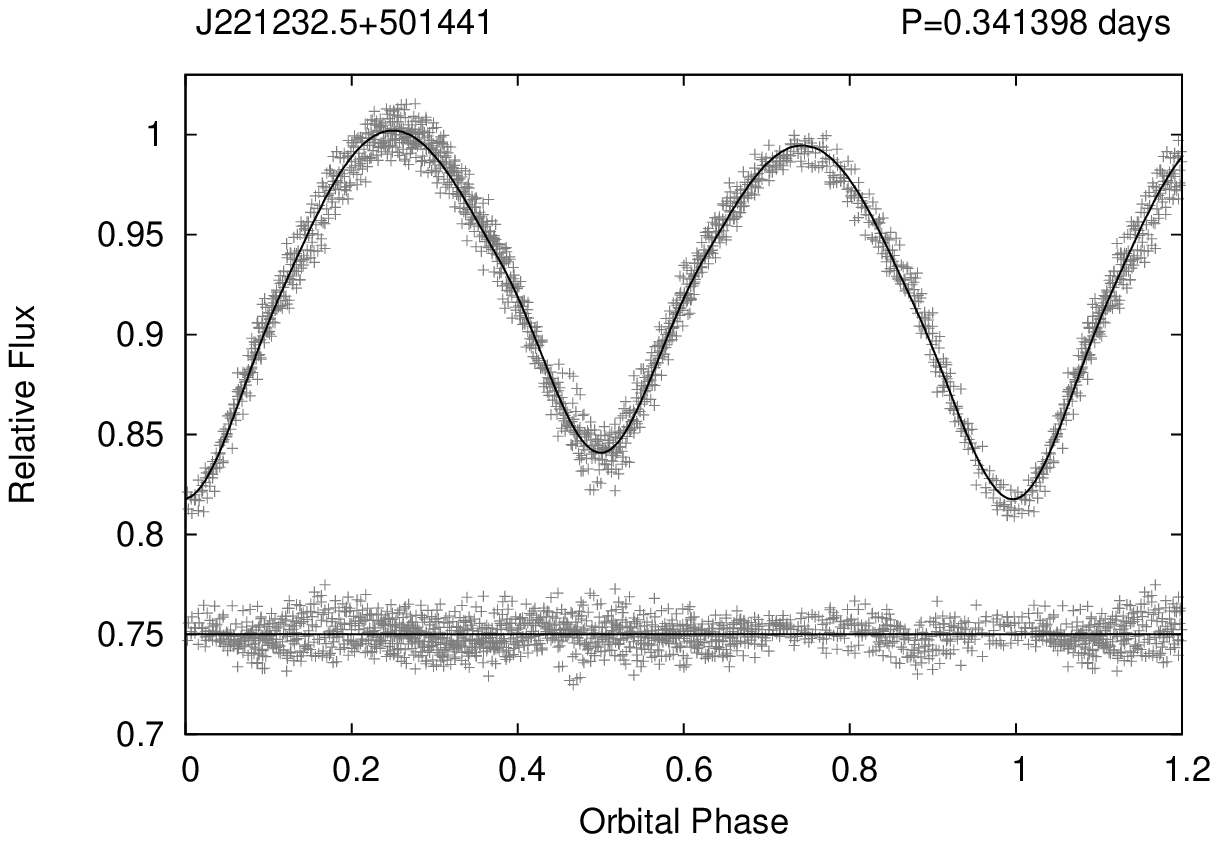} 
\includegraphics[width=70mm]{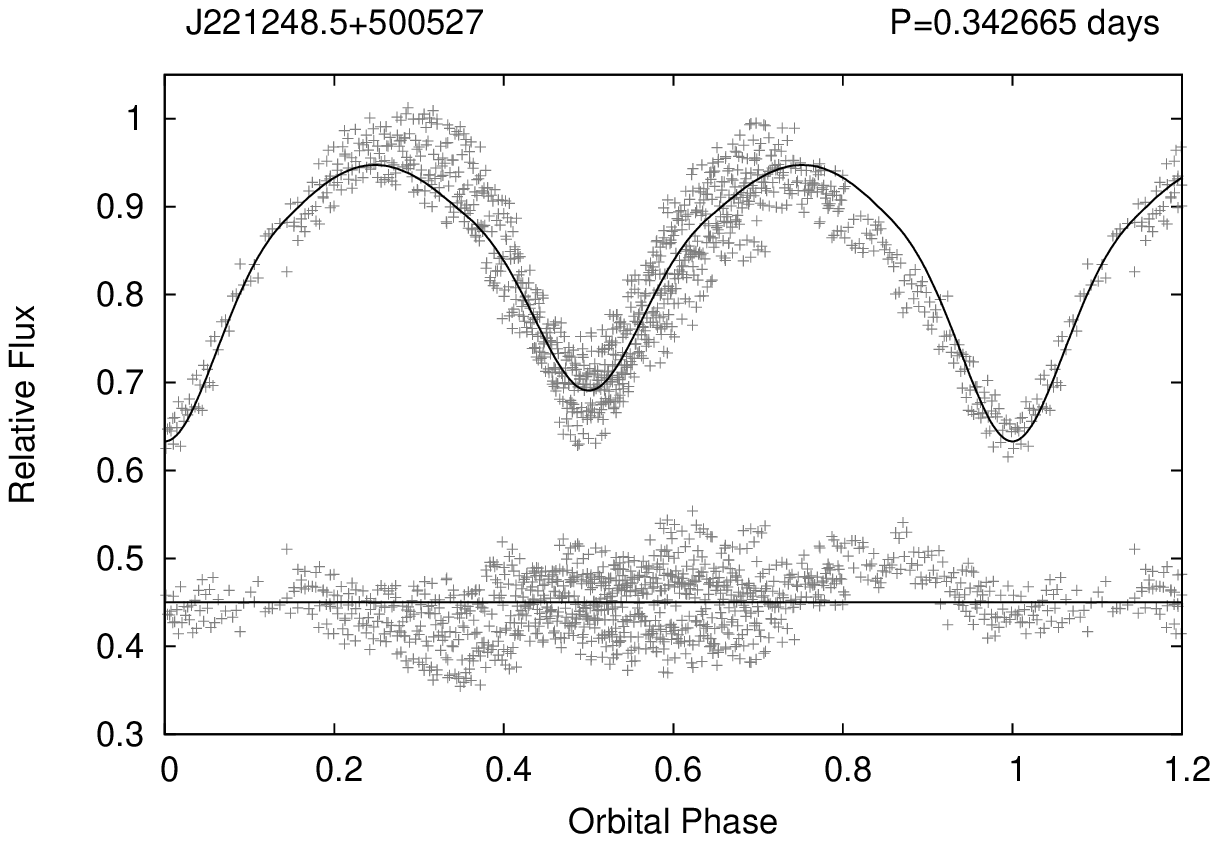}
}
\centerline{
\includegraphics[width=70mm]{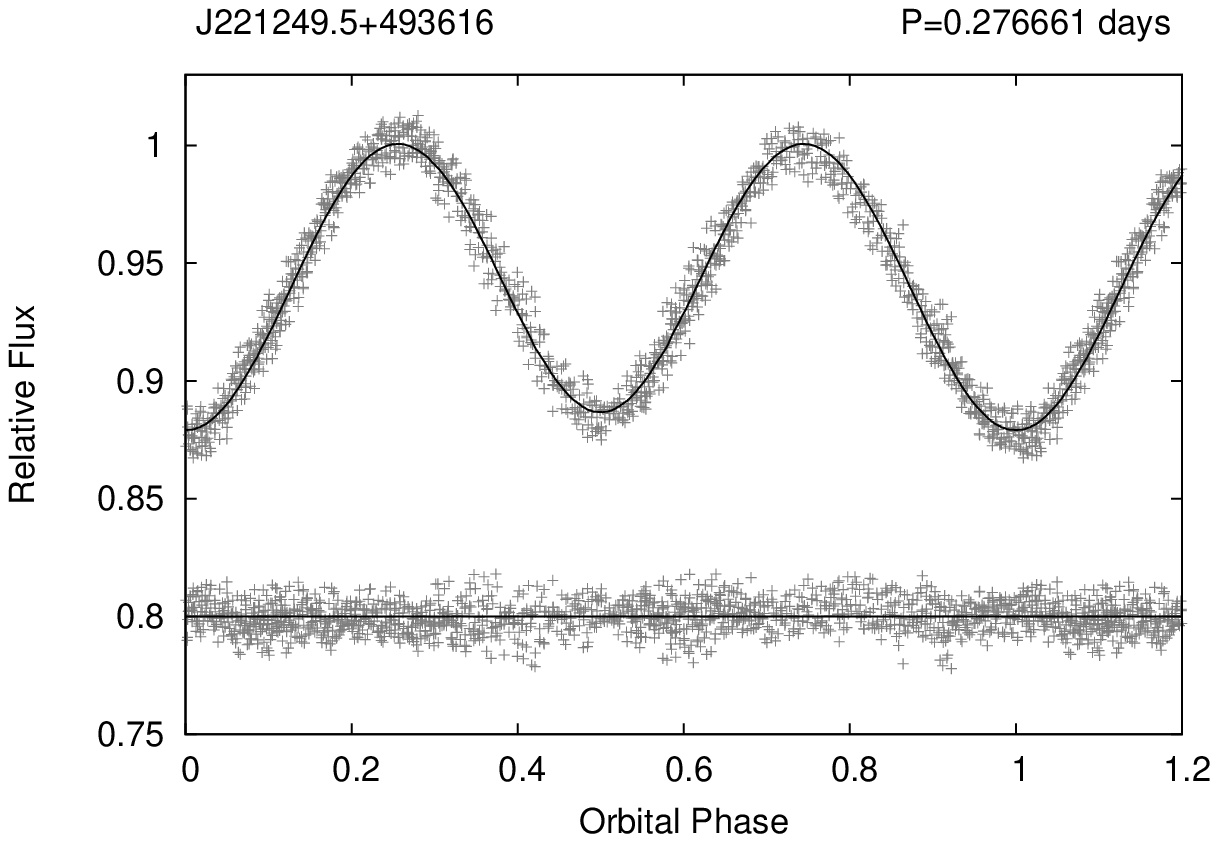} 
\includegraphics[width=70mm]{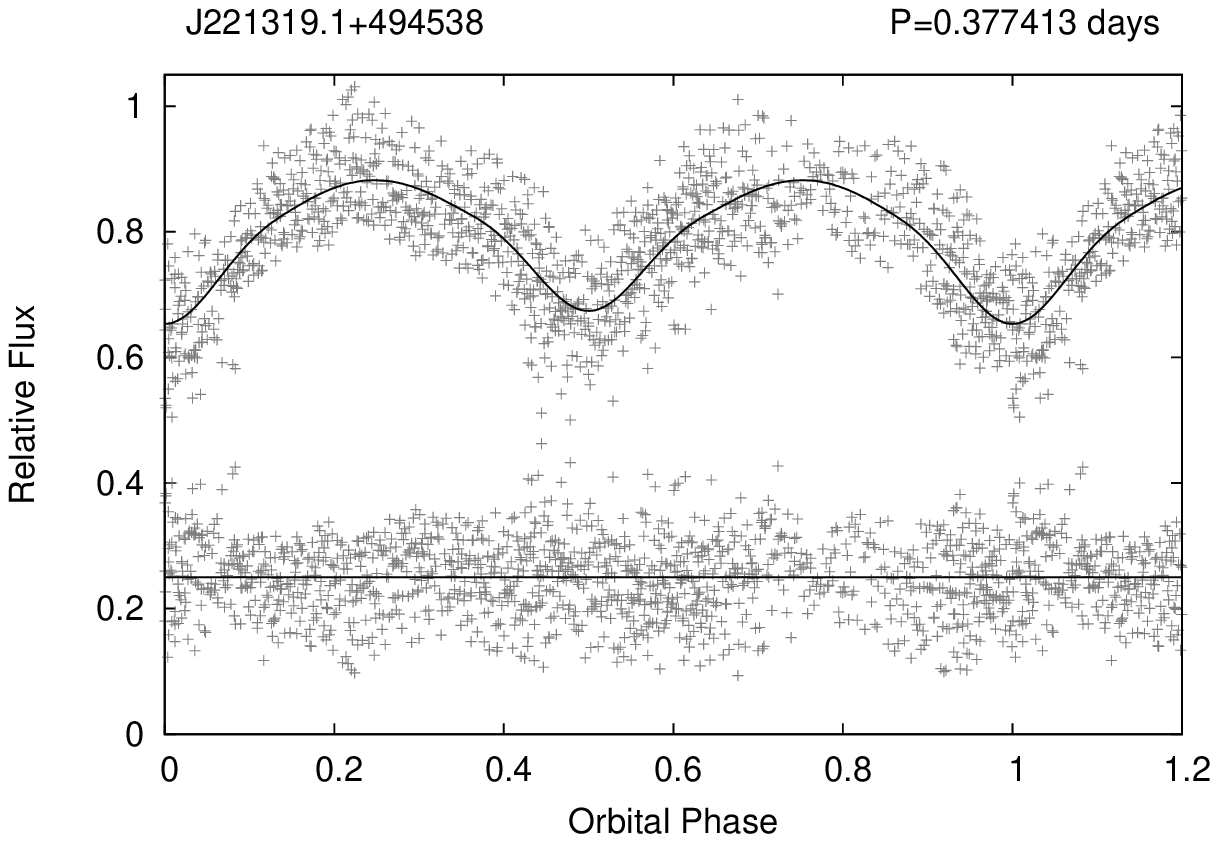} 
}
\centerline{
\includegraphics[width=70mm]{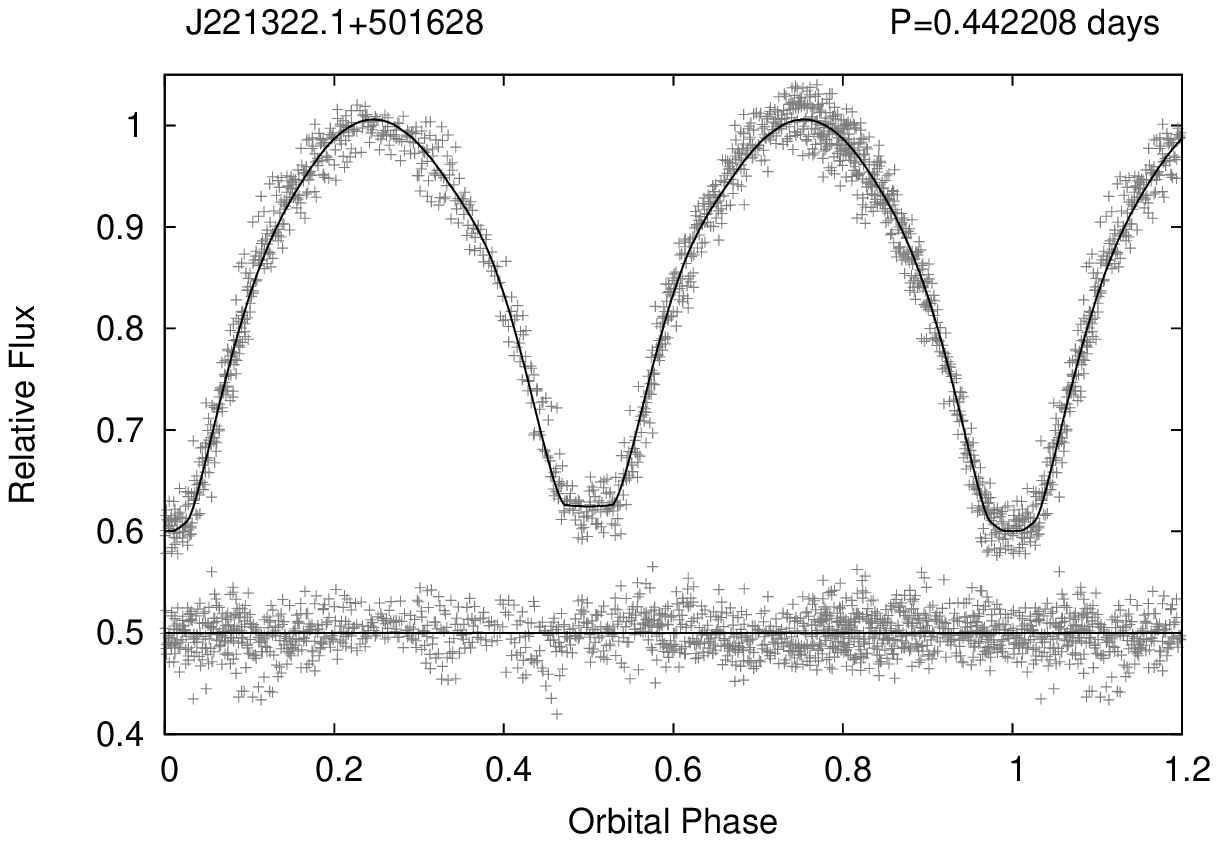}
\includegraphics[width=70mm]{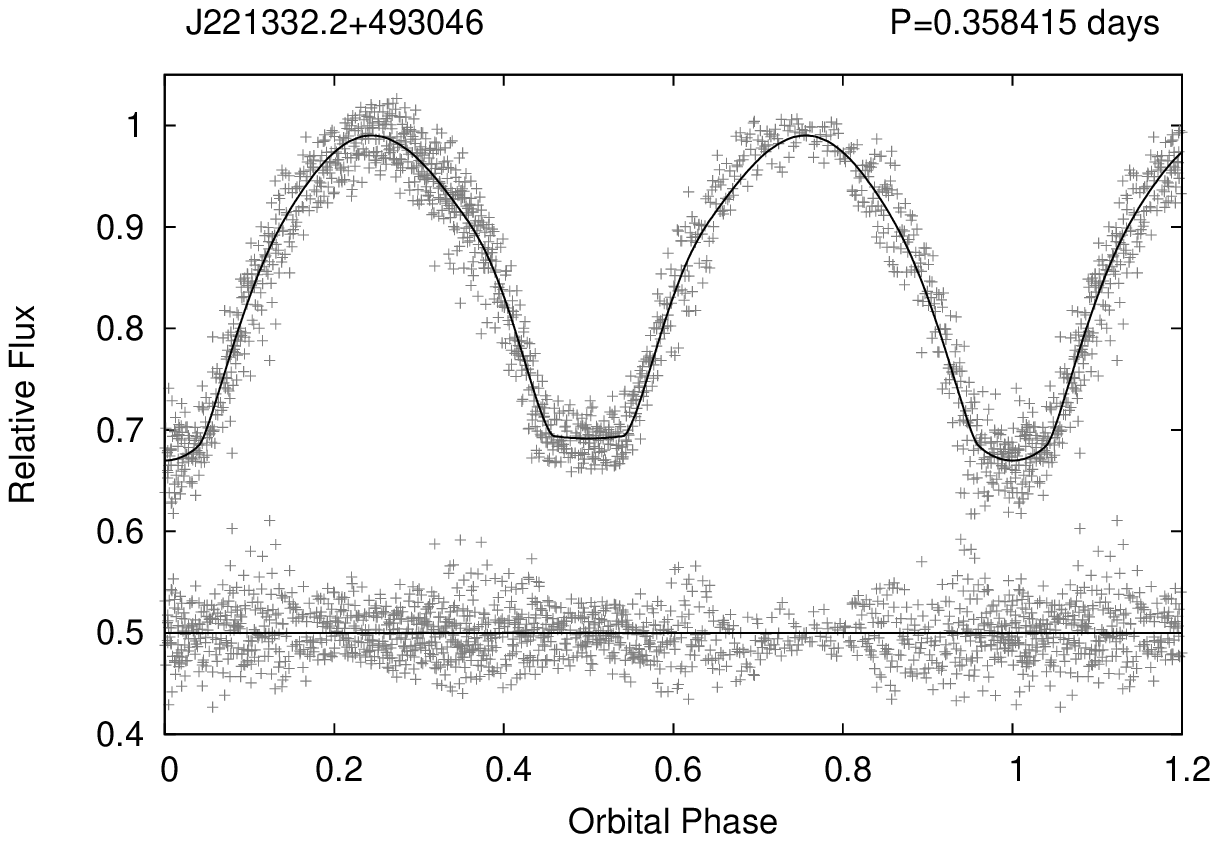} 
}

\caption{}
\label{Figure 1}
\end{figure*}

\begin{figure*}[!h]
\centering
\centerline{
\includegraphics[width=70mm]{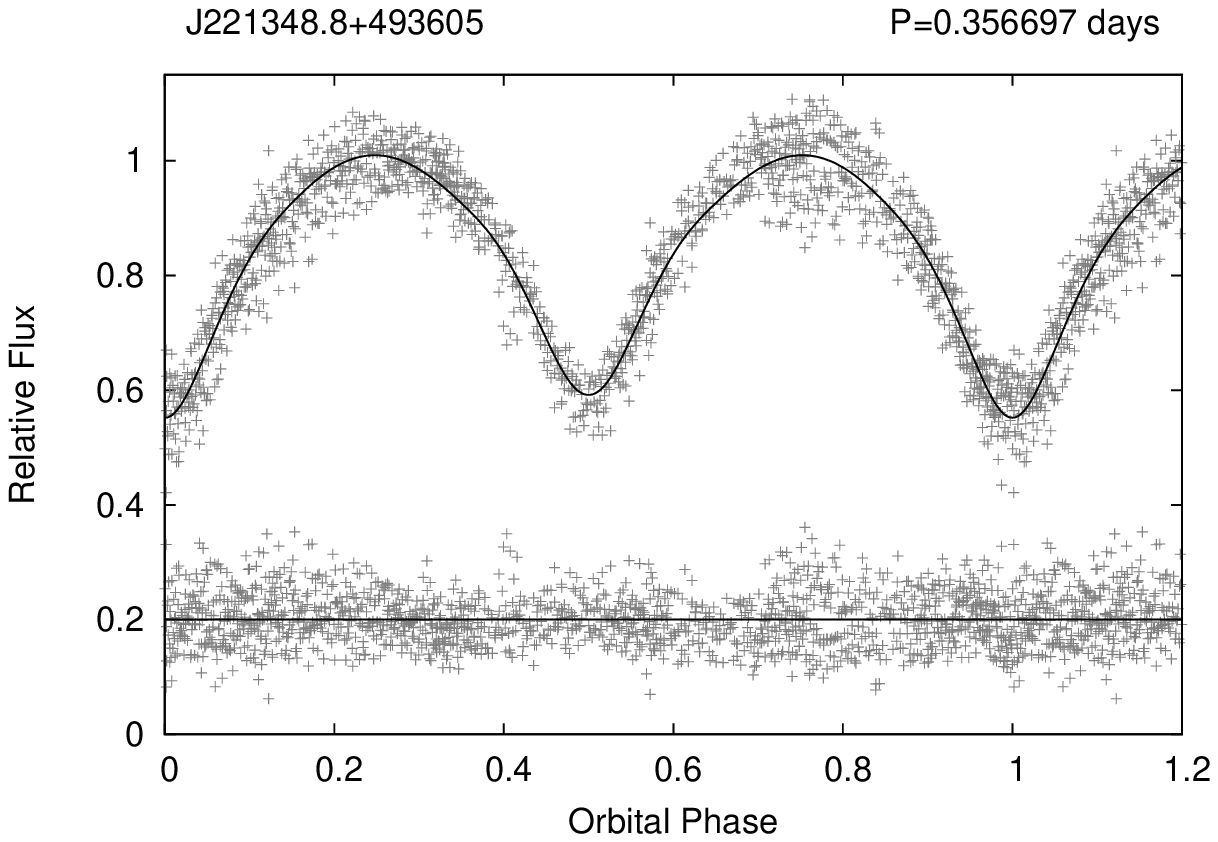} 
\includegraphics[width=70mm]{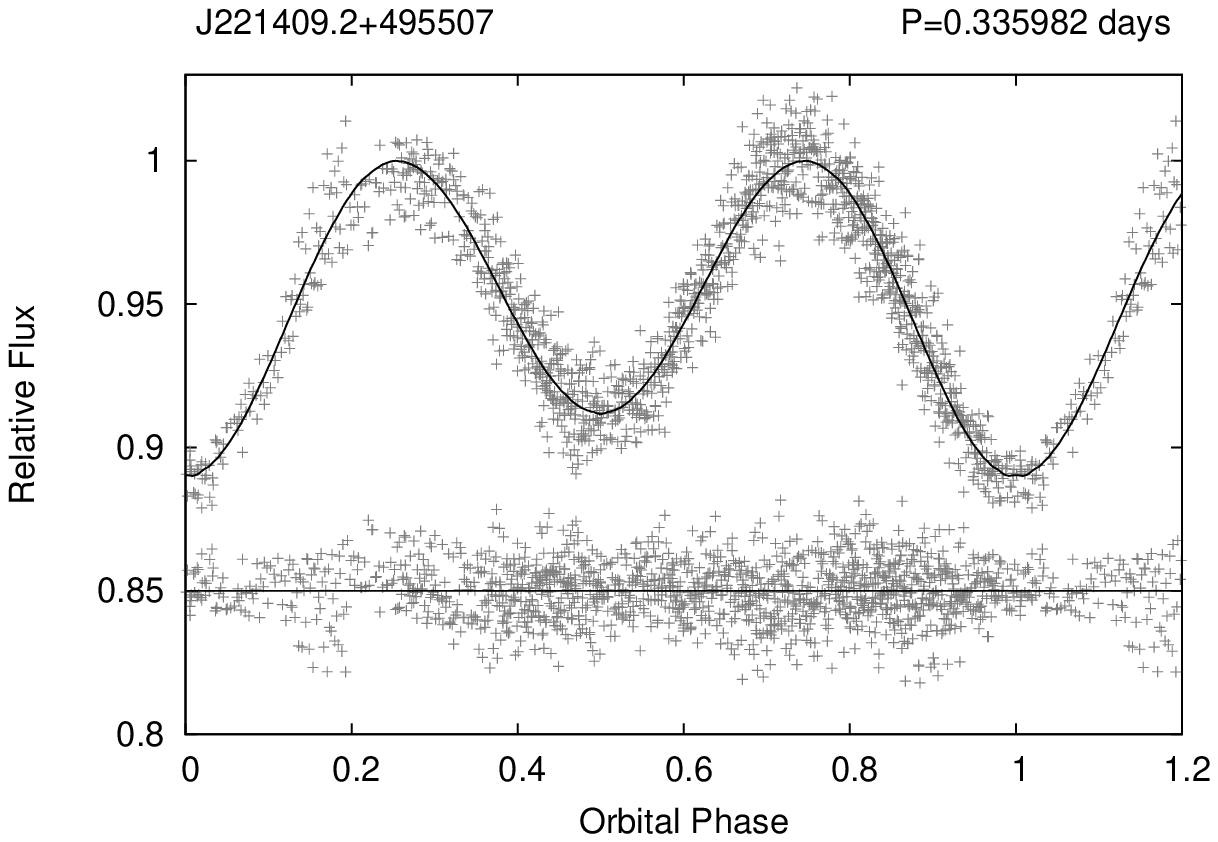}
} 
\centerline{
\includegraphics[width=70mm]{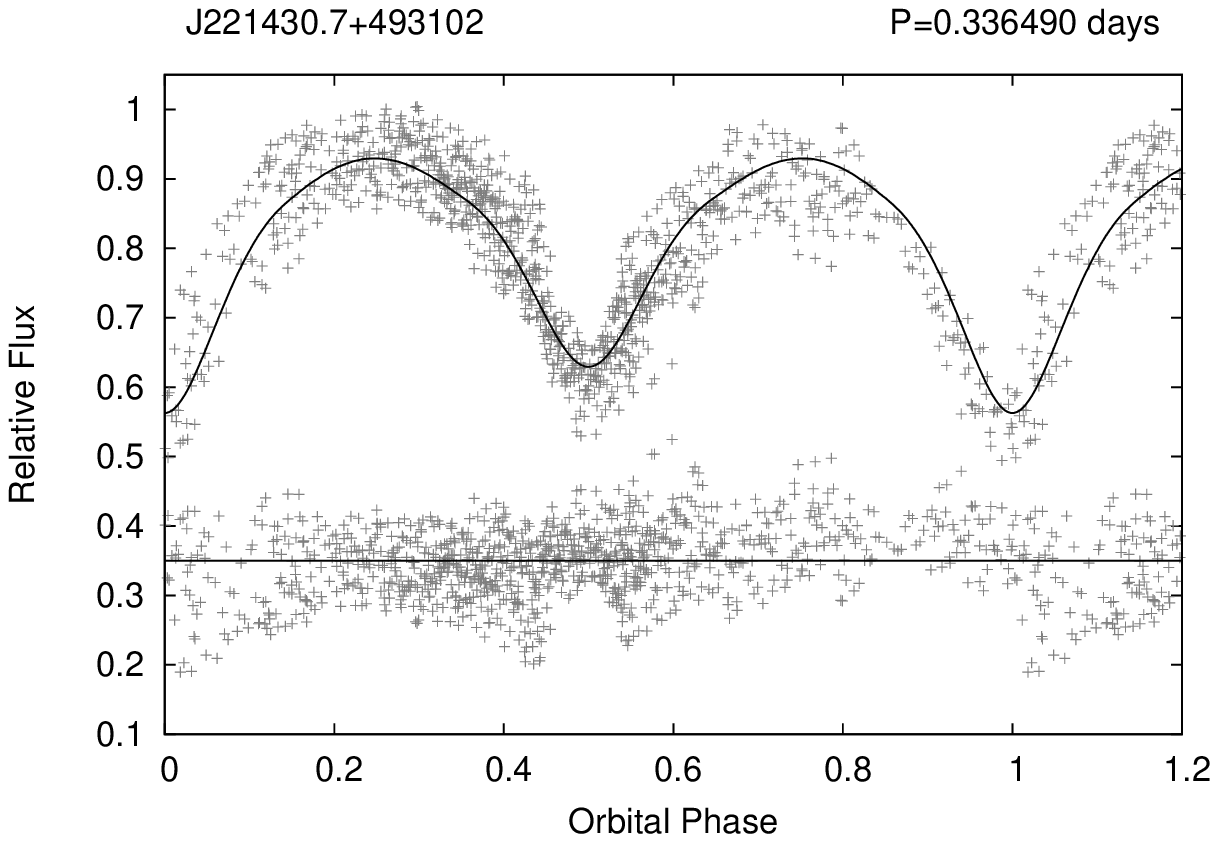} 
\includegraphics[width=70mm]{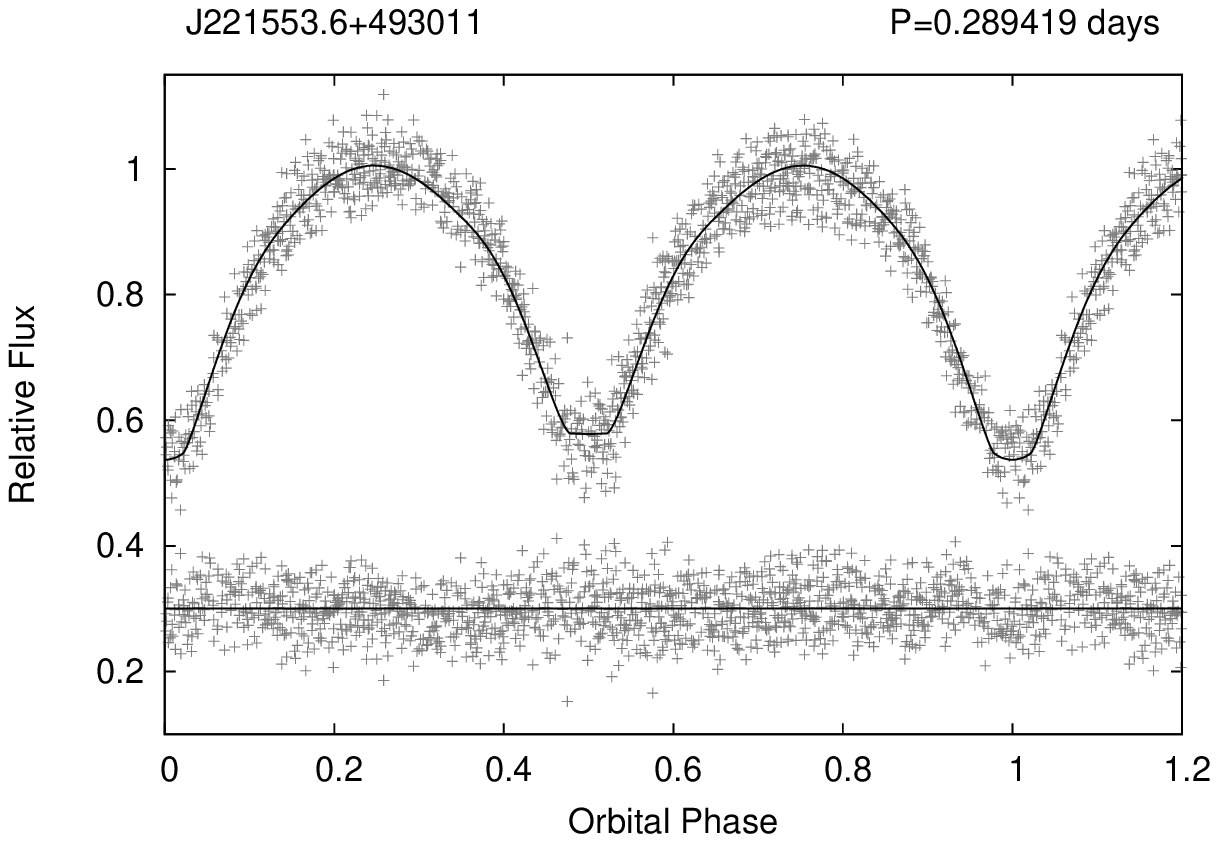} 
}
\centerline{
\includegraphics[width=70mm]{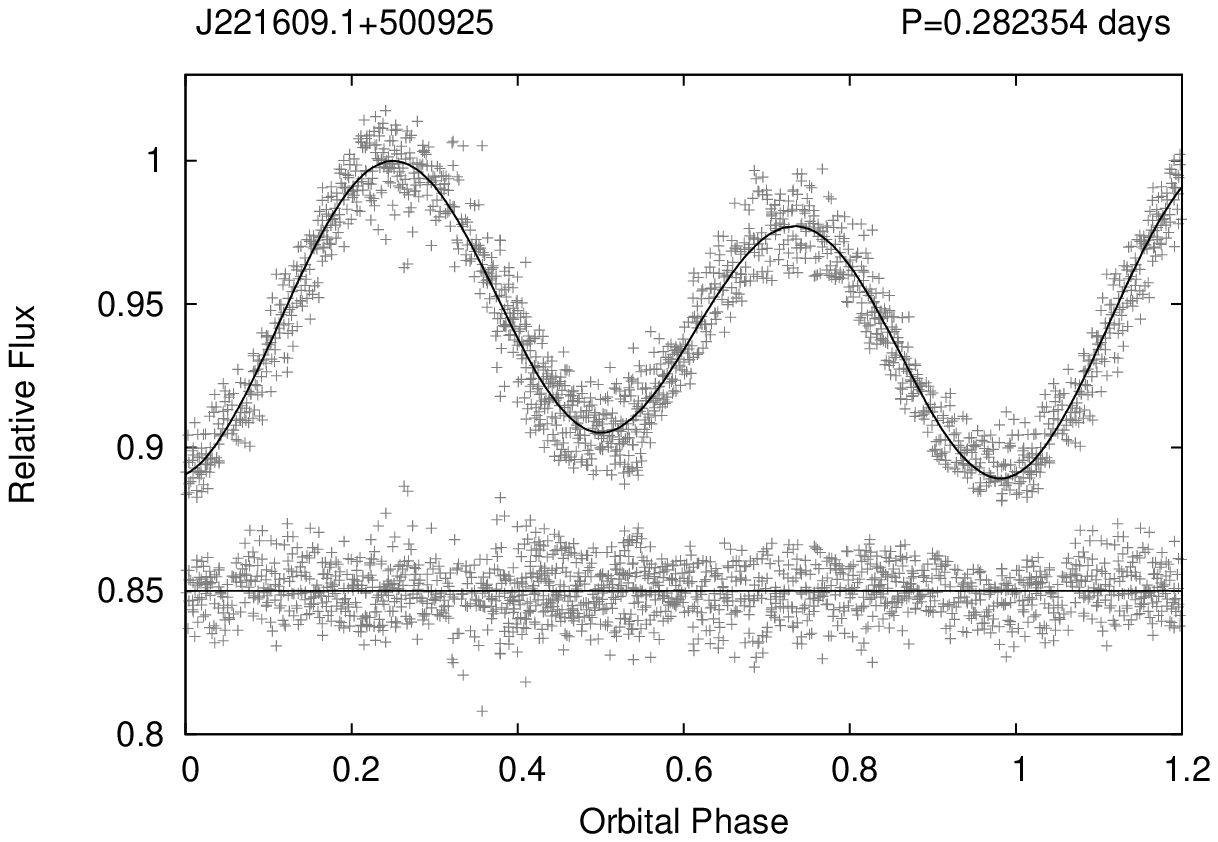} 
\includegraphics[width=70mm]{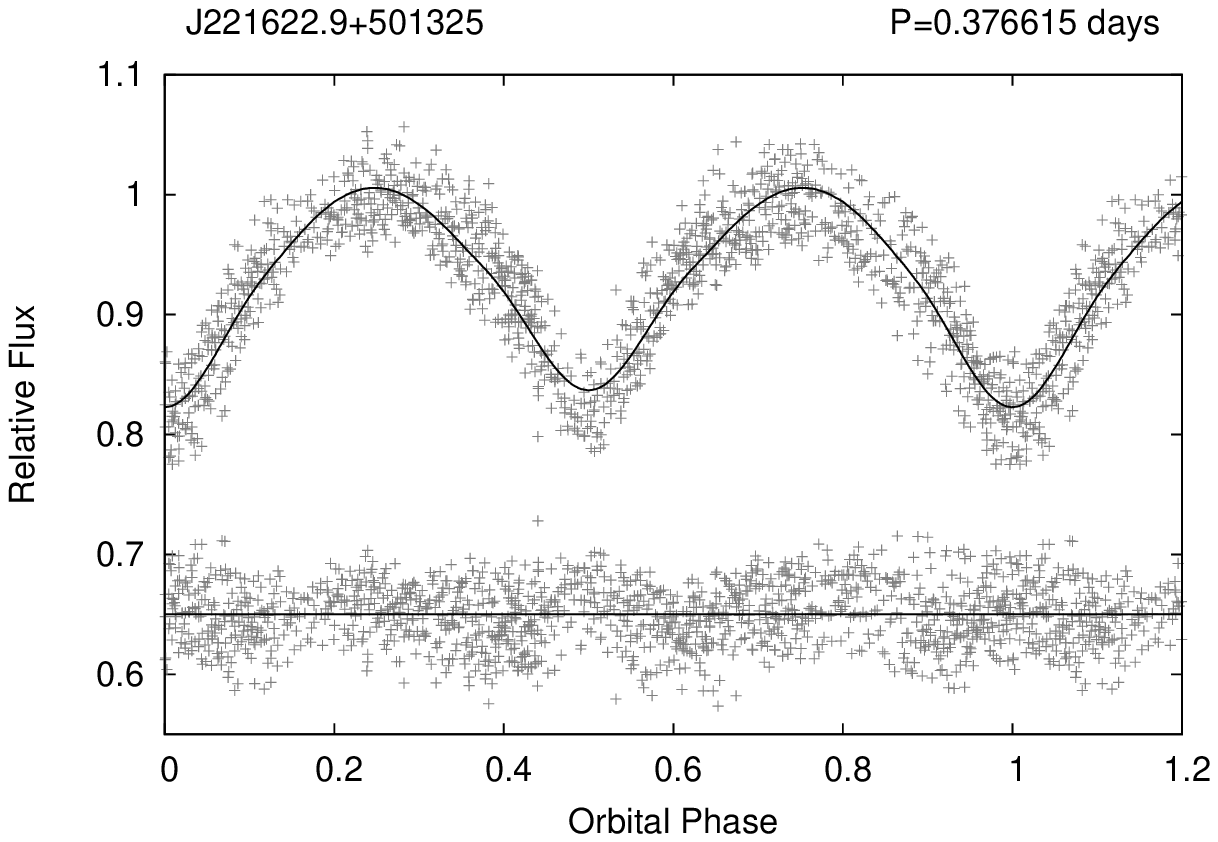}
}
\centerline{
\includegraphics[width=70mm]{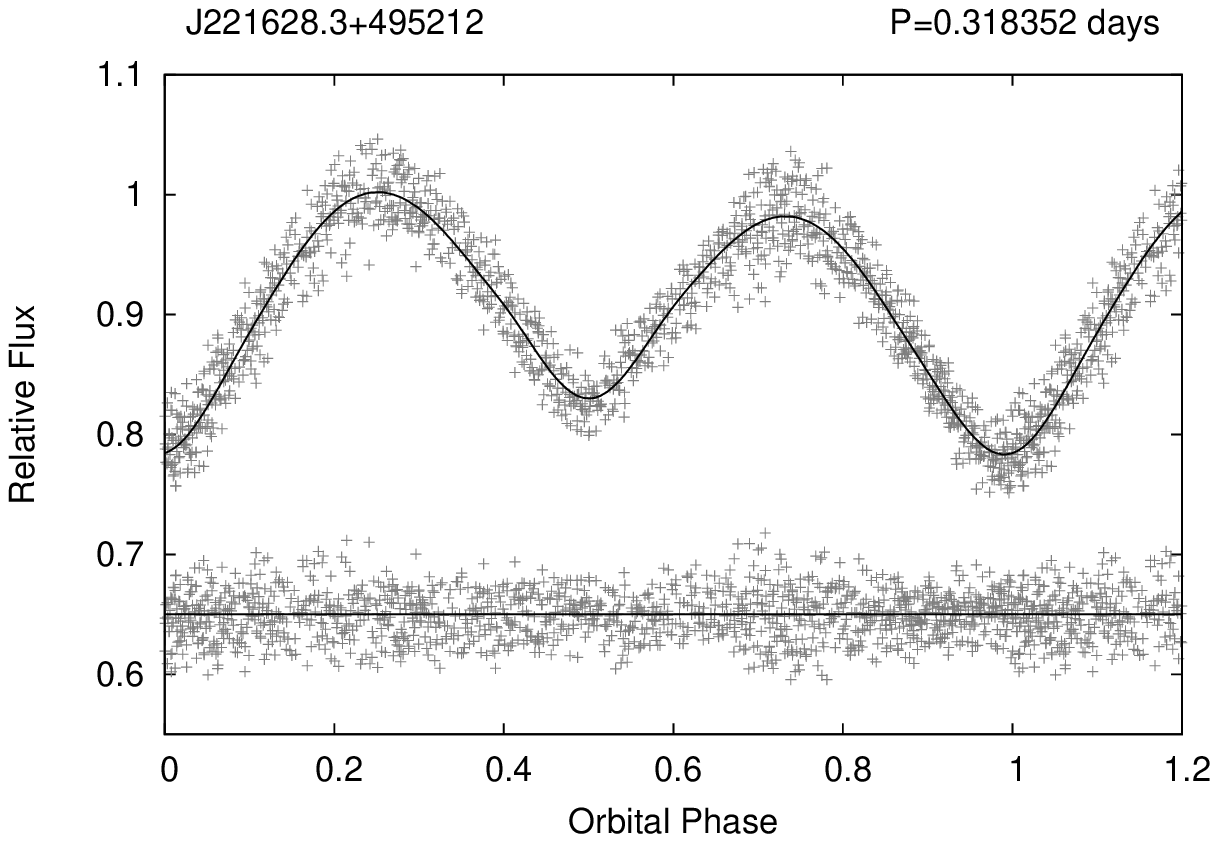}
}
\caption{}
\label{Figure 1}
\end{figure*}

\clearpage
\newpage
\section{An owerview of the periodograms (left) and non-phased one-night-run light curves (right) of the pulsating variable stars}

\begin{figure*}[!h]
\centering
\centerline{
\includegraphics[width=70mm]{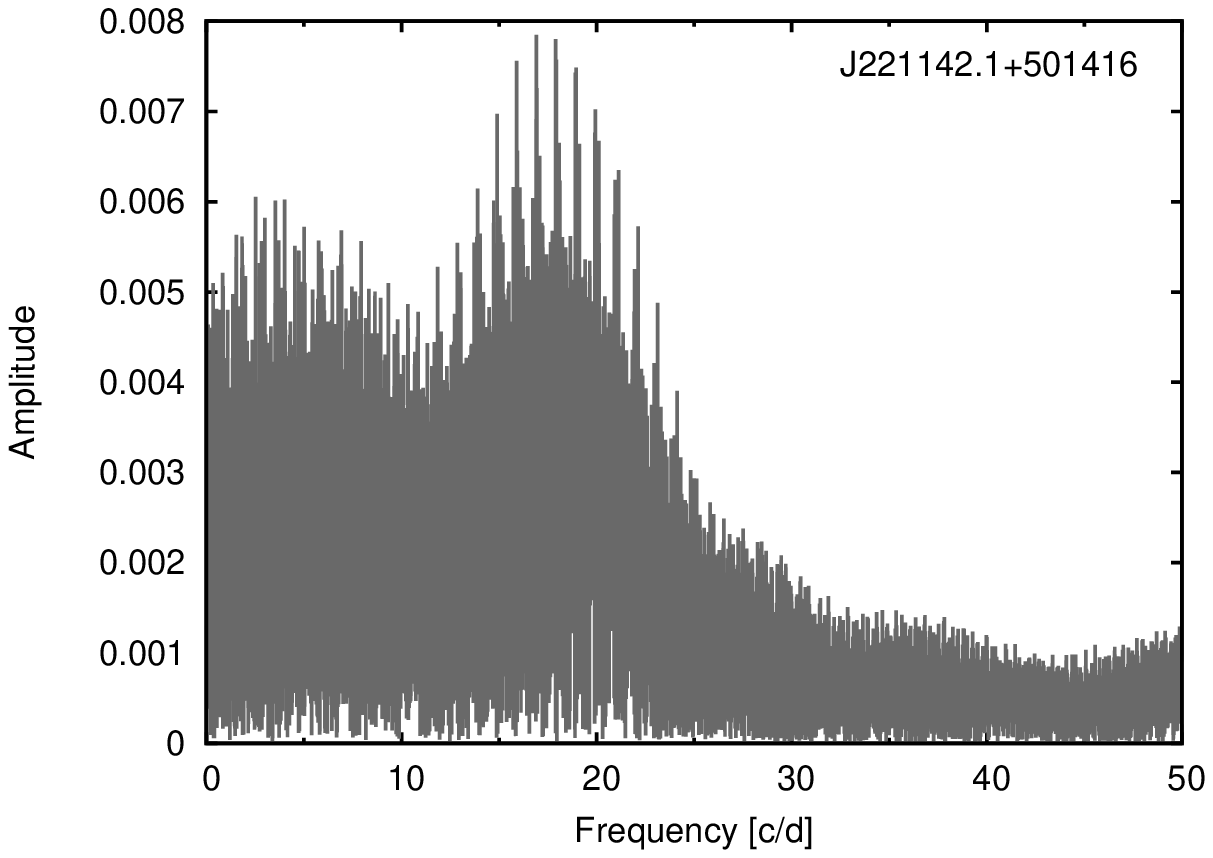}
\includegraphics[width=70mm]{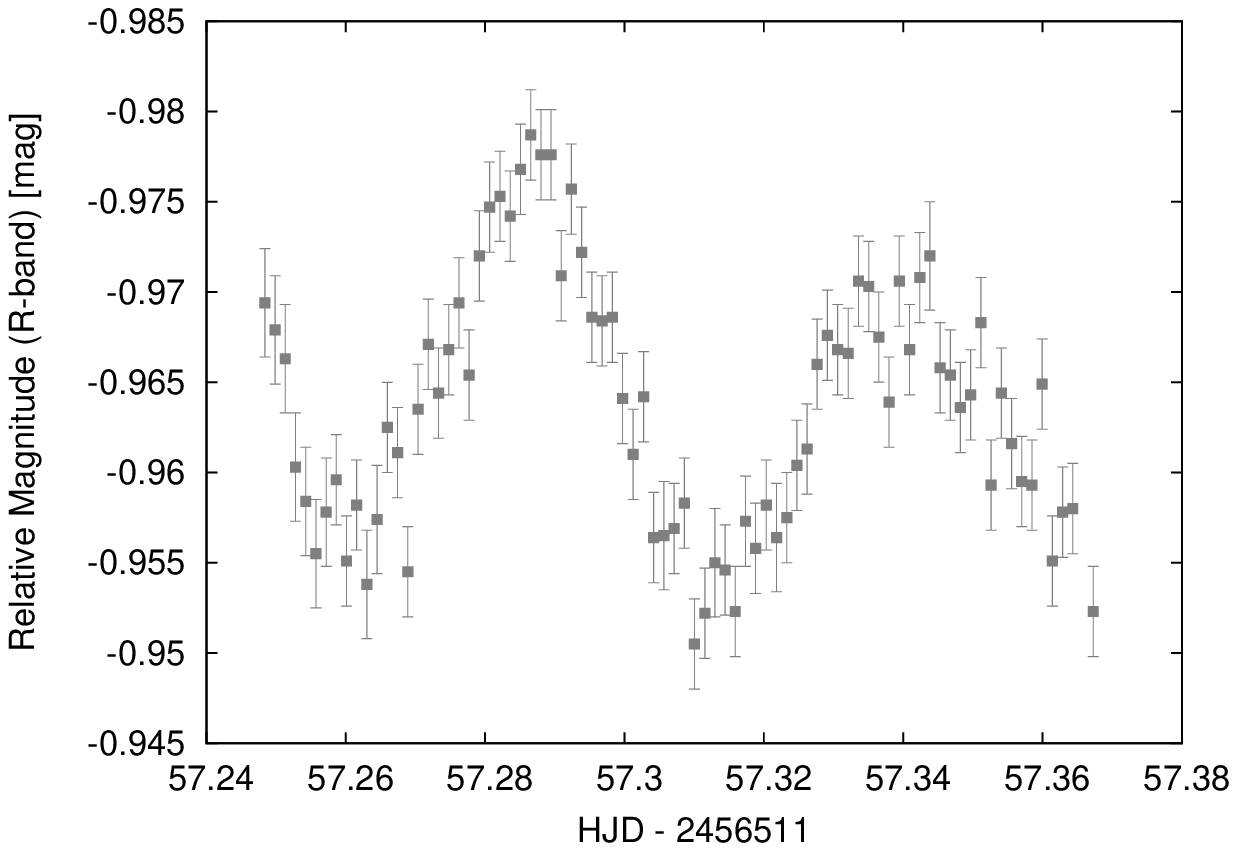}  
}
\centerline{
\includegraphics[width=70mm]{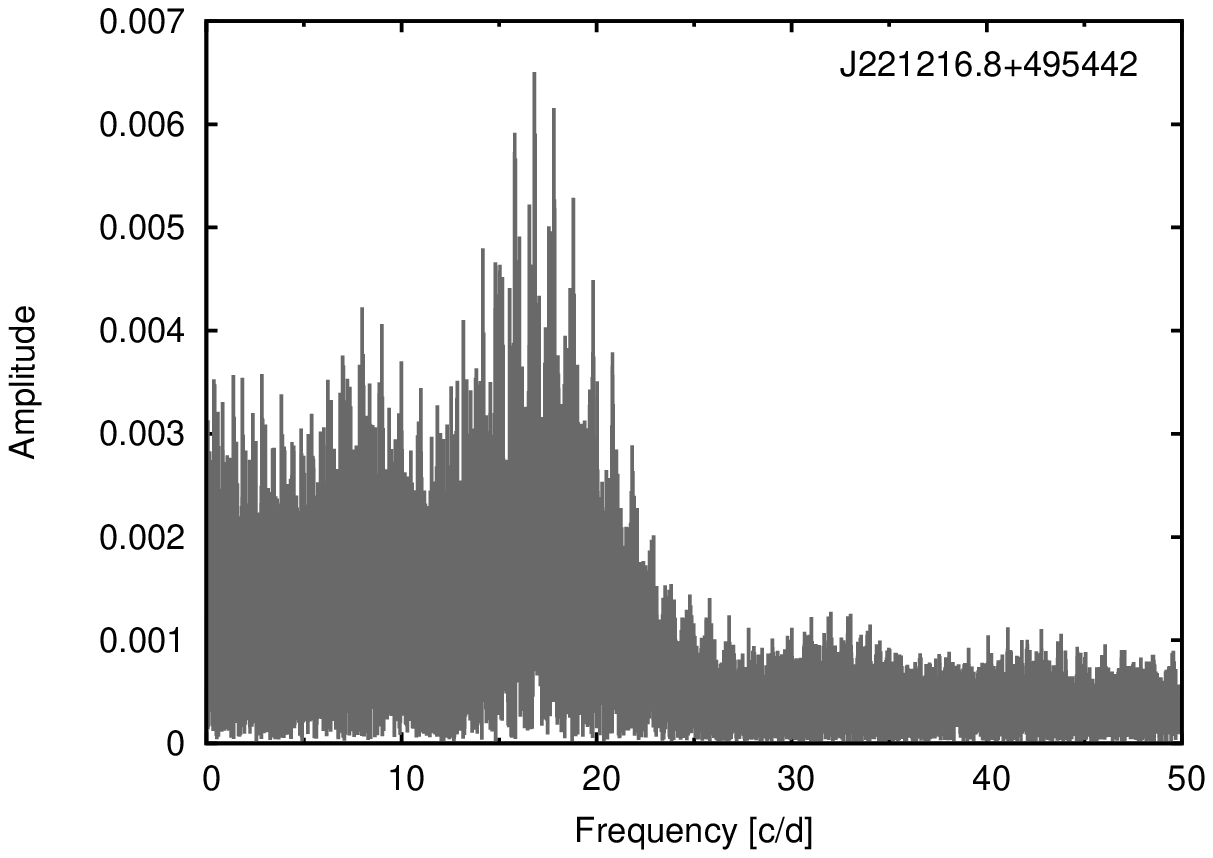}
\includegraphics[width=70mm]{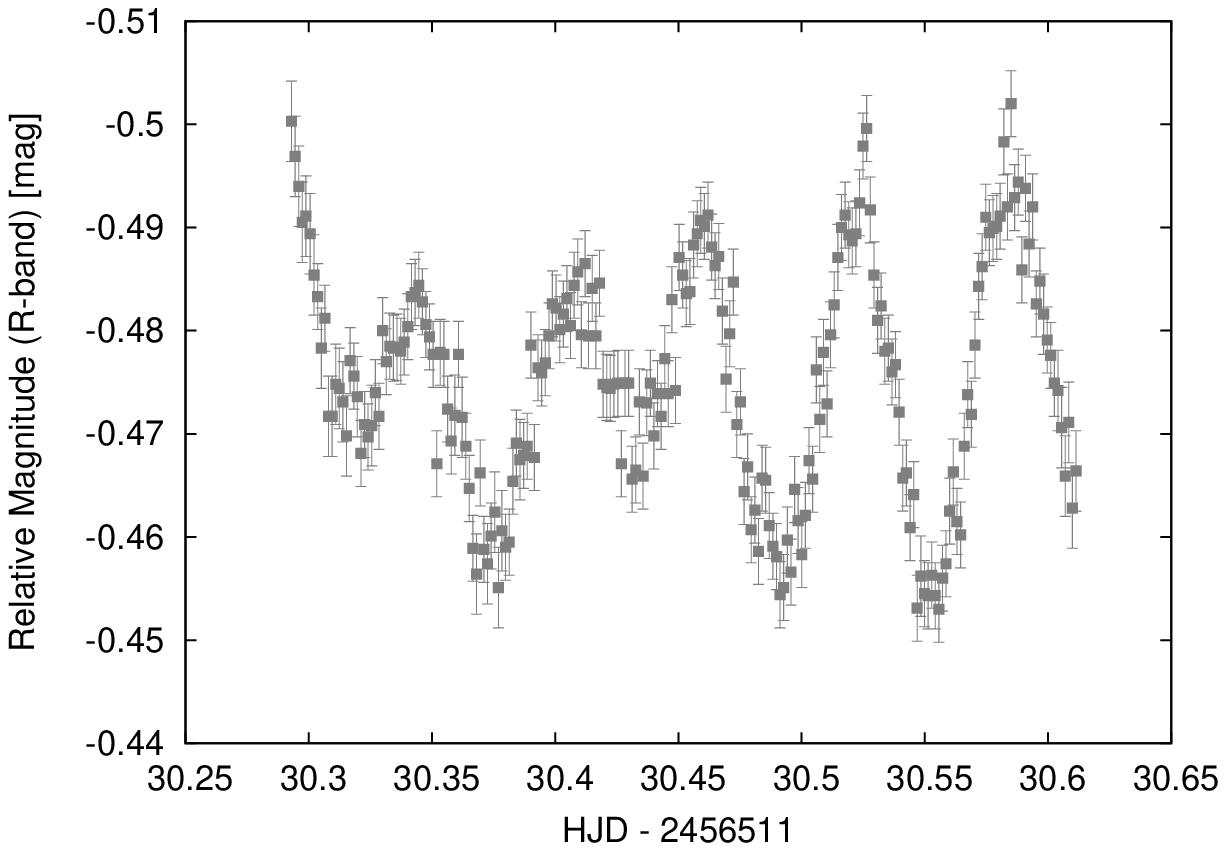}  
}
\centerline{
\includegraphics[width=70mm]{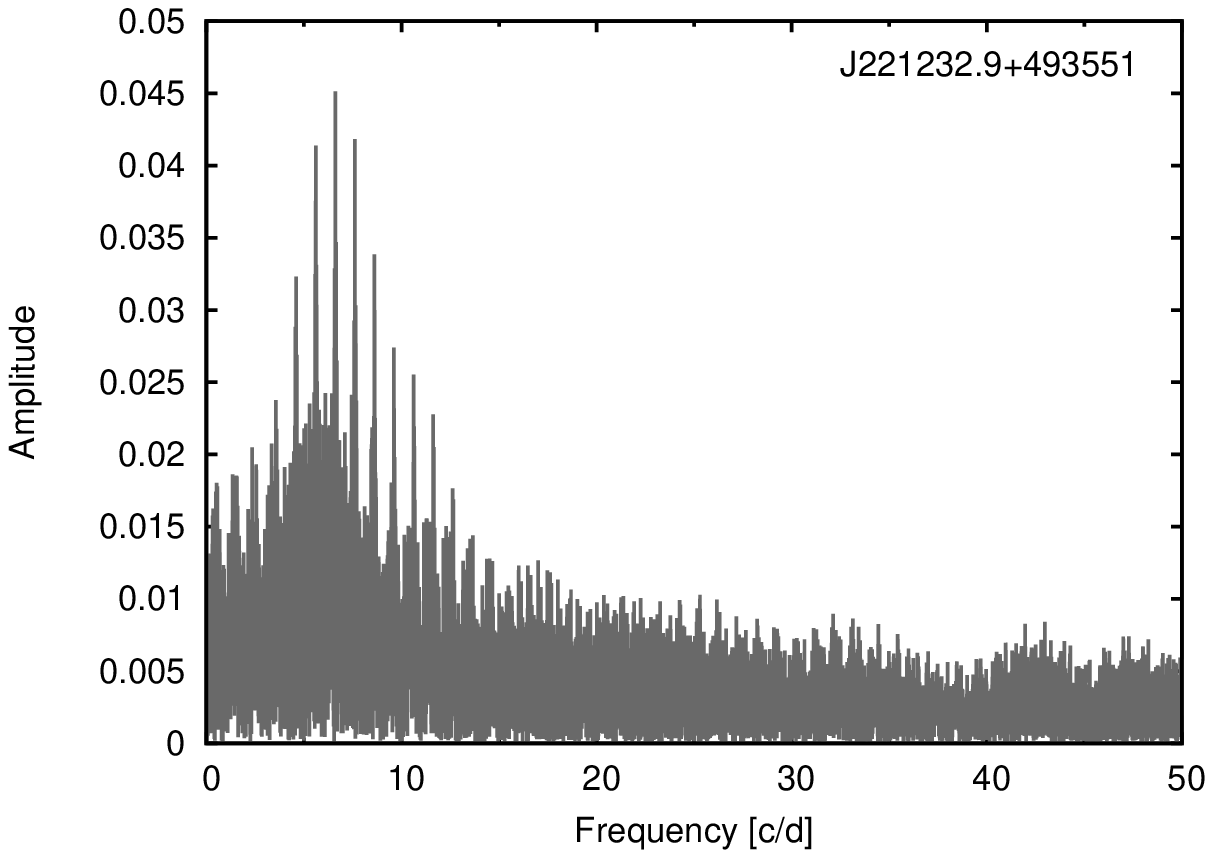}
\includegraphics[width=70mm]{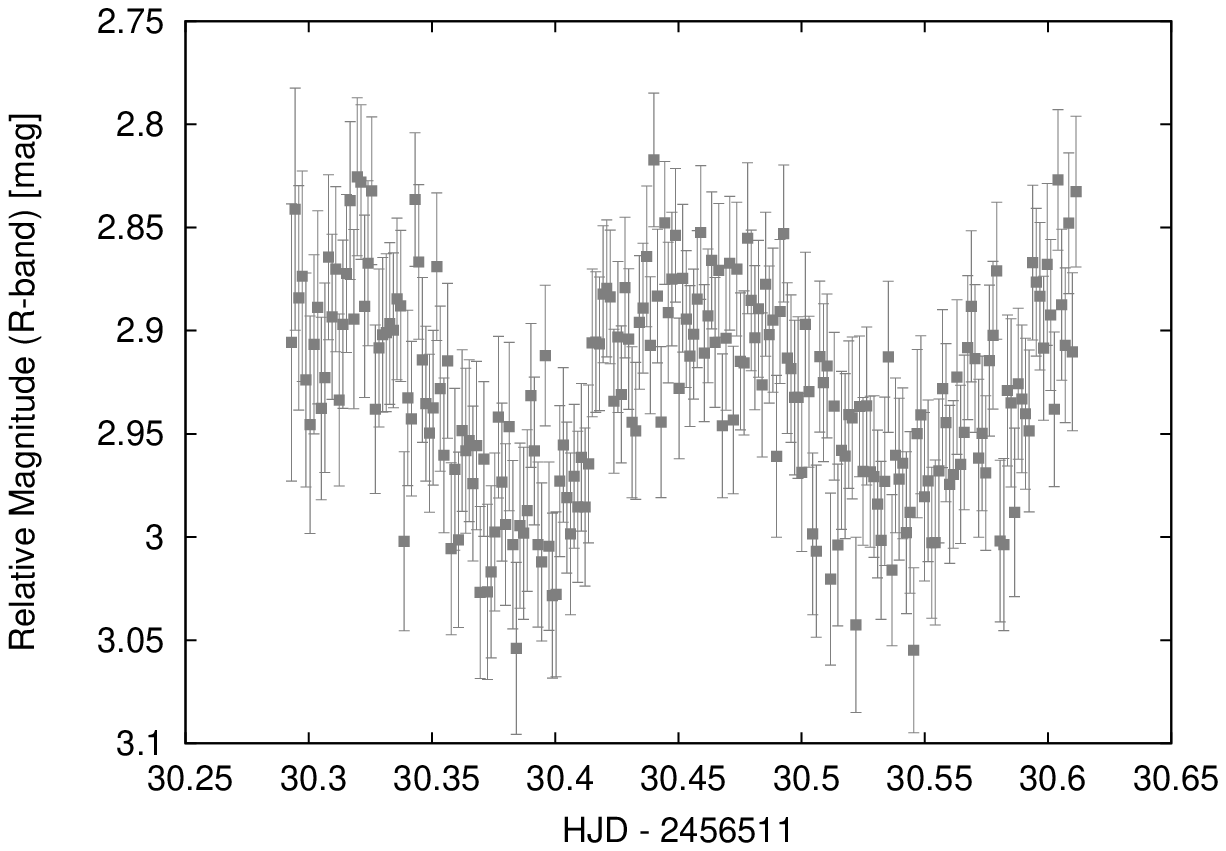}
}
\centerline{
\includegraphics[width=70mm]{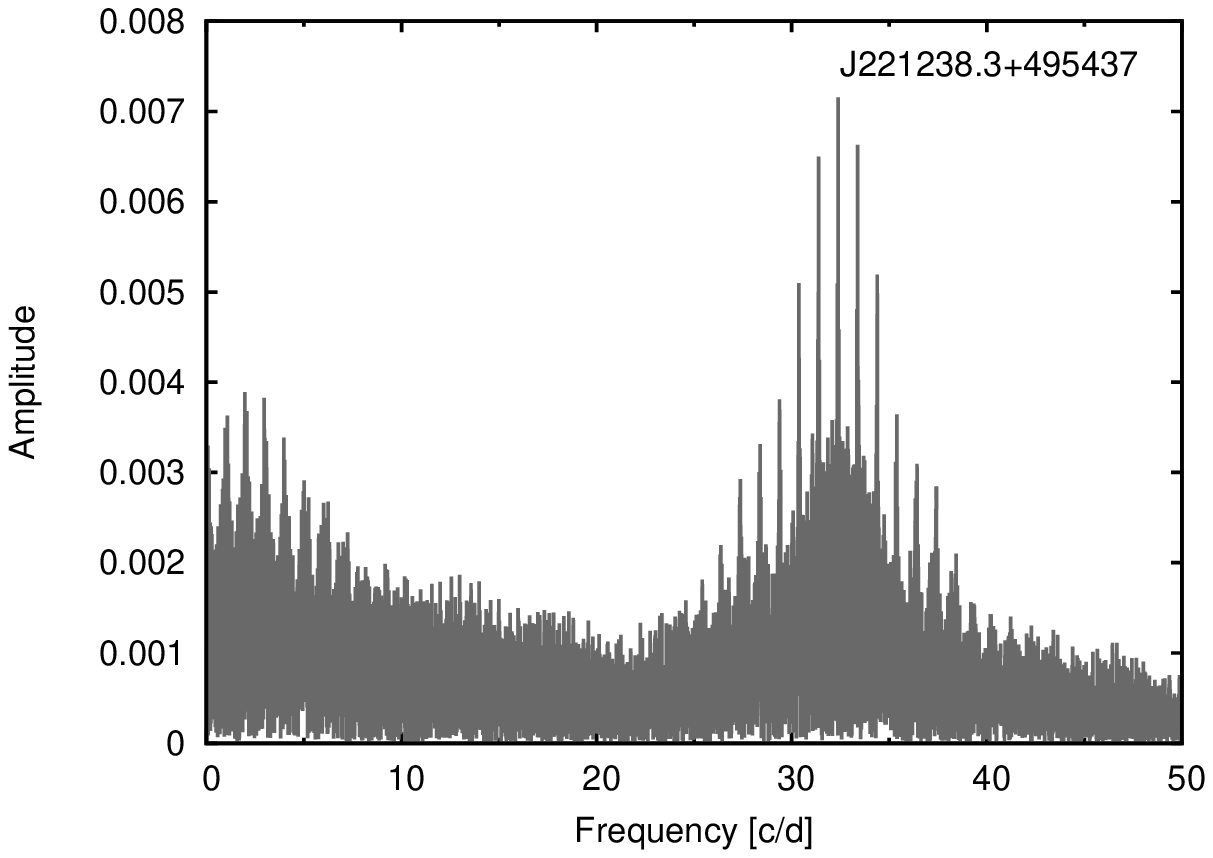}
\includegraphics[width=70mm]{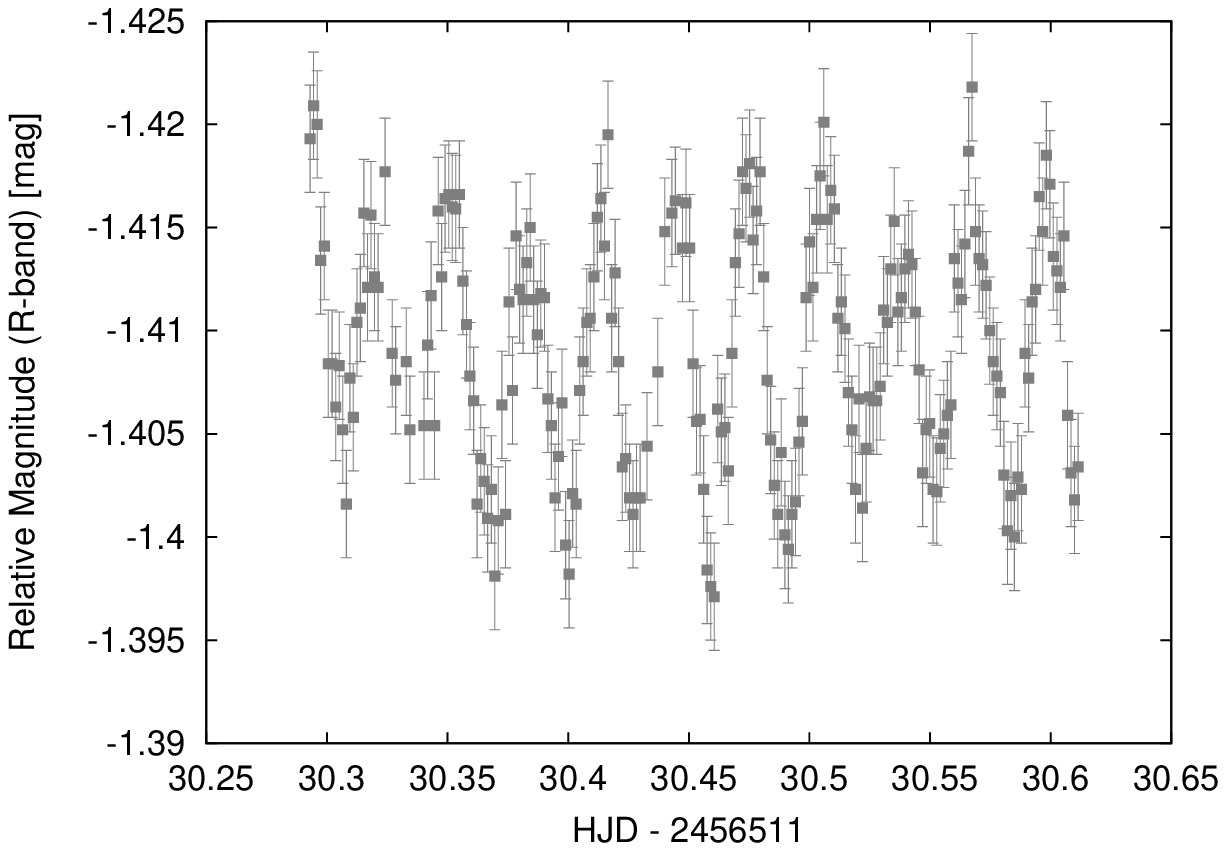}  
}
\caption{}
\label{Fig. 1}
\end{figure*} 

\begin{figure*}[!h]
\centering
\centerline{
\includegraphics[width=70mm]{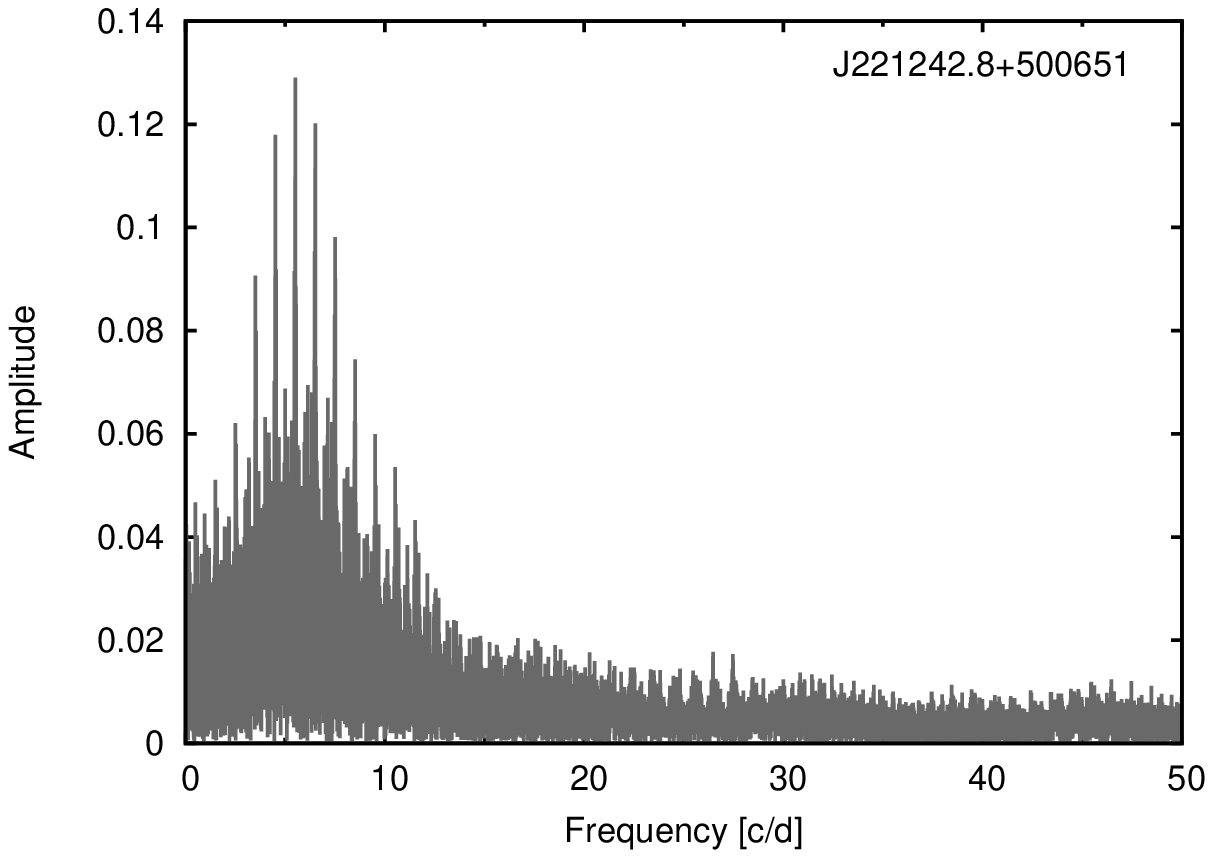}
\includegraphics[width=70mm]{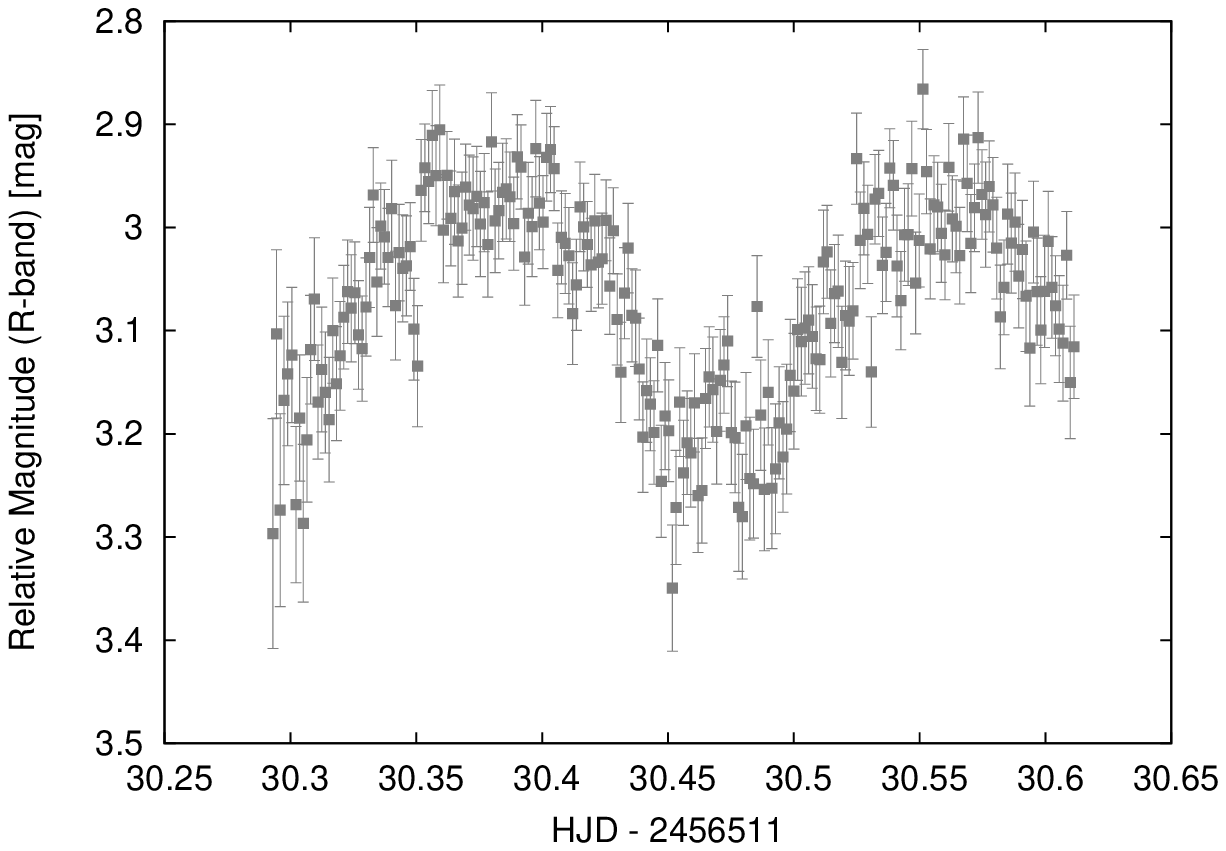}  
}
\centerline{
\includegraphics[width=70mm]{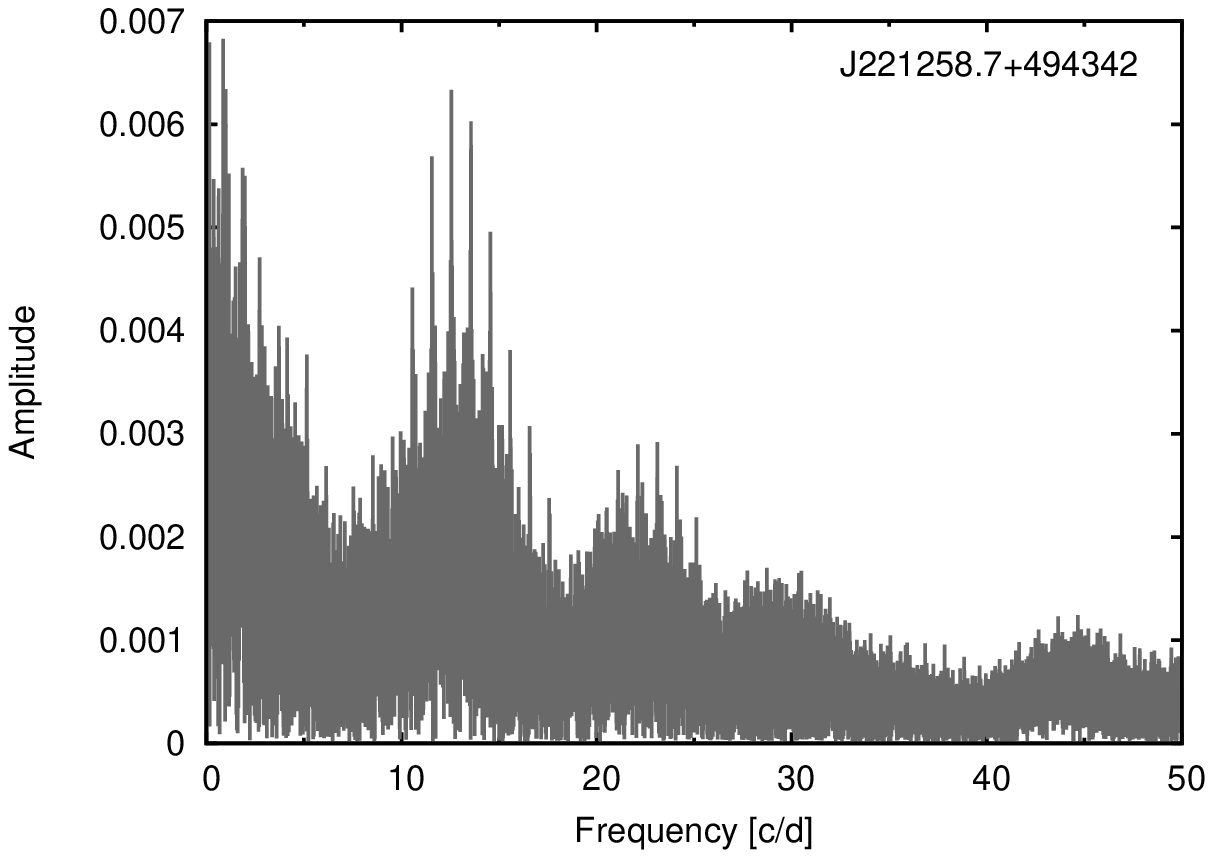}
\includegraphics[width=70mm]{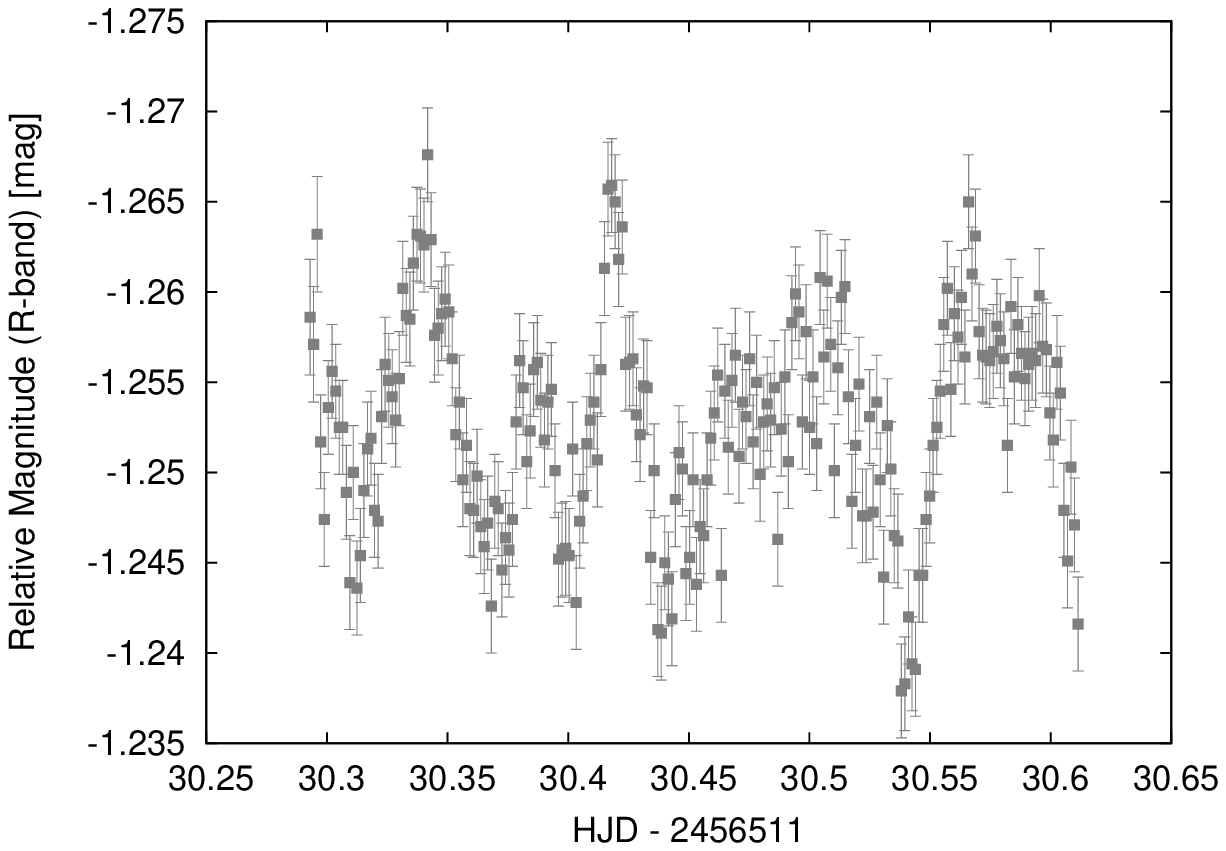}  
}
\centerline{
\includegraphics[width=70mm]{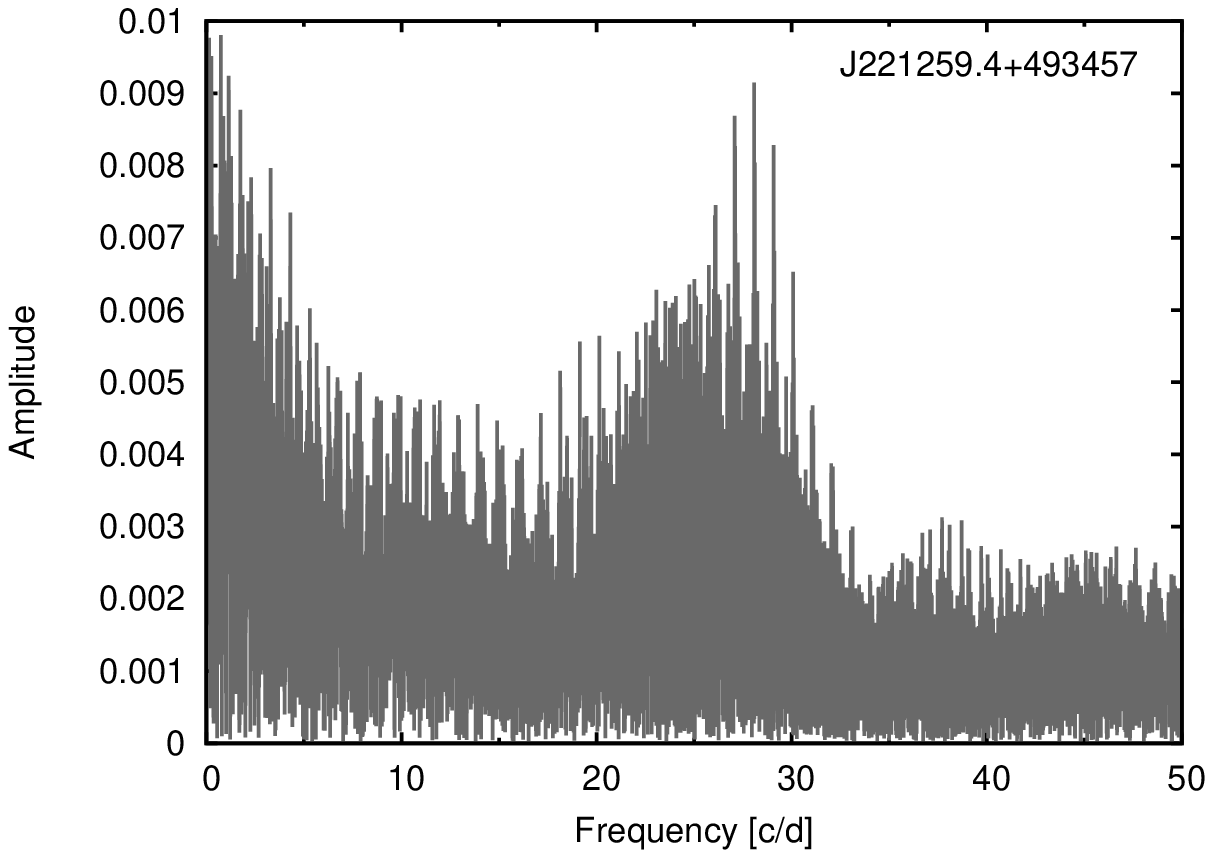}
\includegraphics[width=70mm]{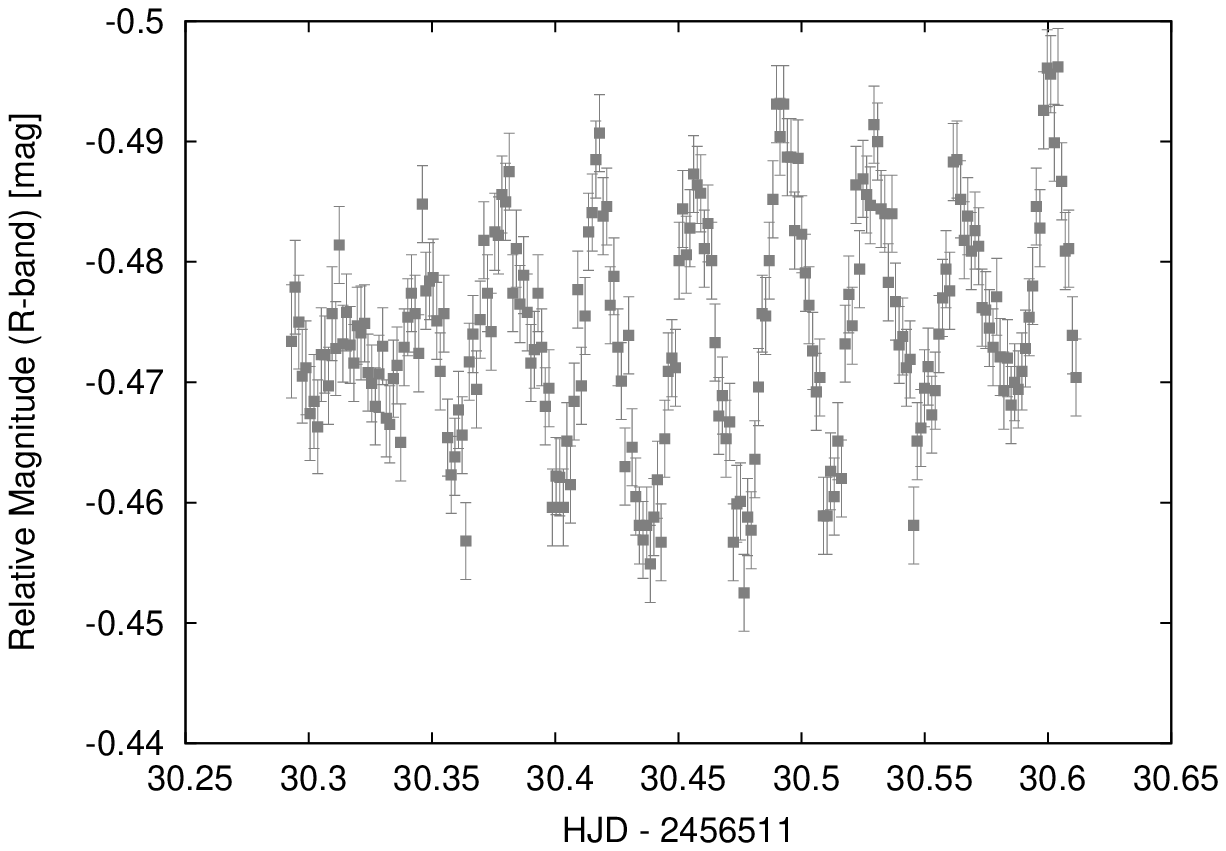}
}
\centerline{
\includegraphics[width=70mm]{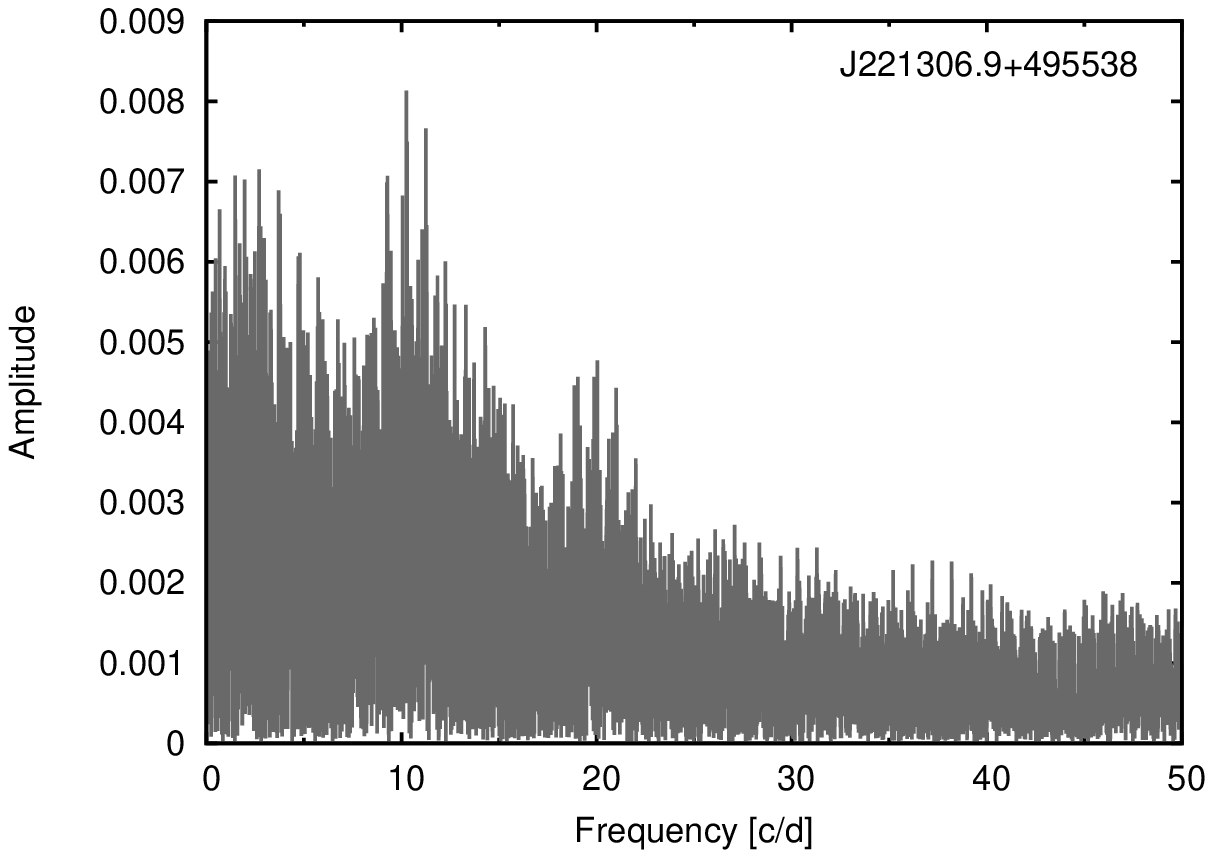}
\includegraphics[width=70mm]{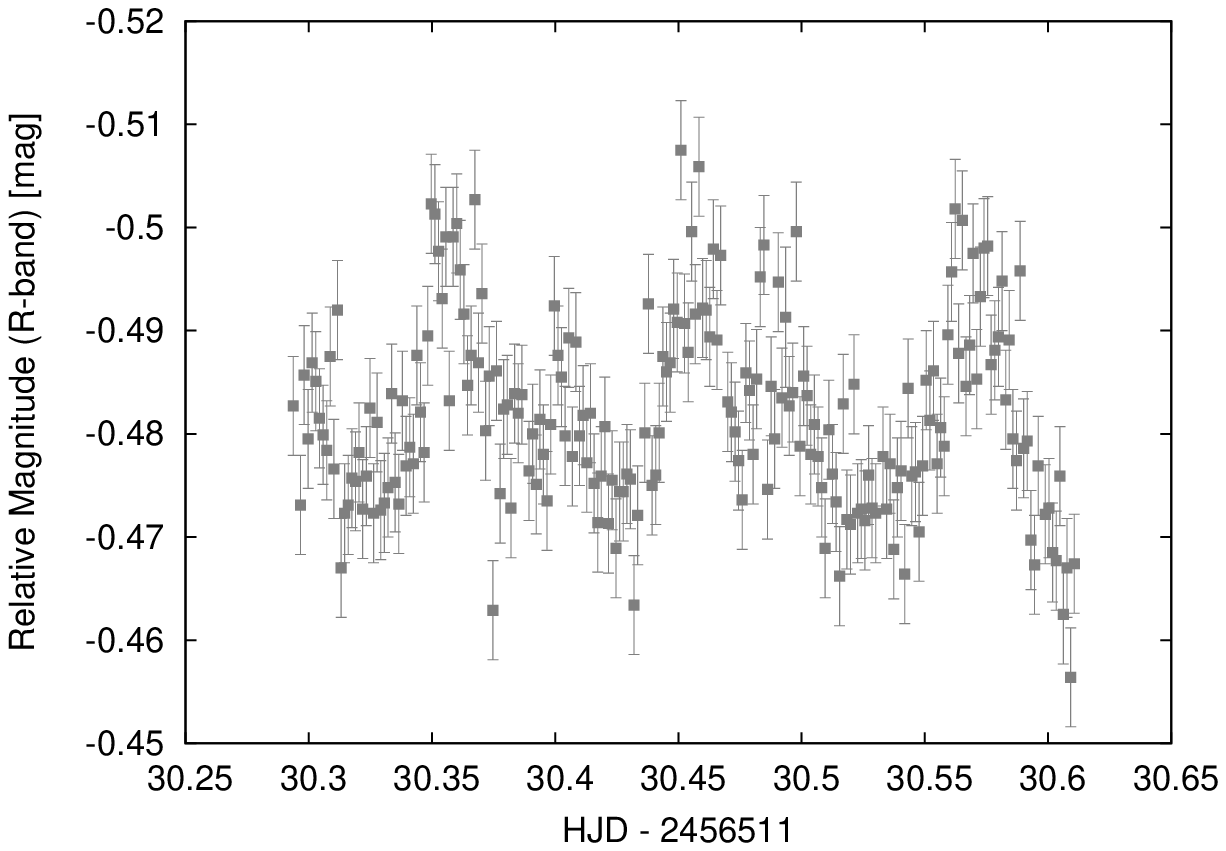}  
}
\caption{}
\label{Fig. 1}
\end{figure*} 

\begin{figure*}[!h]
\centering
\centerline{
\includegraphics[width=70mm]{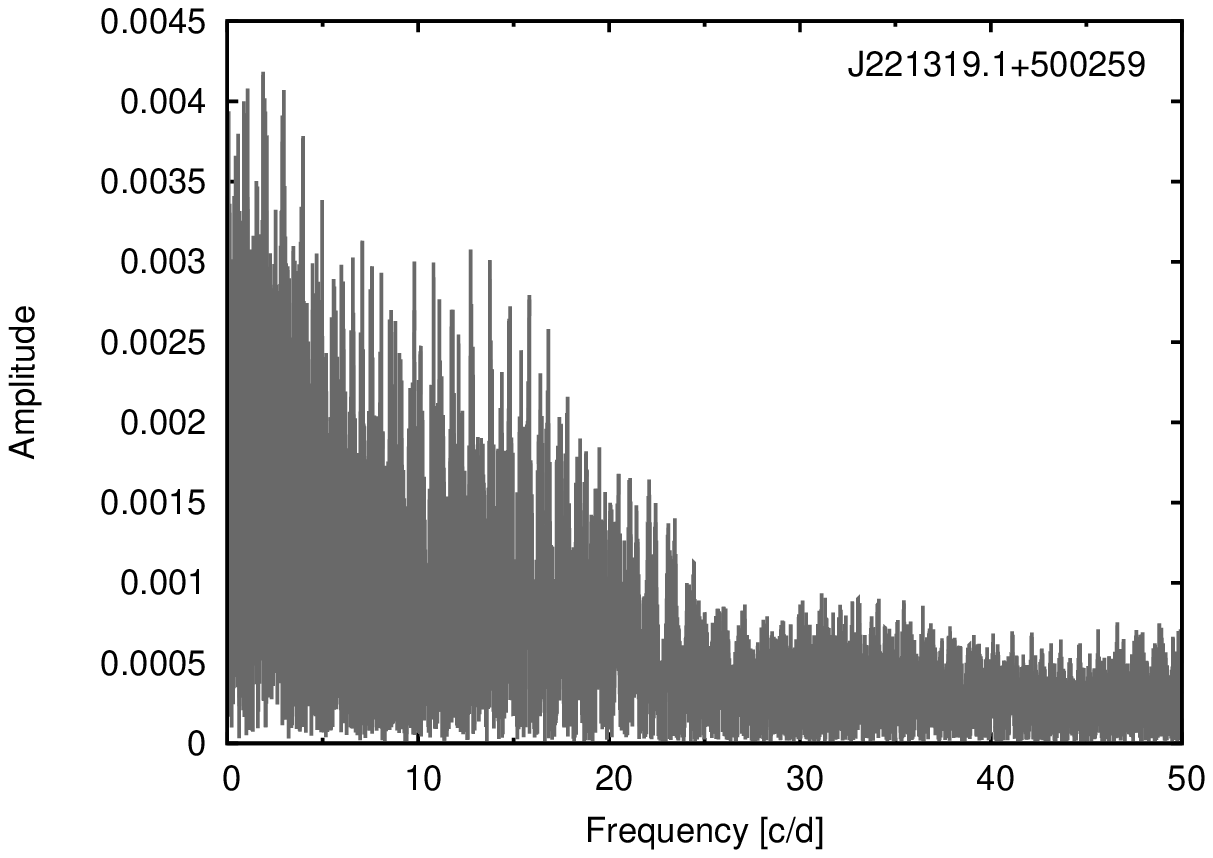}
\includegraphics[width=70mm]{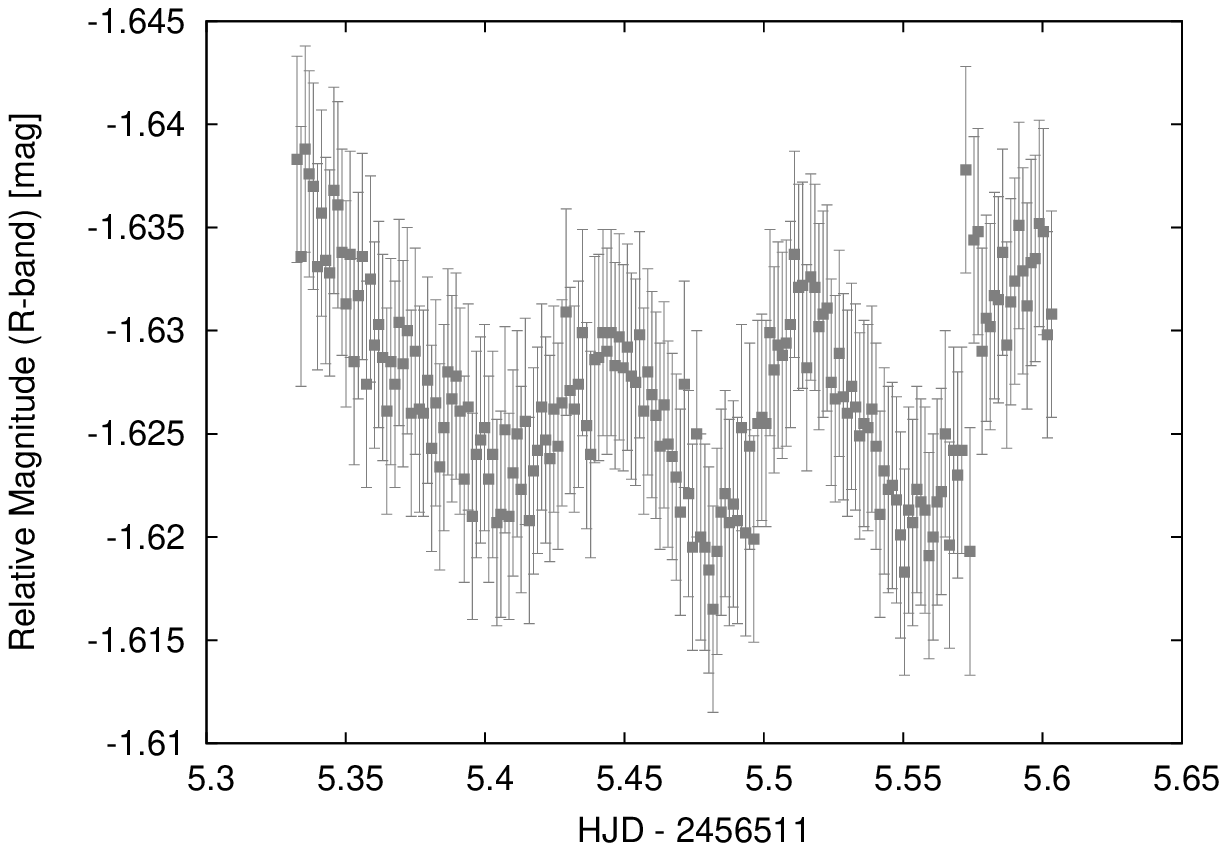}  
}
\centerline{
\includegraphics[width=70mm]{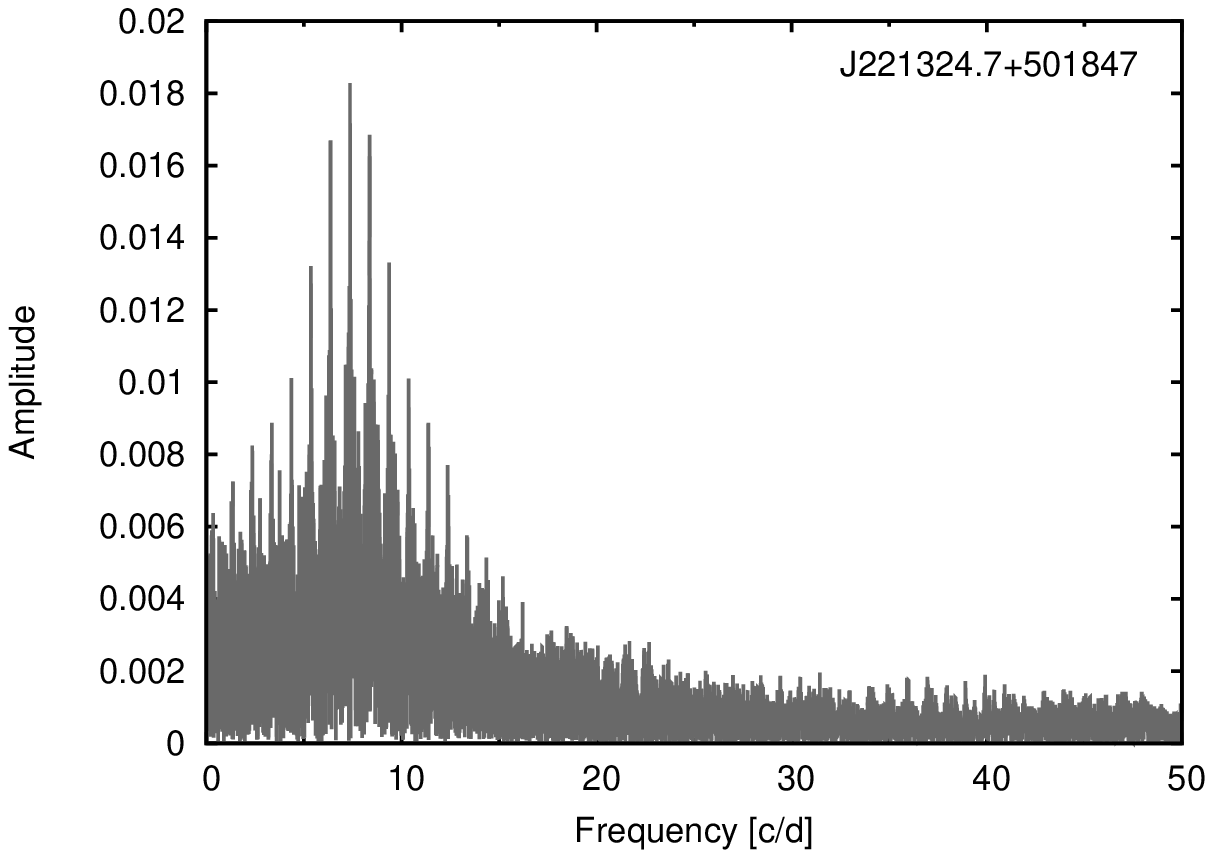}
\includegraphics[width=70mm]{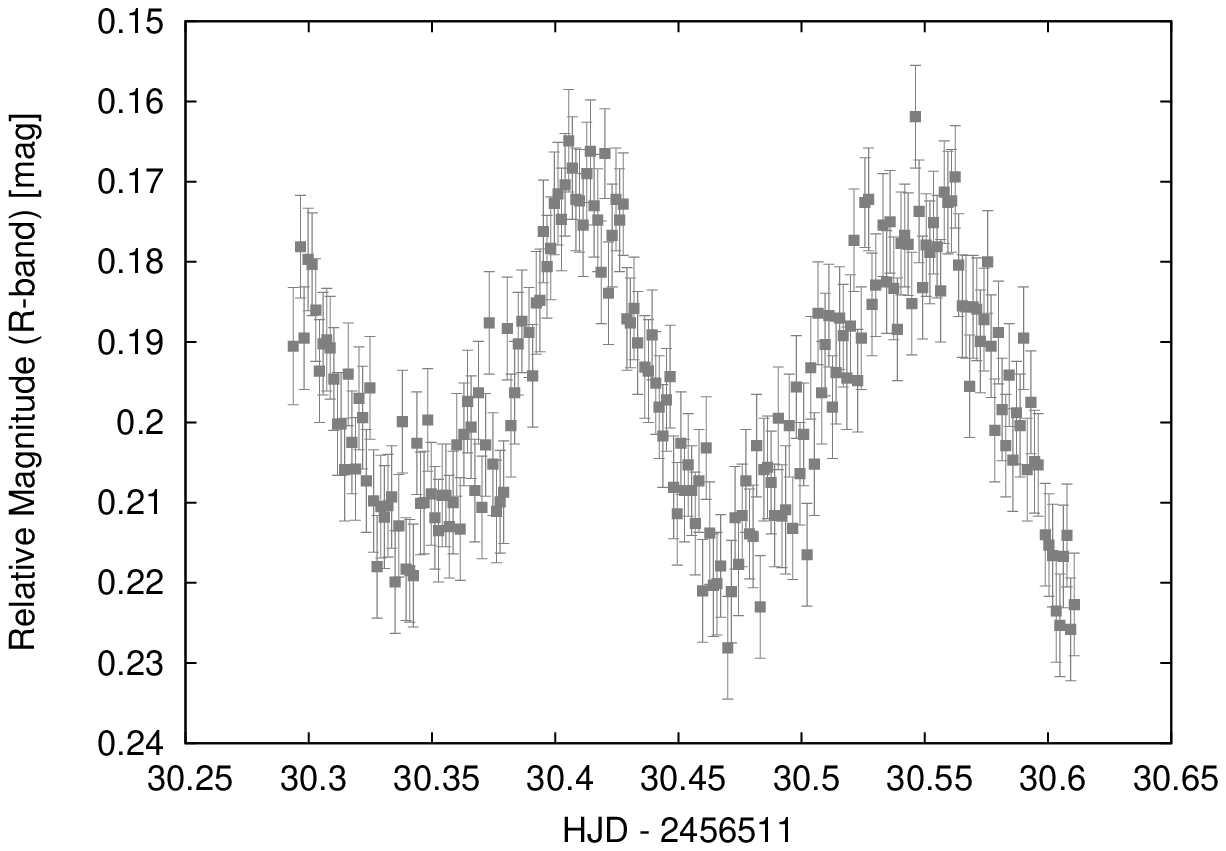}  
}
\centerline{
\includegraphics[width=70mm]{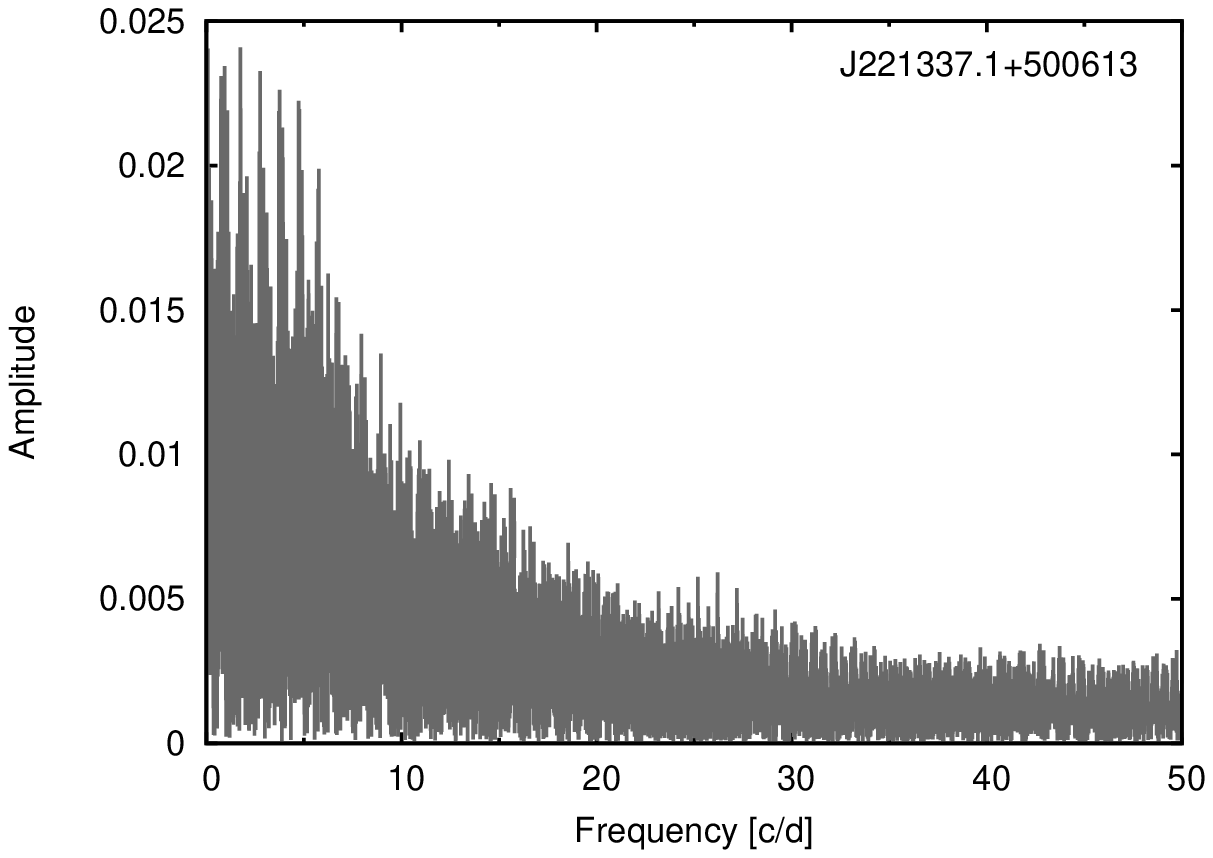}
\includegraphics[width=70mm]{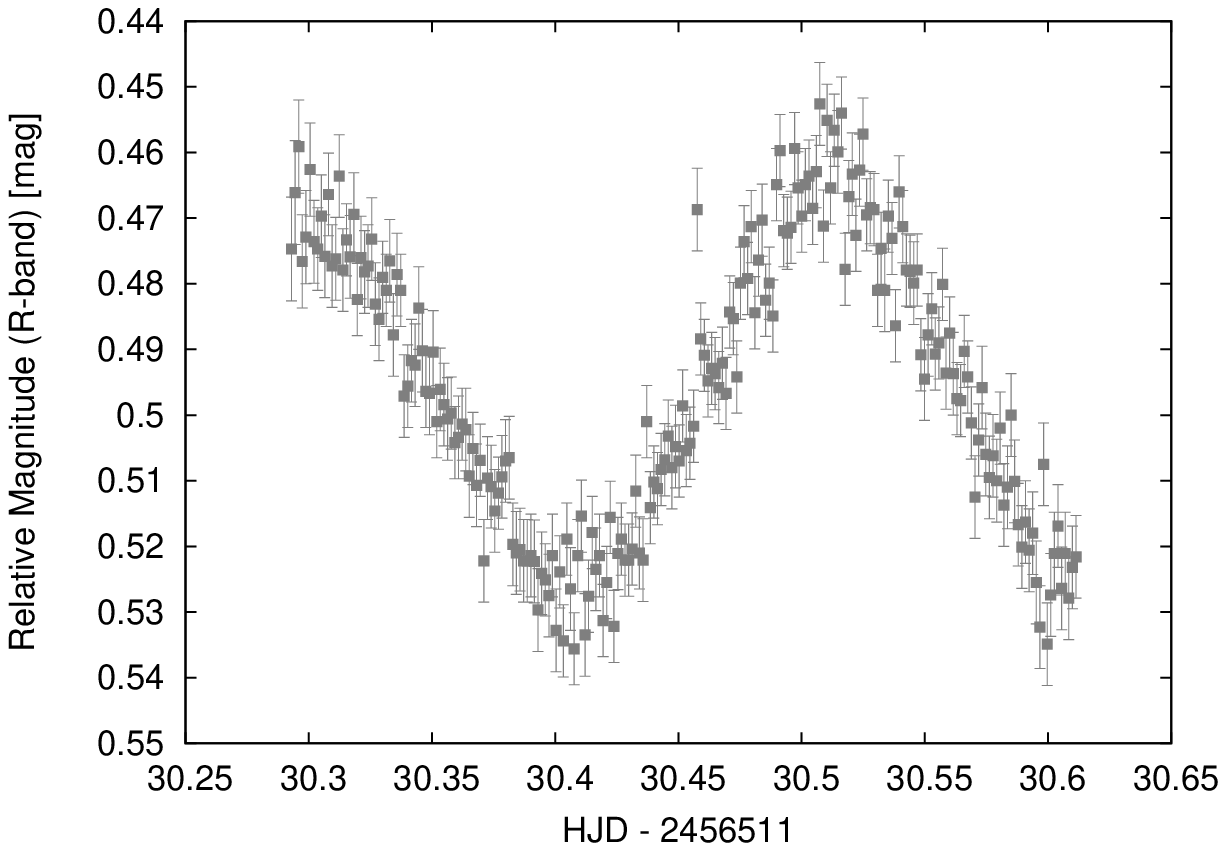}
}
\centerline{
\includegraphics[width=70mm]{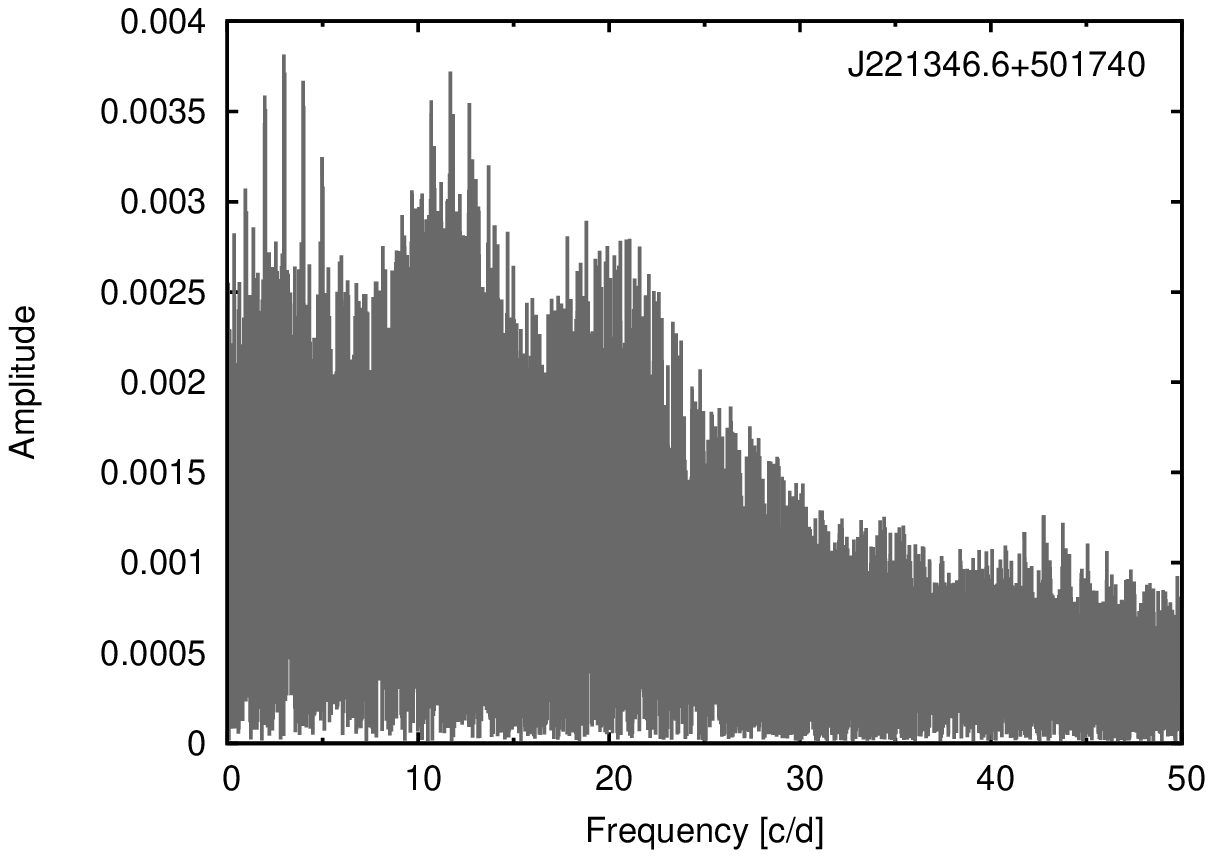}
\includegraphics[width=70mm]{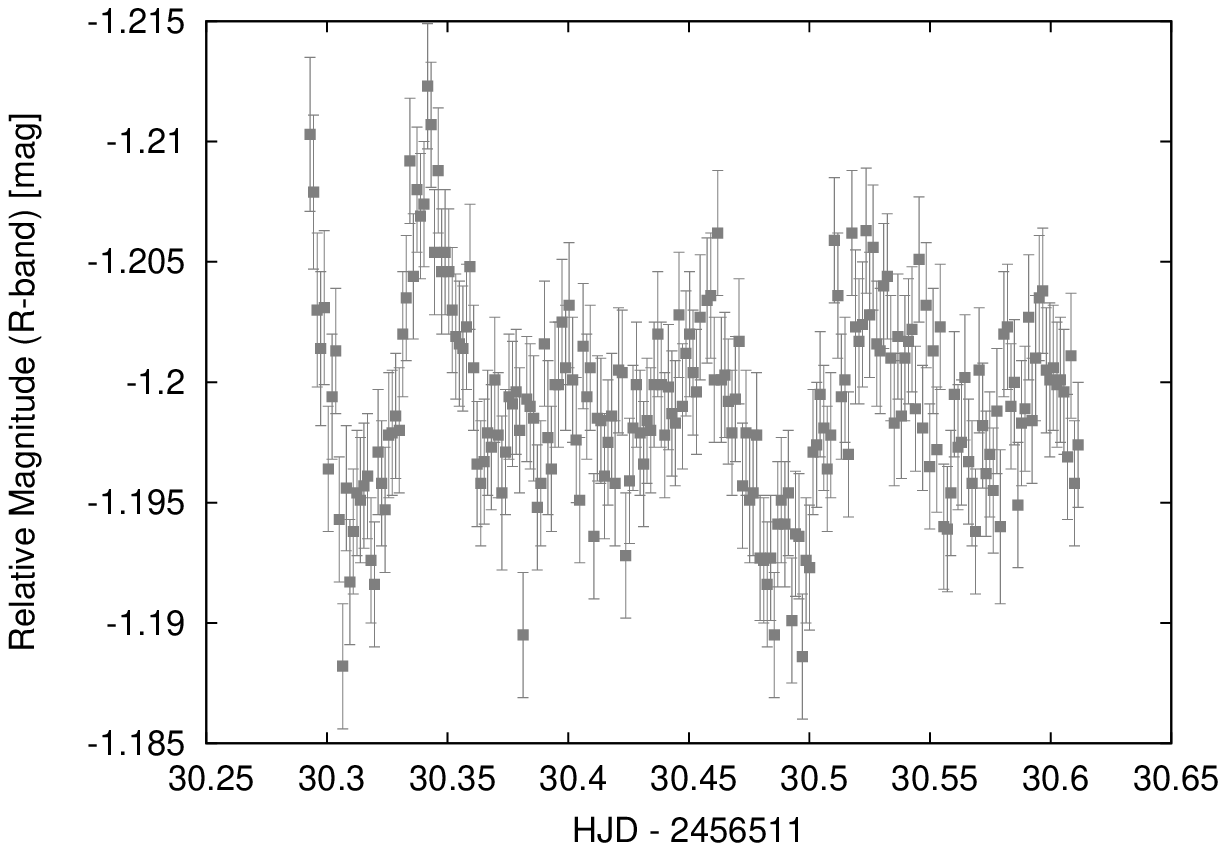}  
}
\caption{}
\label{Fig. 1}
\end{figure*} 

\begin{figure*}[!h]
\centering
\centerline{
\includegraphics[width=70mm]{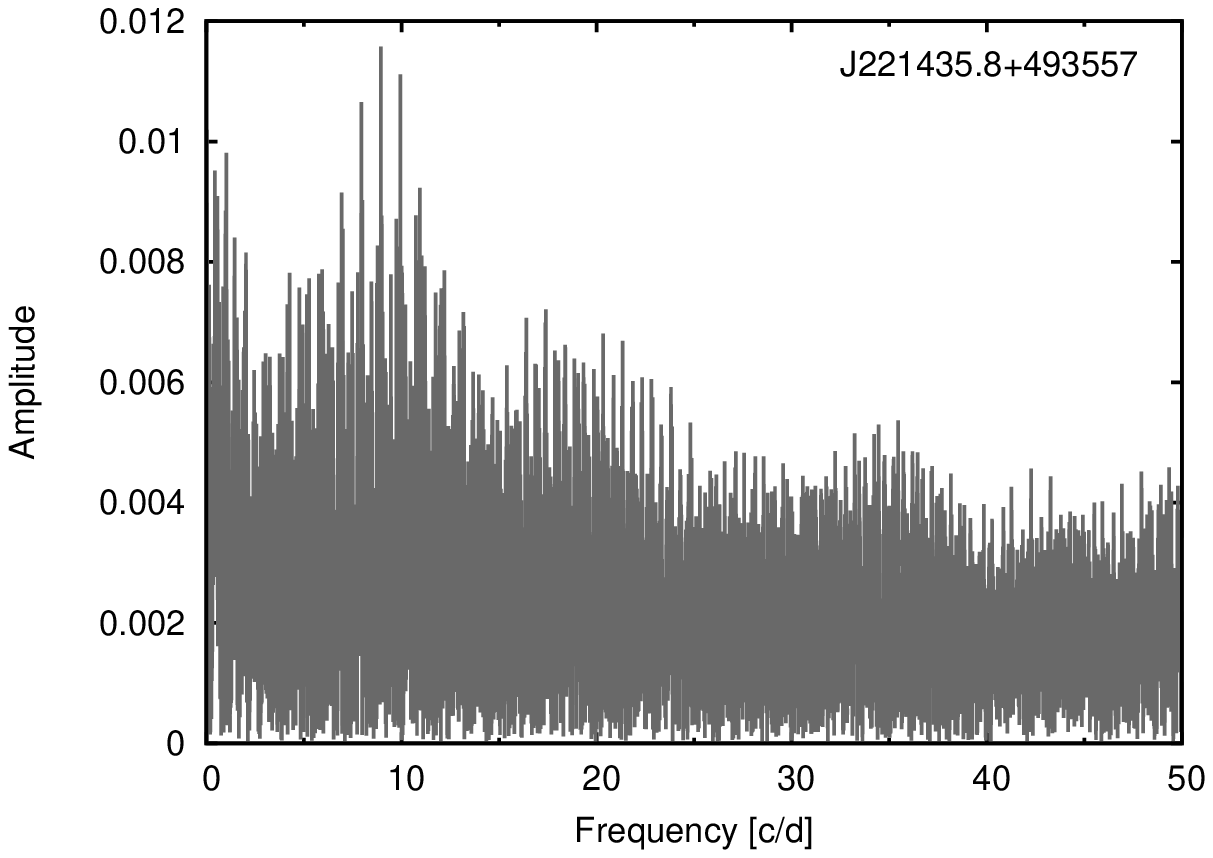}
\includegraphics[width=70mm]{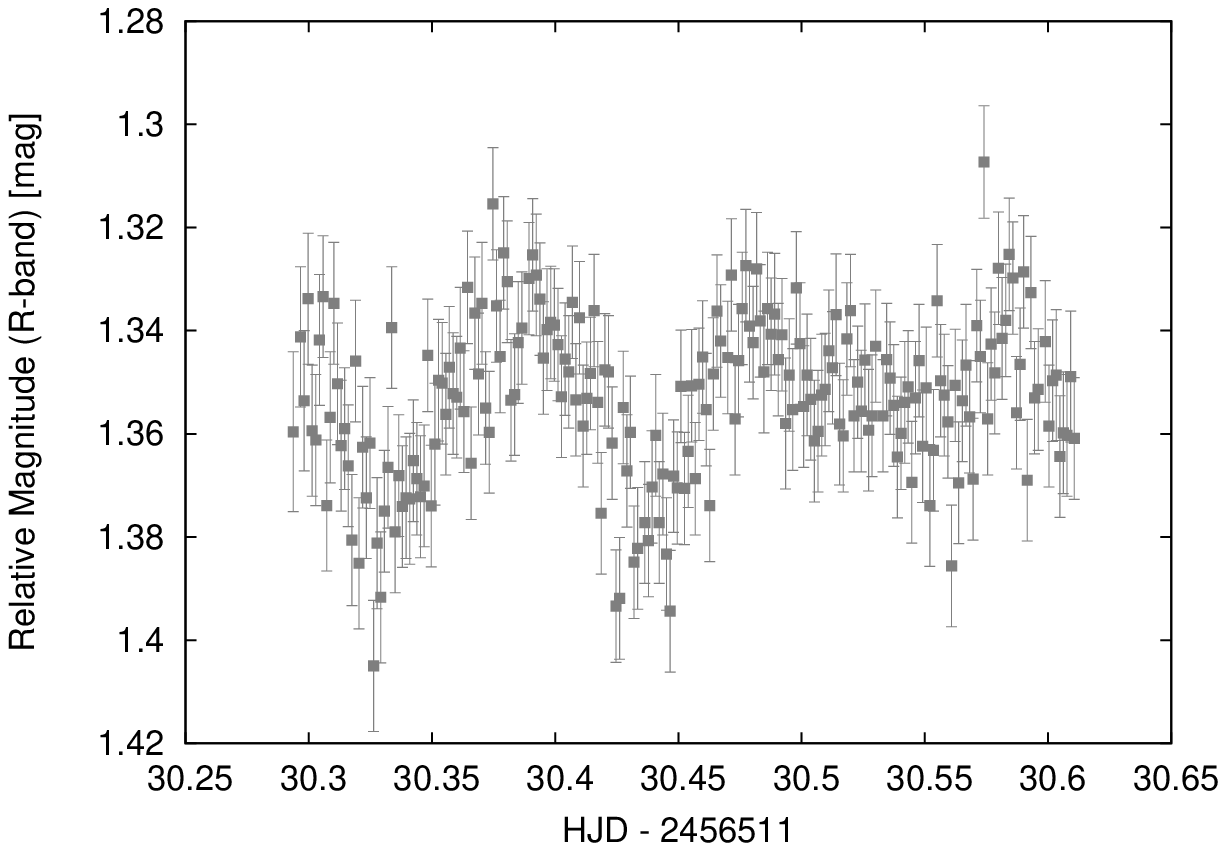}  
}
\centerline{
\includegraphics[width=70mm]{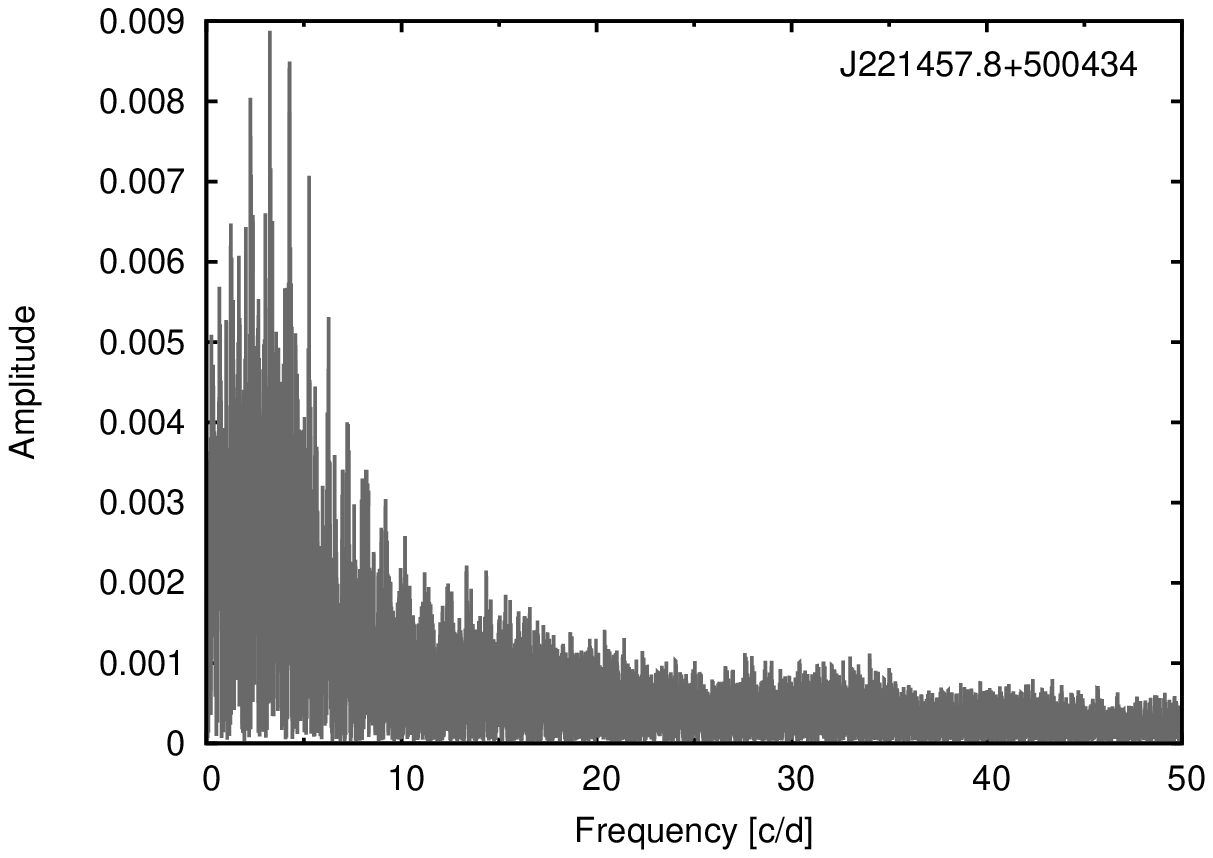}
\includegraphics[width=70mm]{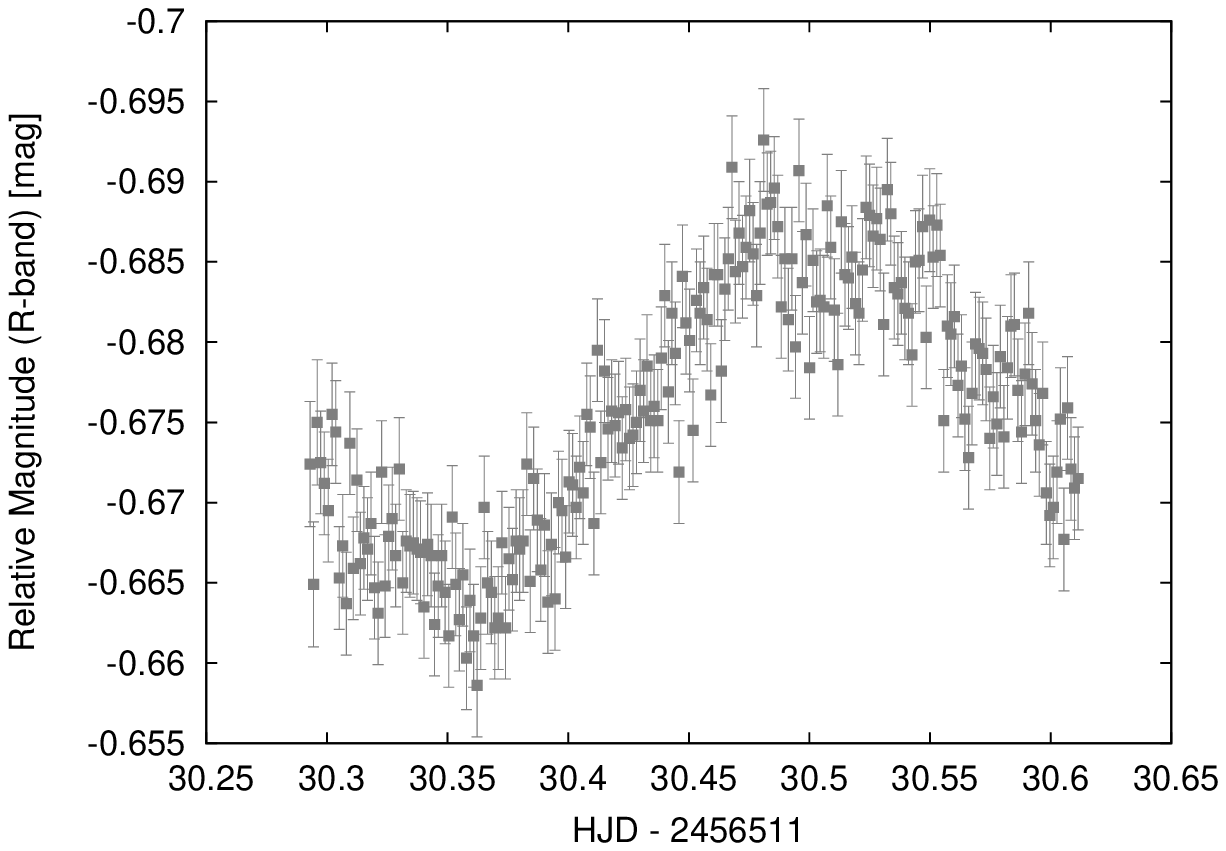}  
}
\centerline{
\includegraphics[width=70mm]{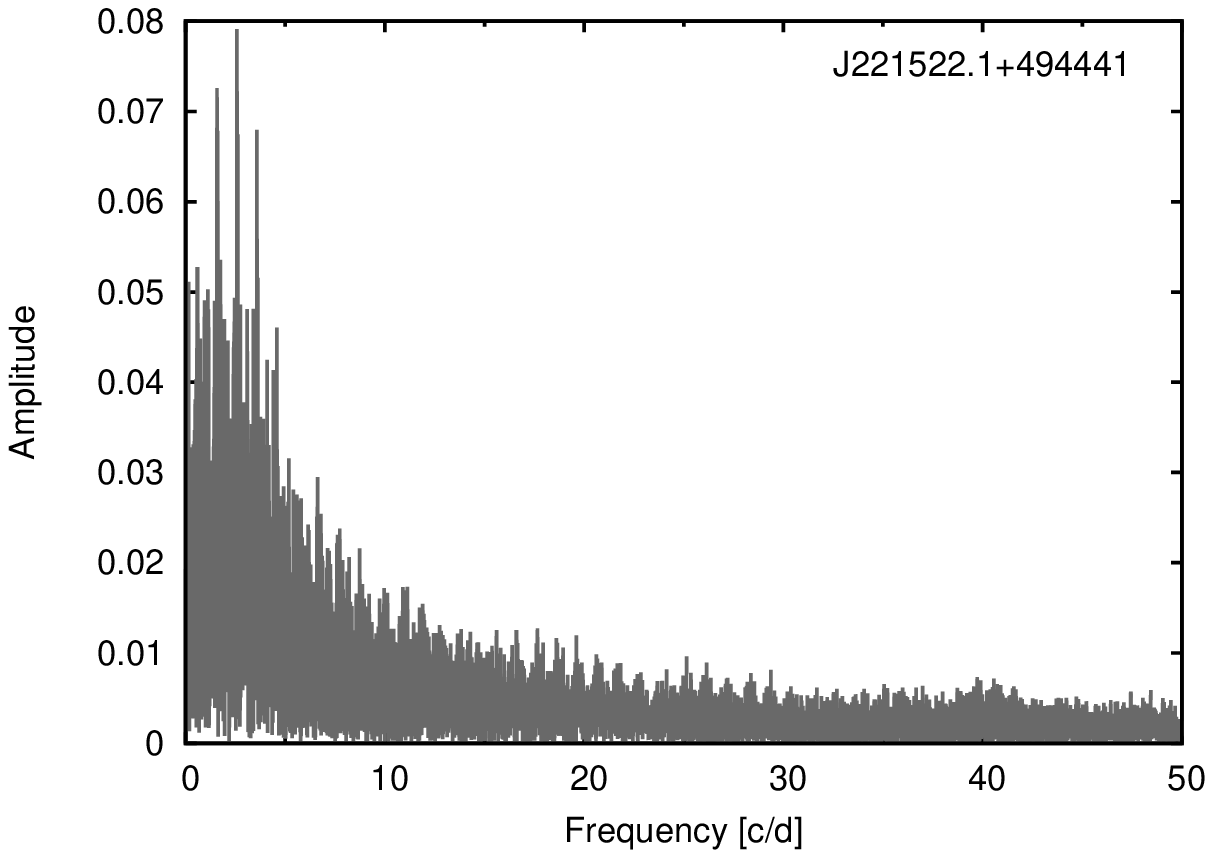}
\includegraphics[width=70mm]{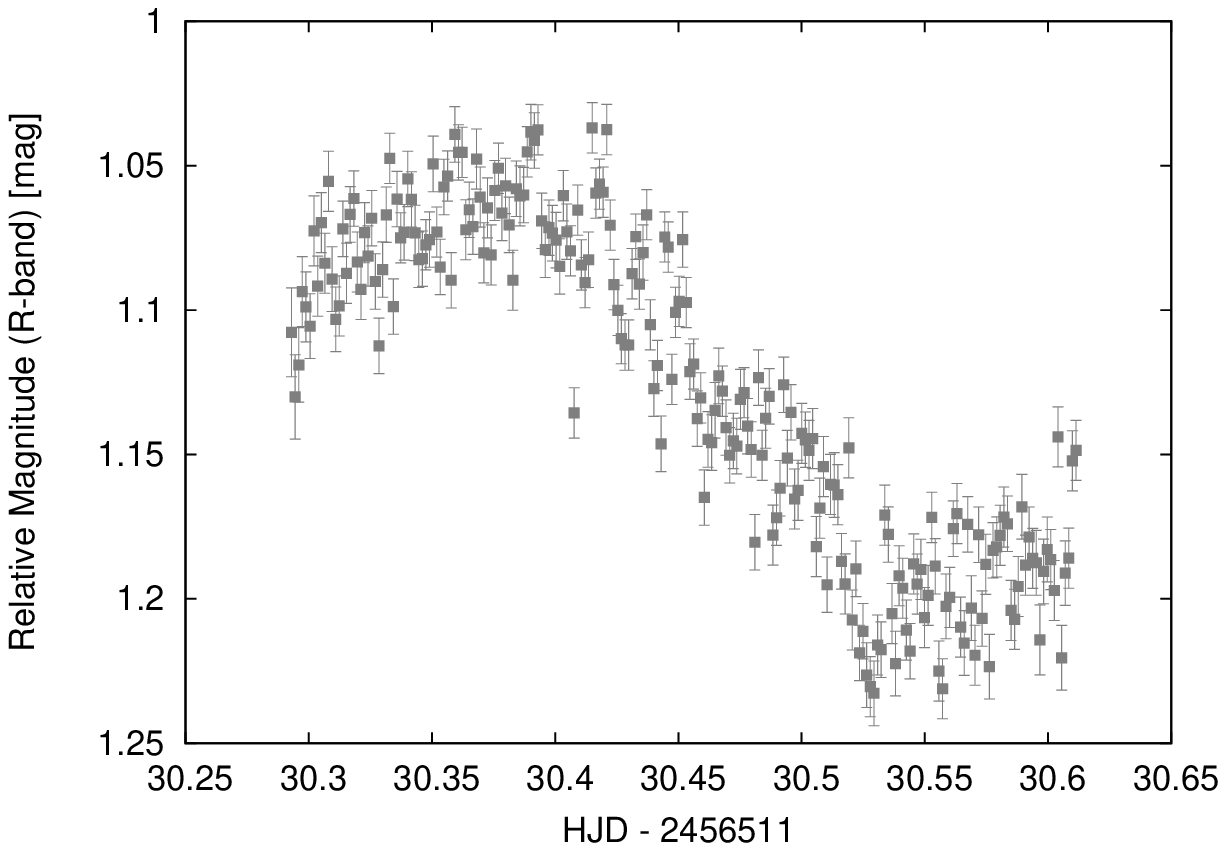}
}
\centerline{
\includegraphics[width=70mm]{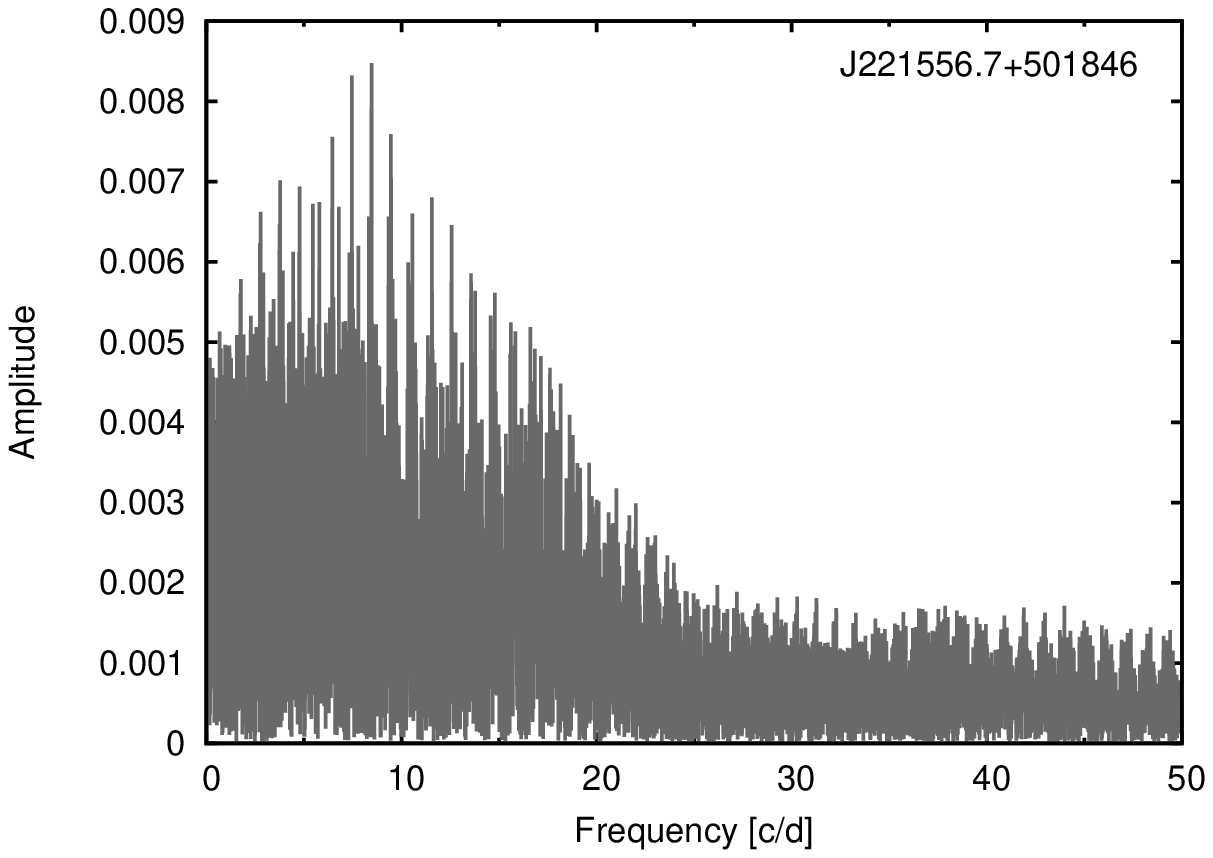}
\includegraphics[width=70mm]{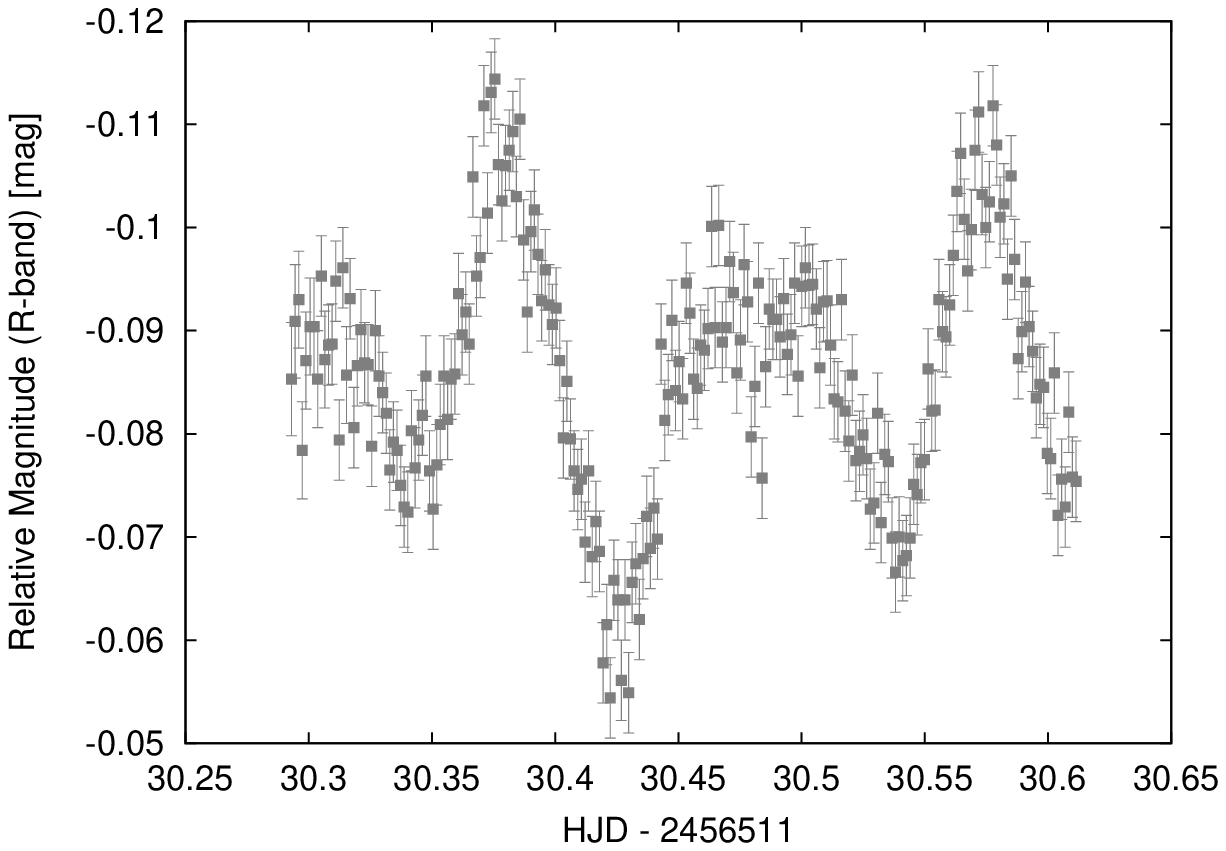}  
}
\caption{}
\label{Fig. 1}
\end{figure*} 

\begin{figure*}[!h]
\centering
\centerline{
\includegraphics[width=70mm]{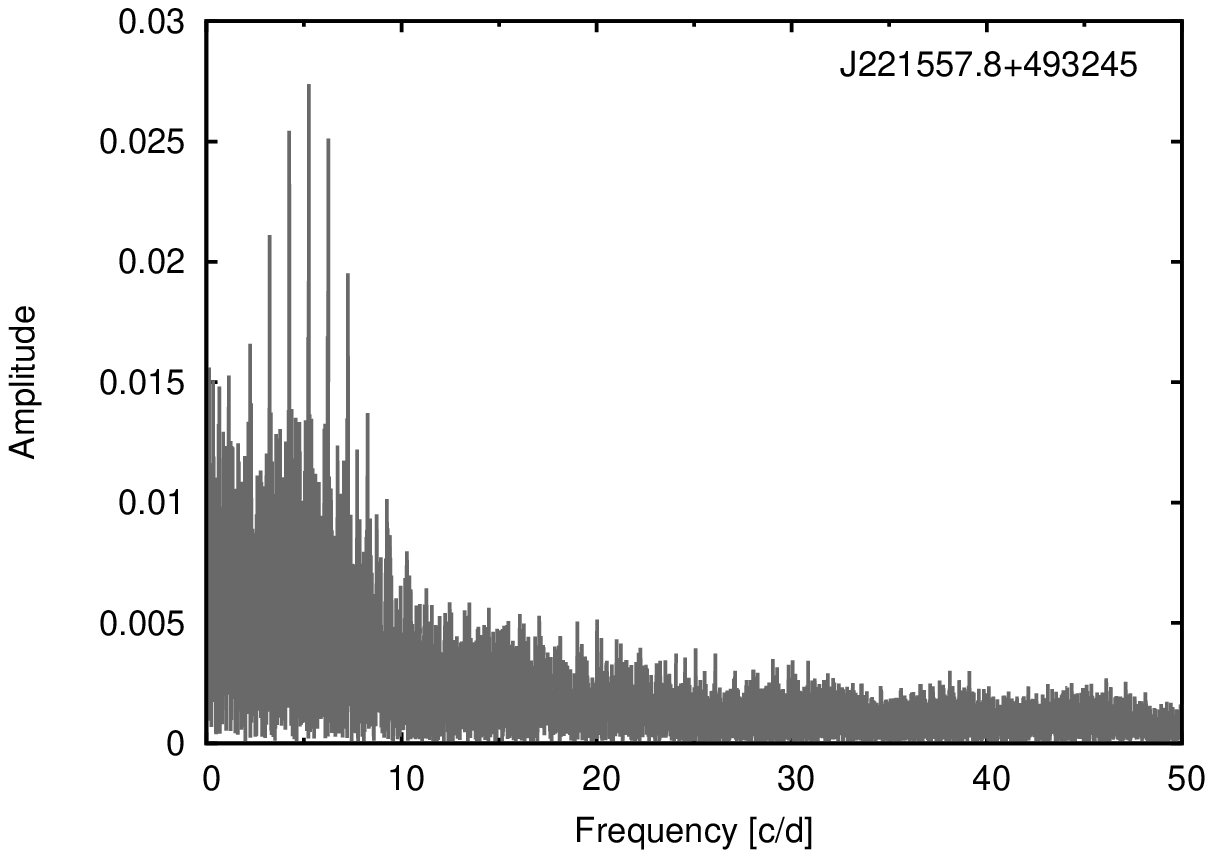}
\includegraphics[width=70mm]{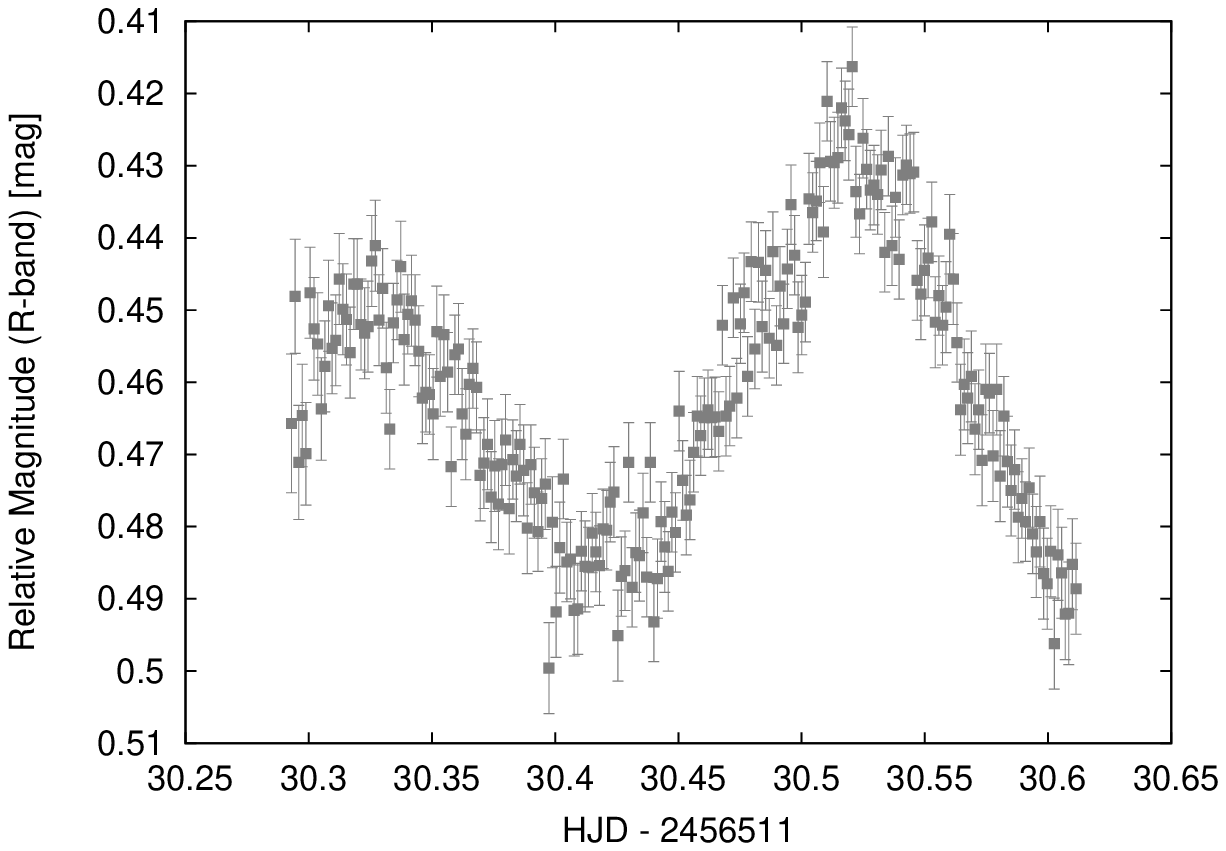}  
}
\centerline{
\includegraphics[width=70mm]{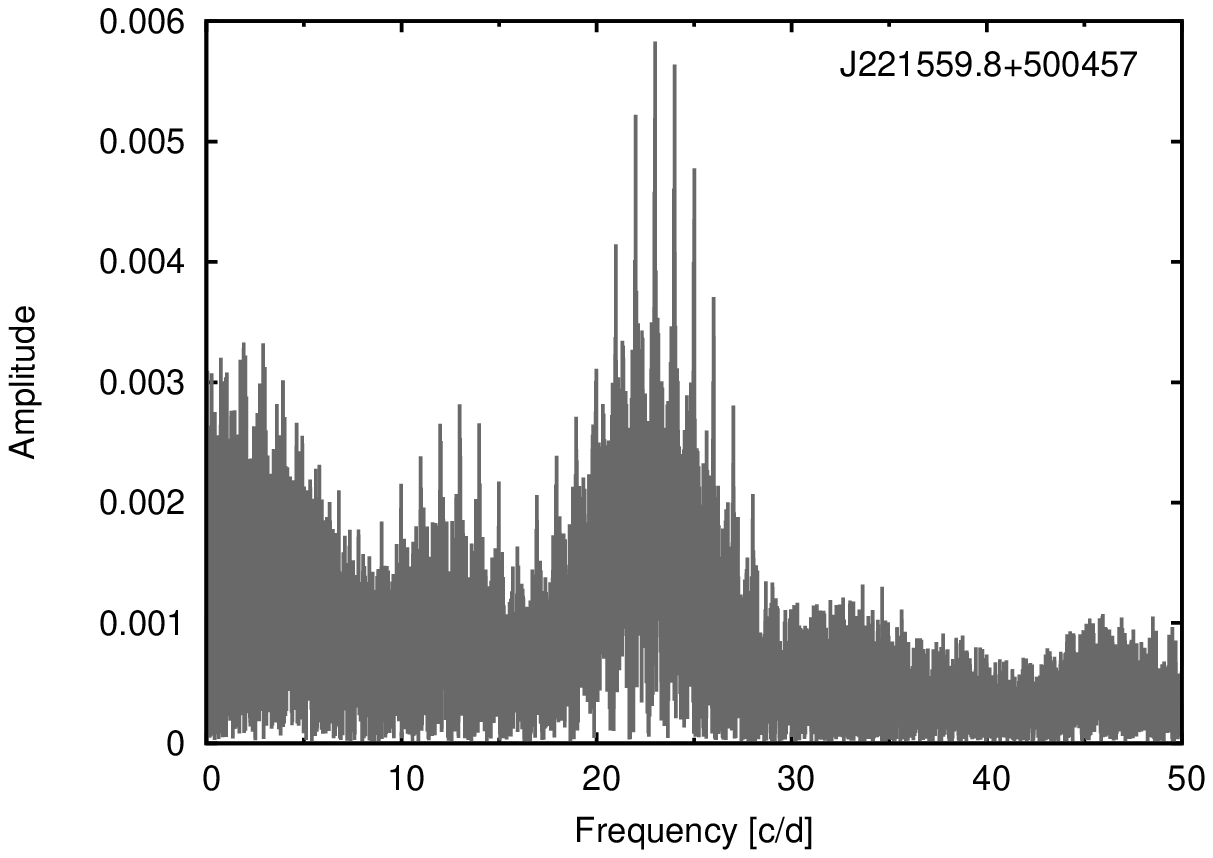}
\includegraphics[width=70mm]{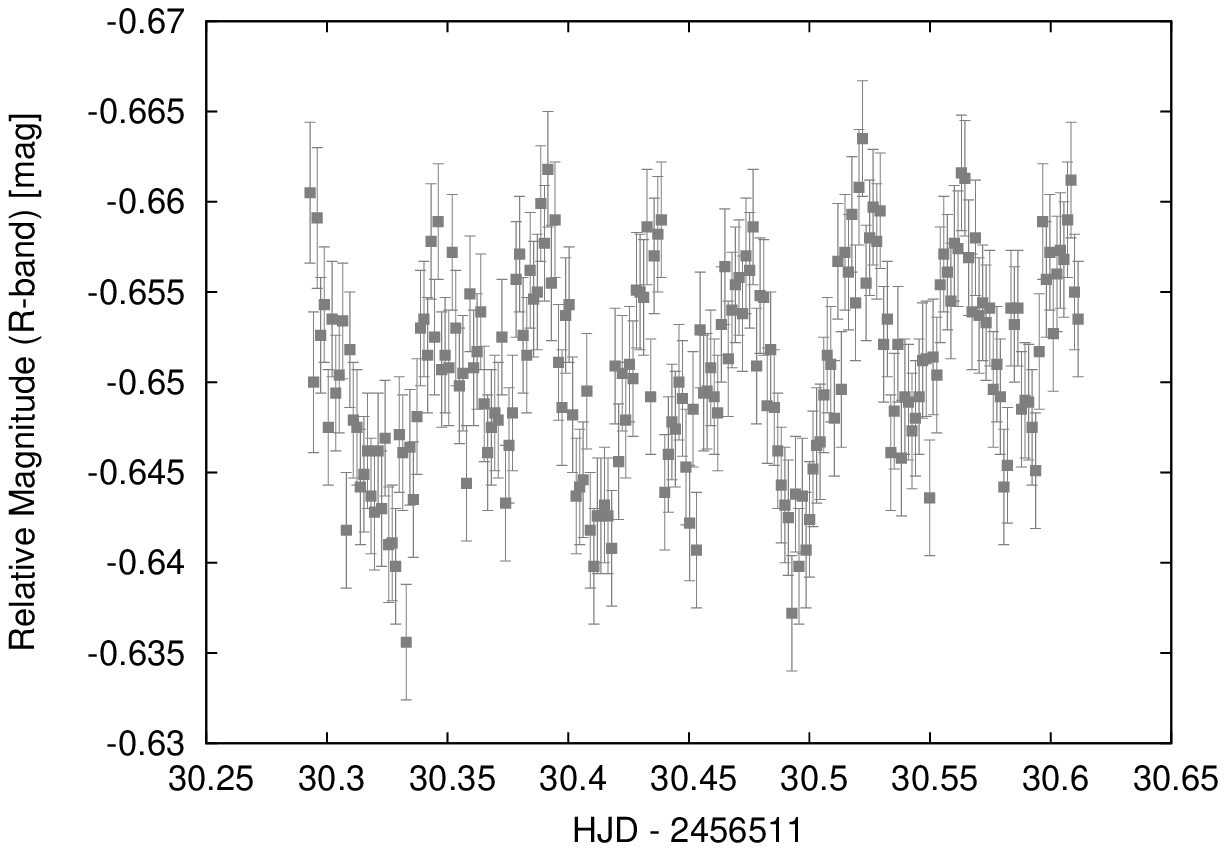}  
}
\centerline{
\includegraphics[width=70mm]{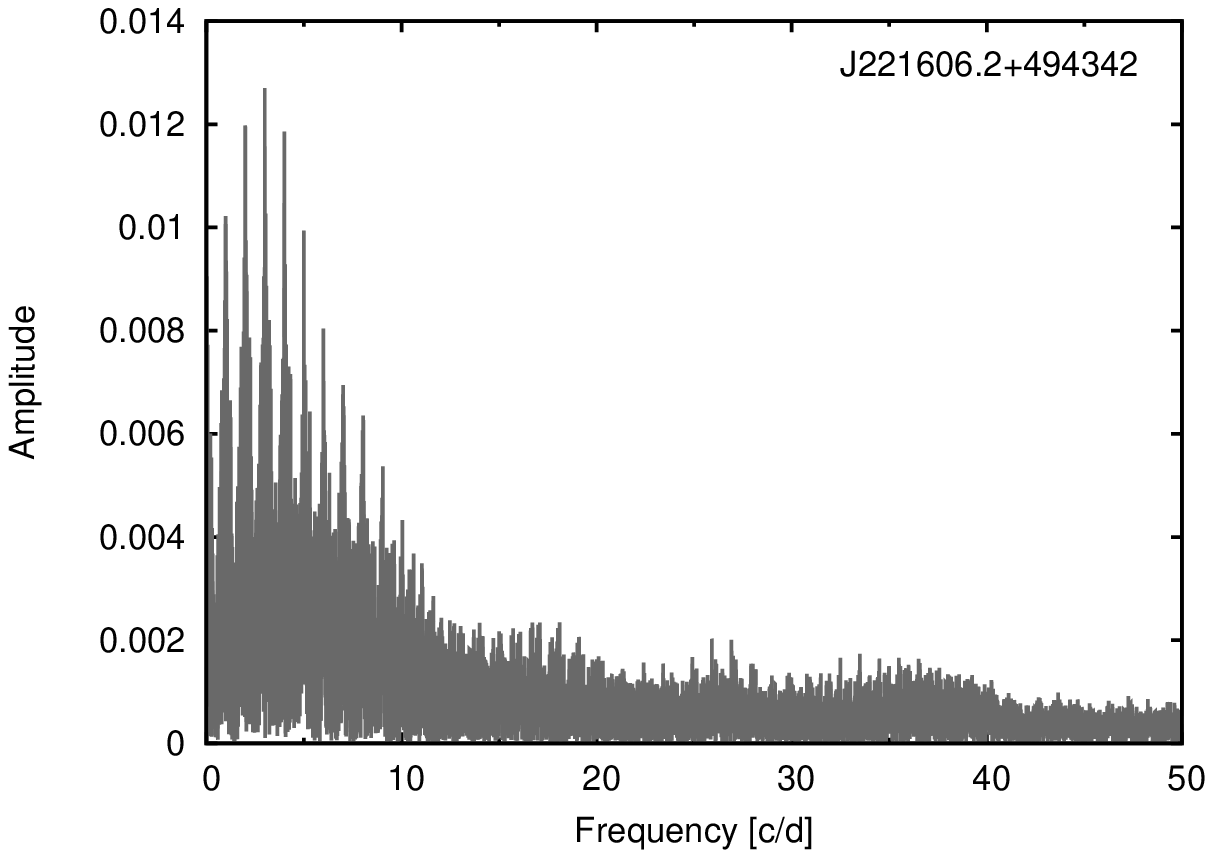}
\includegraphics[width=70mm]{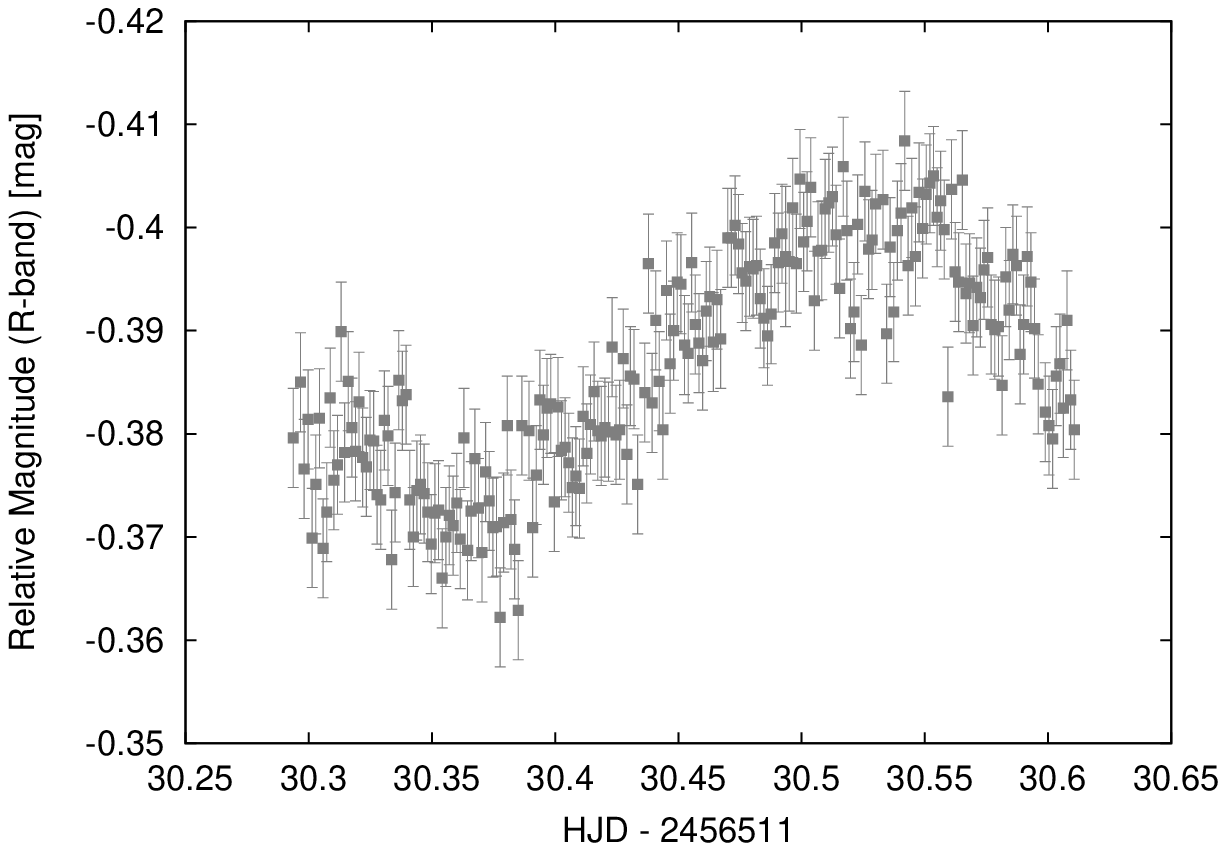}
}
\centerline{
\includegraphics[width=70mm]{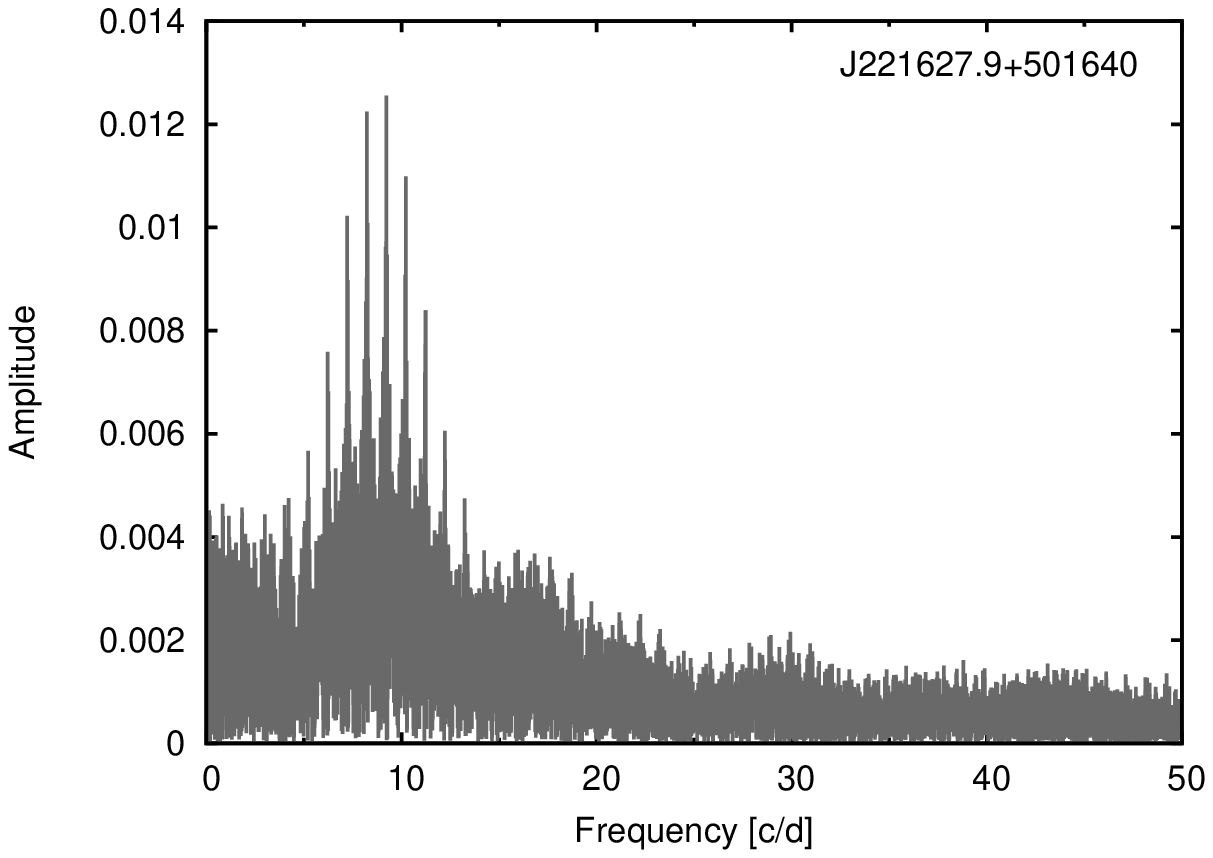}
\includegraphics[width=70mm]{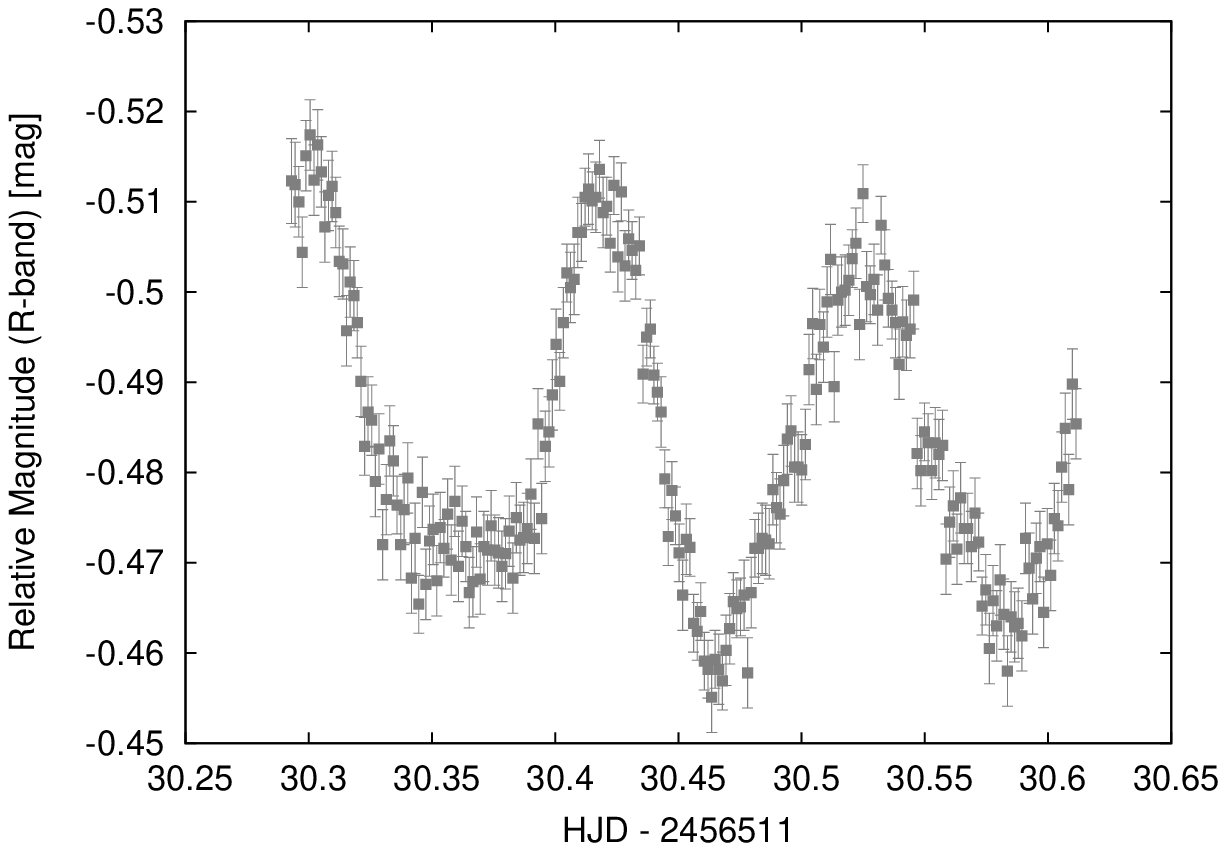}  
}
\caption{}
\label{Fig. 1}
\end{figure*} 

\begin{figure*}[!h]
\centering
\centerline{
\includegraphics[width=70mm]{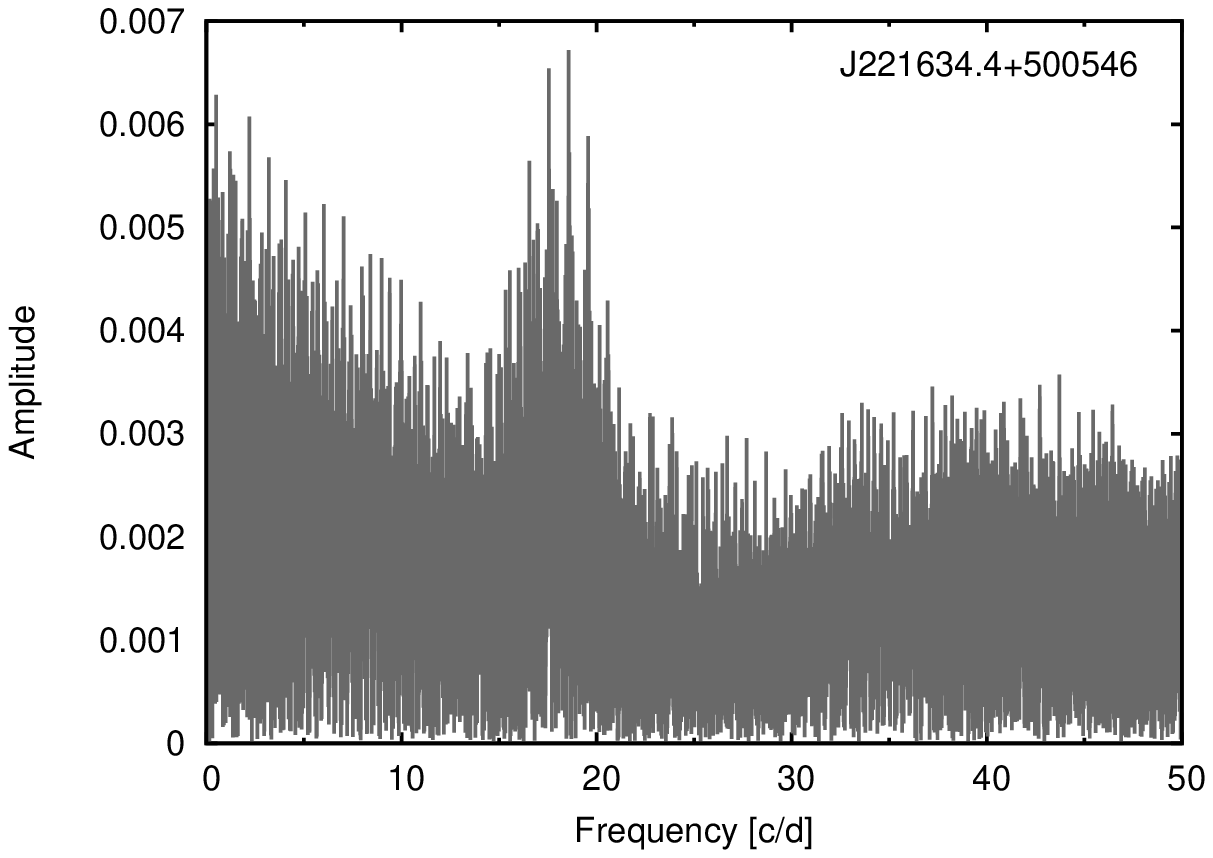}
\includegraphics[width=70mm]{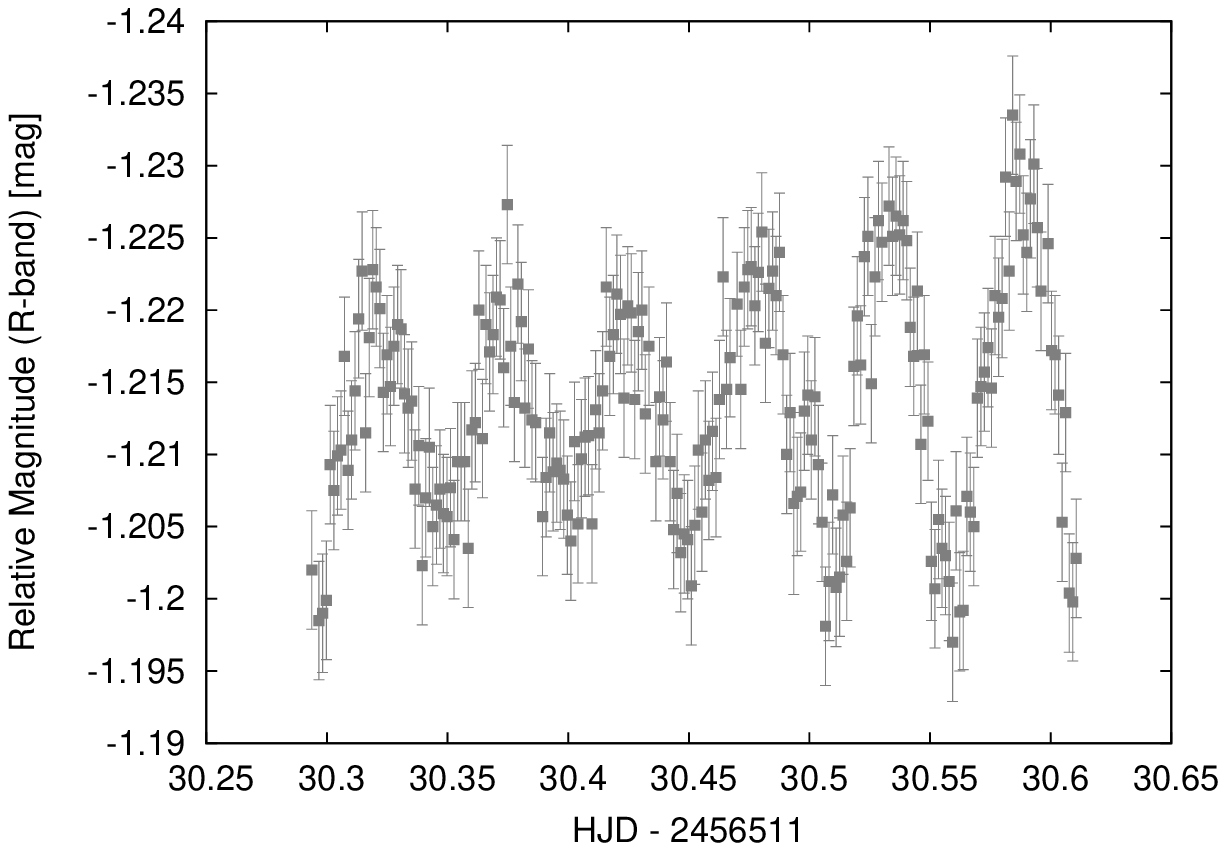}  
}
\centerline{
\includegraphics[width=70mm]{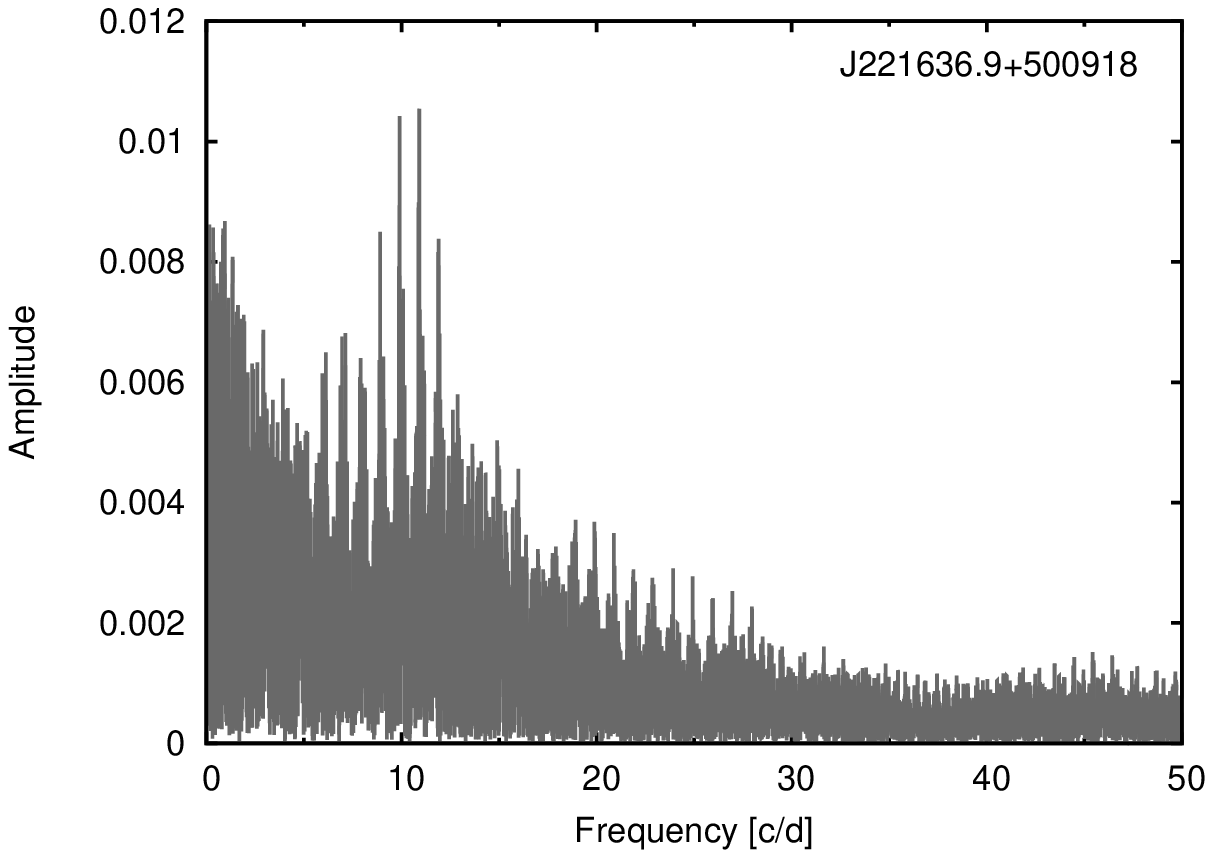}
\includegraphics[width=70mm]{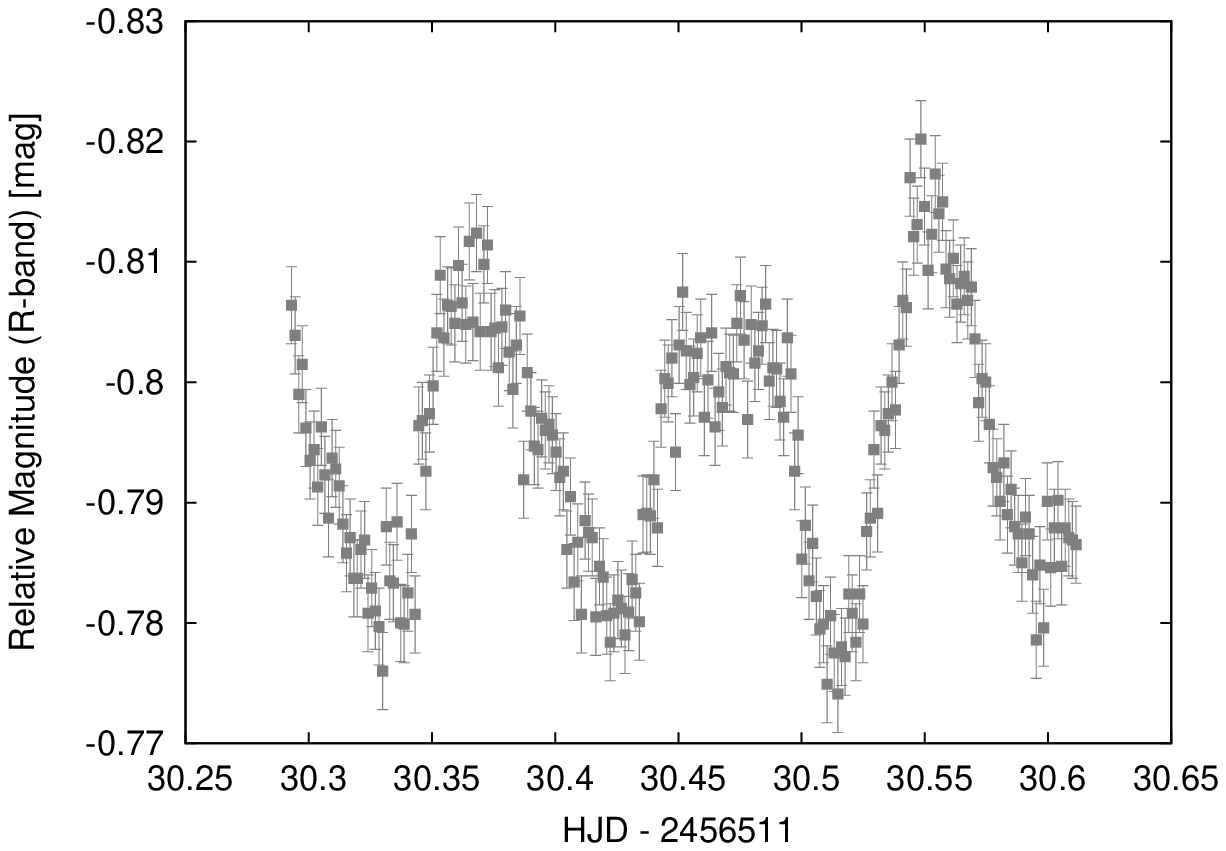}  
}
\centerline{
\includegraphics[width=70mm]{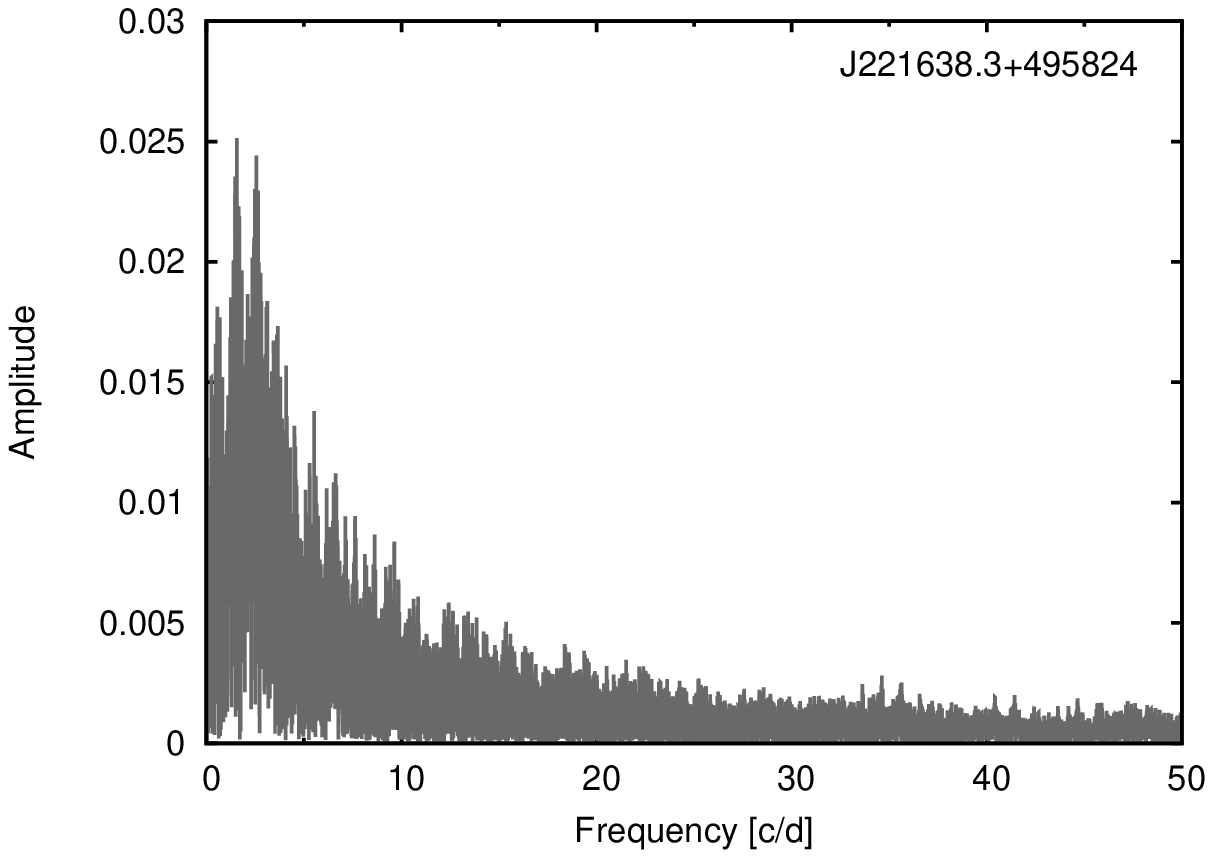}
\includegraphics[width=70mm]{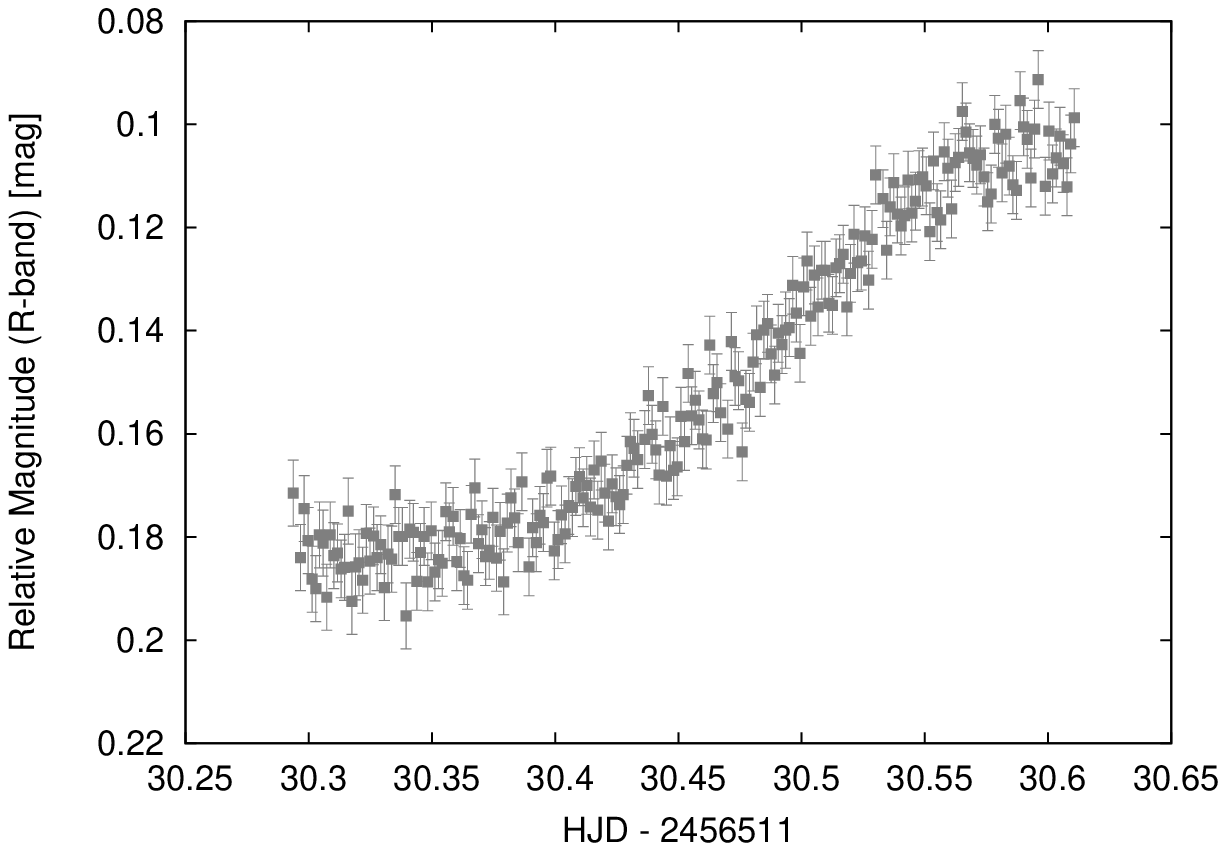}
}
\centerline{
\includegraphics[width=70mm]{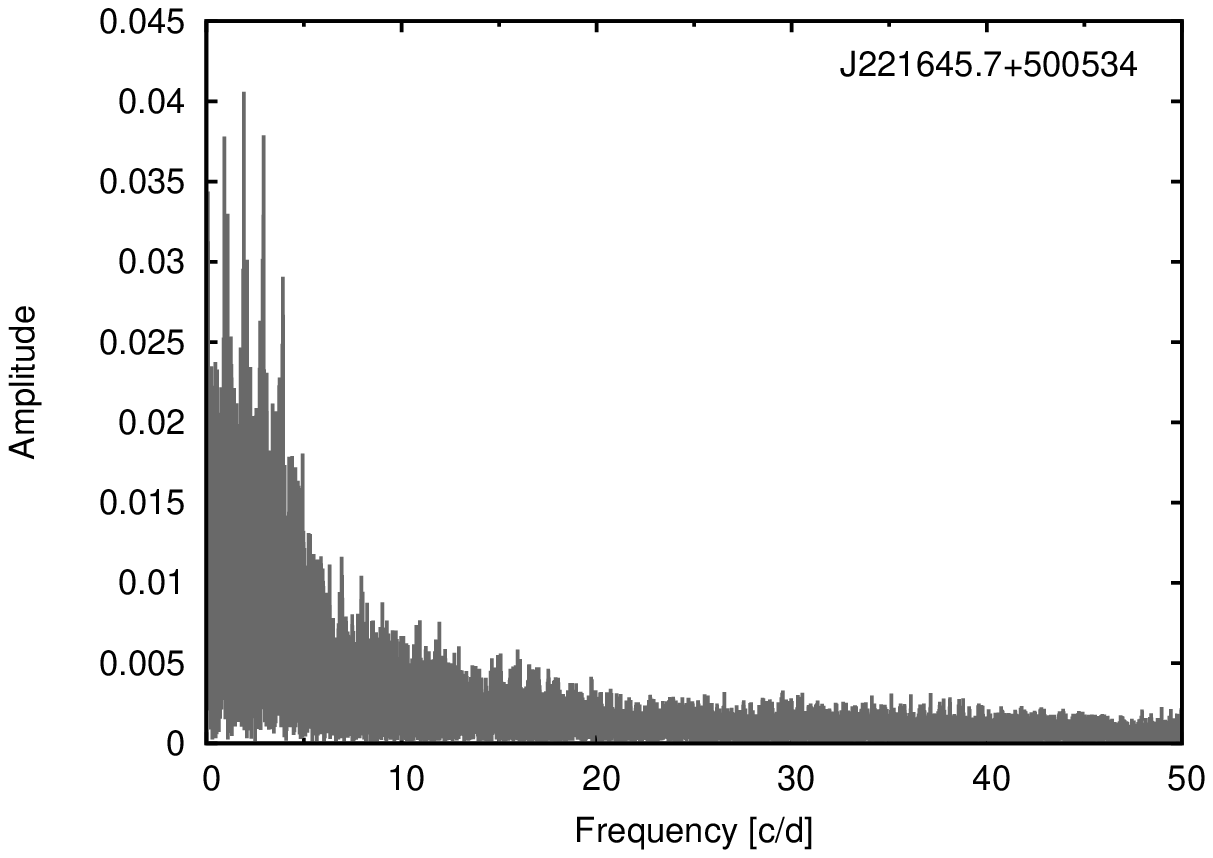}
\includegraphics[width=70mm]{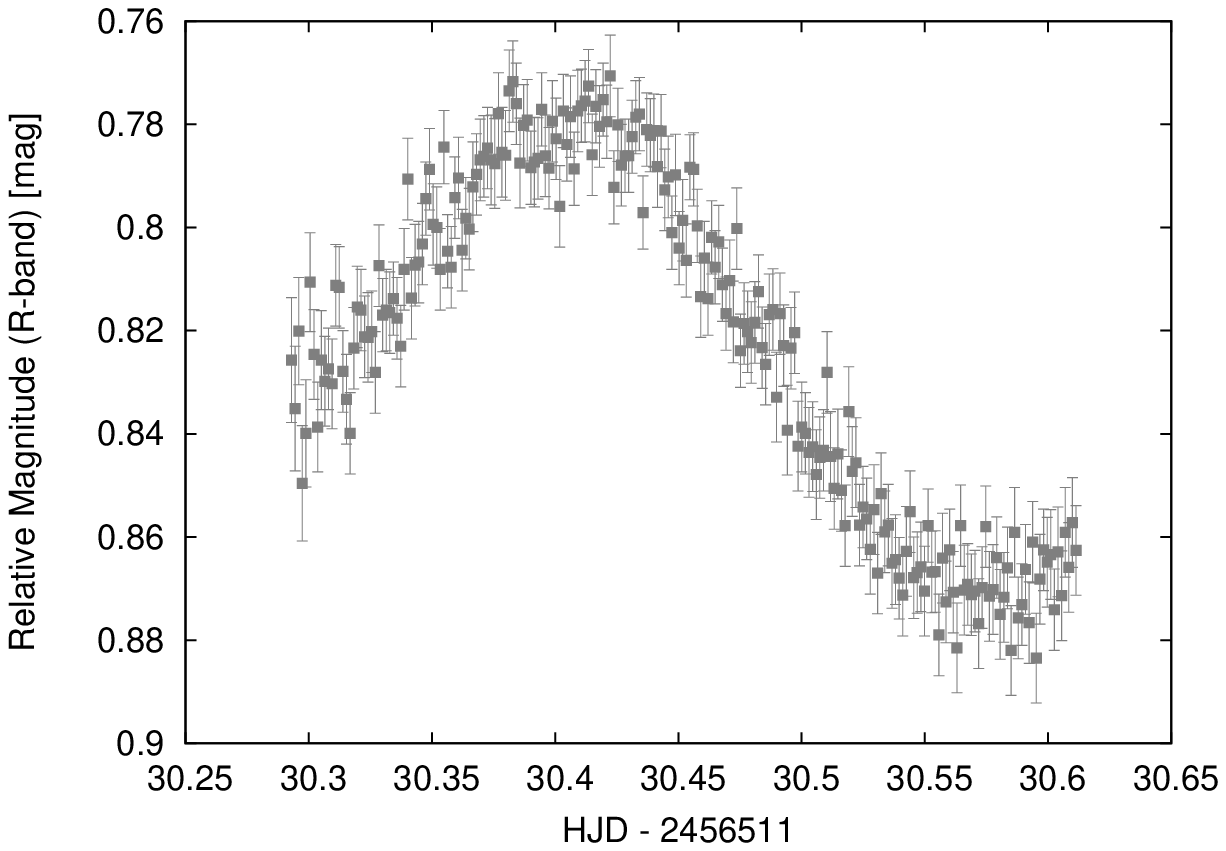}  
}
\caption{}
\label{Fig. 1}
\end{figure*} 

\begin{figure*}[!h]
\centering
\centerline{
\includegraphics[width=70mm]{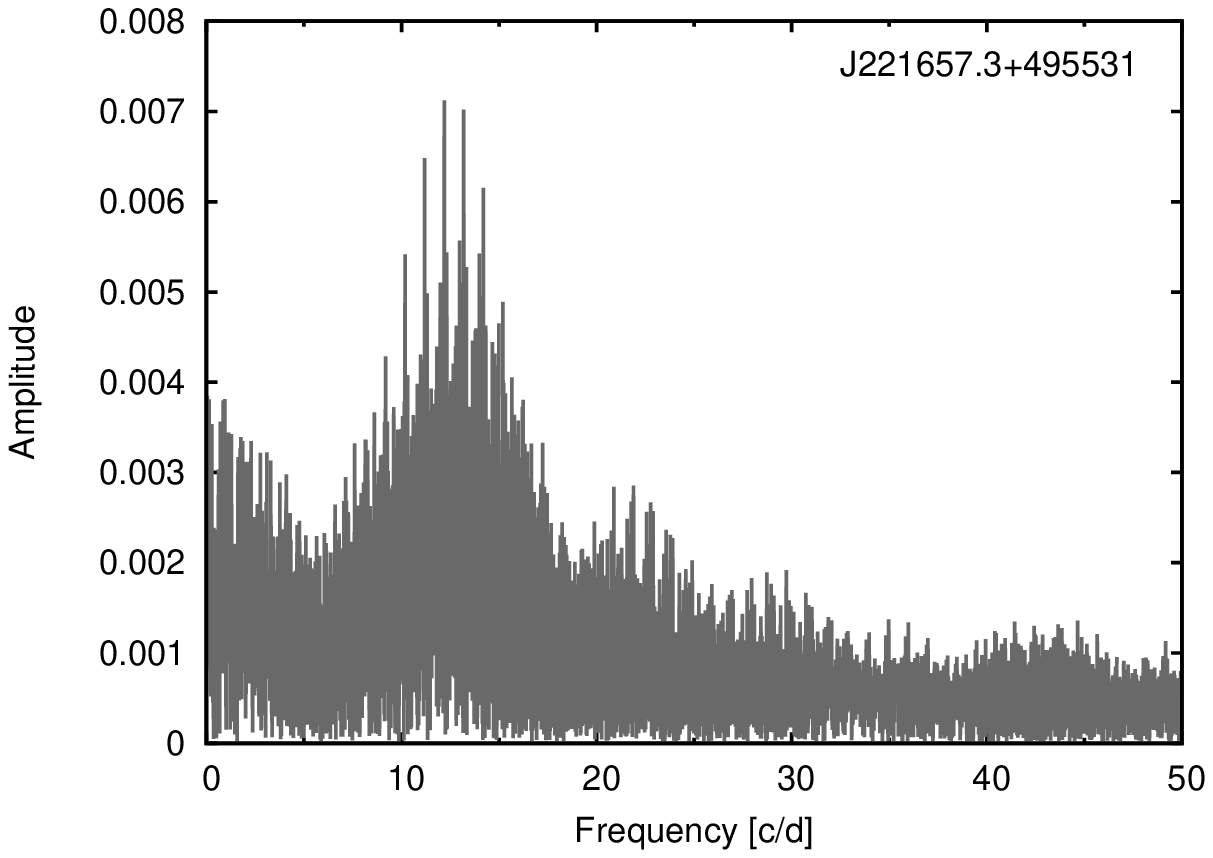}
\includegraphics[width=70mm]{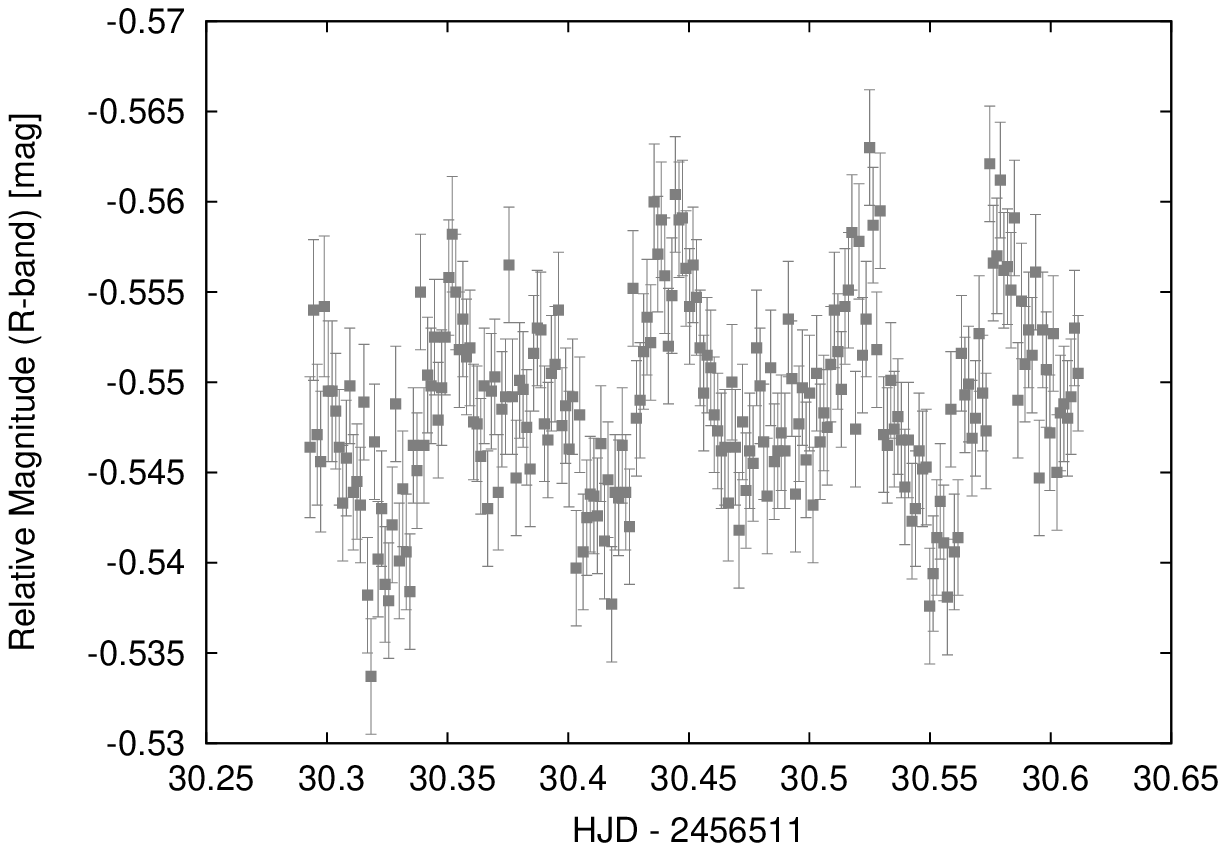}  
}
\centerline{
\includegraphics[width=70mm]{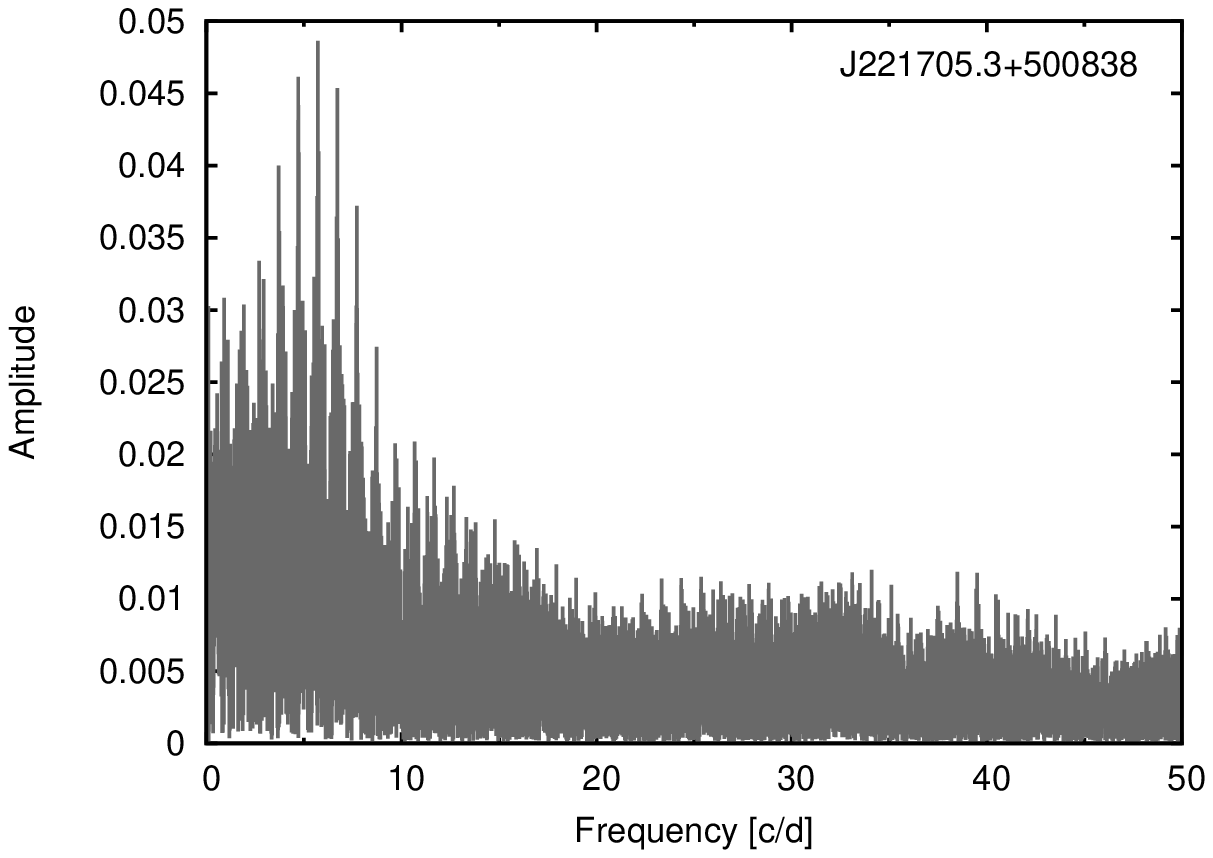}
\includegraphics[width=70mm]{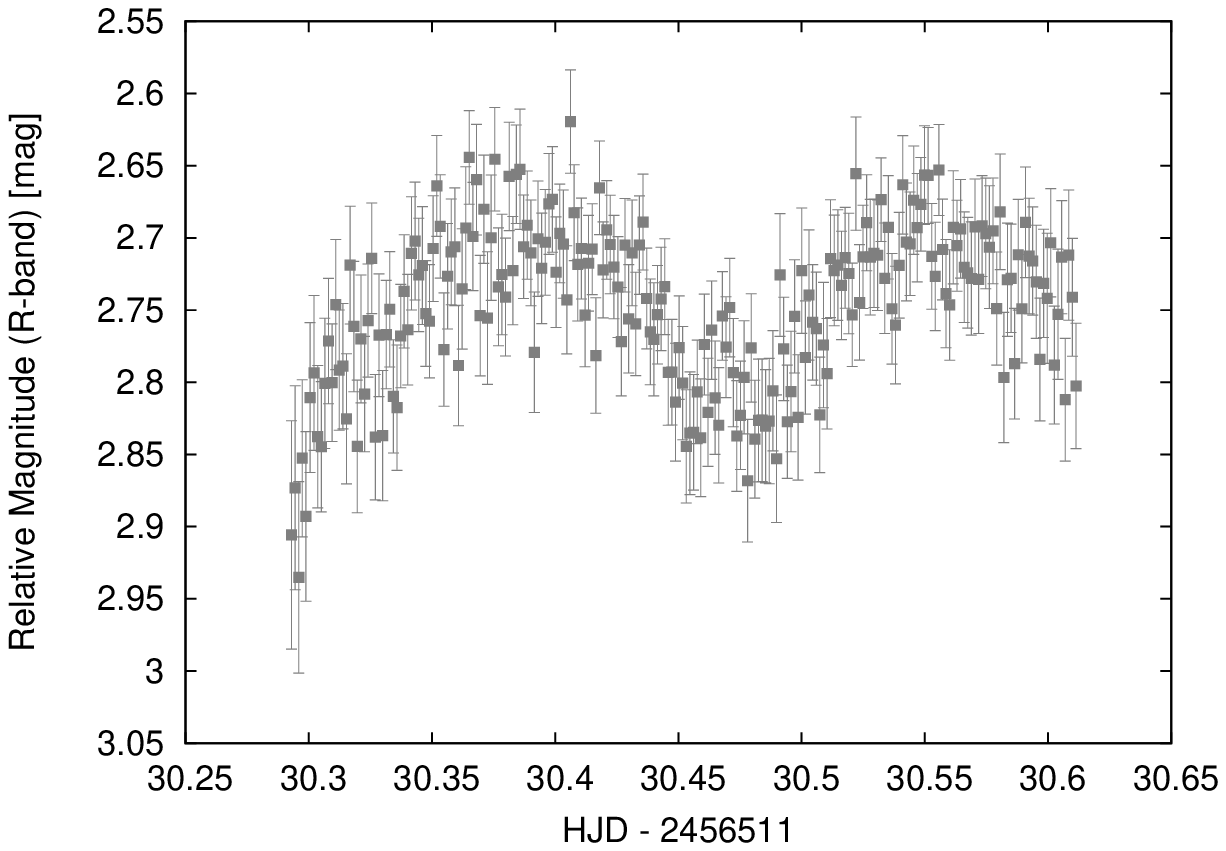}  
}
\caption{}
\label{Fig. 1}
\end{figure*} 

\end{document}